\documentclass[
 reprint,
superscriptaddress,
nofootinbib,
 amsmath,amssymb,
 aps,
prd,
floatfix,
longbibliography
]{revtex4-2}

\usepackage{orcidlink}
\usepackage[english]{babel}
\usepackage{xcolor}
\usepackage{physics}
\usepackage{cancel}
\usepackage[normalem]{ulem}
\usepackage{graphicx}
\usepackage{dcolumn}
\usepackage{bm}
\usepackage[separate-uncertainty=true,multi-part-units=single]{siunitx}
\usepackage{hyperref}
\hypersetup{
     colorlinks=true,
     breaklinks=true,
     linkcolor=blue,
     filecolor=blue,
     citecolor=blue,      
     urlcolor=cyan,
     pdfencoding=auto
     }

\usepackage{subfigure}
\usepackage{multirow}
\usepackage{booktabs}
\usepackage{tabularx}

\newcommand{\LCDM}{\Lambda\mathrm{CDM}}
\newcommand{\cdm}{\mathrm{cdm}}
\newcommand{\YHe}{Y_{\mathrm{He}}}

\AtBeginDocument{\RenewCommandCopy\qty\SI}

\DeclareSIUnit[quantity-product = {}]\parsec{\text{pc}}
\DeclareSIUnit[quantity-product = {}]\year{\text{yr}}
\DeclareSIUnit[quantity-product = {}]\hmpc{\,\textit{h}\,\text{Mpc}^{-1}}
\DeclareSIUnit[quantity-product = {}]\mpch{\,\textit{h}^{-1}\,\text{Mpc}}

\graphicspath{{./}{figures/}}

\begin{document}

\title{Signatures of Very Early Dark Energy in the Matter Power Spectrum}

\author{Alexander C. Sobotka\,\orcidlink{0000-0002-7576-5417}}
\email{asobotka@live.unc.edu}
\affiliation{Department of Physics and Astronomy, University of North Carolina at Chapel Hill,\\
Phillips Hall CB3255, Chapel Hill, North Carolina 27599, USA }

\author{Adrienne L. Erickcek\,\orcidlink{0000-0002-0901-3591}}
\email{erickcek@physics.unc.edu}
\affiliation{Department of Physics and Astronomy, University of North Carolina at Chapel Hill,\\
Phillips Hall CB3255, Chapel Hill, North Carolina 27599, USA }

\author{Tristan L. Smith\,\orcidlink{0000-0003-2685-5405}}
\email{tsmith2@swarthmore.edu}
\affiliation{Department of Physics and Astronomy, Swarthmore College, Swarthmore,\\ Pennsylvania 19081, USA}

\begin{abstract}

Axion-like scalar fields can induce temporary deviations from the standard expansion history of the universe. The scalar field’s contribution to the energy density of the universe grows while the field is held constant by Hubble friction, but when the scalar field starts to evolve, its energy density decreases faster than the radiation density for some potentials. We explore the observational signatures of such a scalar field that becomes dynamical between big bang nucleosynthesis and matter-radiation equality, which we call very Early Dark Energy (vEDE). If vEDE momentarily dominates the energy density of the universe, it generates a distinctive feature in the matter power spectrum that includes a bump on scales that enter the horizon just after the scalar field starts to evolve.  For $k \gtrsim 10\,h\,\text{Mpc}^{-1}$, the amplitude of this bump can exceed the amplitude of the standard matter spectrum.  The power on scales on either side of this peak is suppressed relative to the standard power spectrum, but only scales that are within the horizon while the scalar field makes a significant contribution to the total energy density are affected.  We determine how vEDE scenarios are constrained by observations of the cosmic microwave background, measurements of the primordial deuterium abundance, and probes of the late-time expansion history. We find that these observations are consistent with vEDE scenarios that enhance power on scales $k \gtrsim 30\,h\,\text{Mpc}^{-1}$ and nearly double the amplitude of the matter power spectrum around $200\,h\,\text{Mpc}^{-1}$. These scenarios also suppress power on scales between $0.3\,h\,\text{Mpc}^{-1}$ and $30\,h\,\text{Mpc}^{-1}$.

\end{abstract}

\maketitle

\section{Introduction} \label{sec:Introduction}

String theory generally includes the existence of an ensemble of light axion-like scalar fields that span a wide range of masses and comprise a ``string axiverse" \cite{svrcek_cosmological_2006,svrcek_axions_2006, arvanitaki_string_2010, marsh_axion_2016}. For a given axion, the expansion of the universe (i.e.\ Hubble friction) keeps the field fixed at its initial value until the expansion rate drops below a threshold dependent on the axion mass, at which point the field becomes dynamical and its energy density decreases rapidly. While frozen, the energy density of the field acts as a cosmological constant and can come to dominate the energy density of the universe, triggering a period of accelerated expansion. Therefore, an ensemble of axion-like fields with different masses could give rise to multiple periods of accelerated expansion throughout cosmic time - a picture that potentially alleviates the coincidence problem \cite{dodelson_solving_2000,griest_toward_2002,kamionkowski_dark_2014,karwal_early_2016}.

The string axiverse, including scalar fields with generalized axion-like potentials, has received a great deal of interest due to its ability to address the Hubble tension: a discrepancy between local measurements of the present-day expansion rate ($H_0$) \cite{yuan_consistent_2019,wong_h0licow_2020,blakeslee_hubble_2021,riess_comprehensive_2022} and the value inferred from observations of the cosmic microwave background (CMB) \cite{planck_collaboration_parameters}. Early Dark Energy (EDE), which becomes dynamical around the time of matter-radiation equality \cite{karwal_early_2016, poulin_cosmological_2018-1}, has been proposed as a solution to the Hubble tension \cite{poulin_early_2019, smith_oscillating_2020, murgia_early_2021, smith_hints_2022}. A rotating axion field whose kinetic energy exceeds its potential energy around the time of recombination can slightly reduce the Hubble tension while also providing a baryogenesis mechanism when coupled to right-handed neutrinos \cite{co_axion_2024}.

Furthermore, recent baryon acoustic oscillation (BAO) measurements from the DESI collaboration suggest that the dark energy that dominates the universe today may be thawing, with the dark energy equation of state increasing from $w=-1$ \cite{desi_collaboration_desi_2024-1}, but see also \cite{cortes_interpreting_2024-1, colgain_does_2024-1, wolf_underdetermination_2023, wolf_scant_2024, colgain_desi_2024}. This thawing behavior is expected of a scalar field that is starting to become dynamical \cite{berghaus_quantifying_2024}. If a scalar field’s contribution to the energy density in the early universe is the resolution to the Hubble tension and if present-day dark energy is sourced by a scalar field, it seems likely that other scalar fields would be present in the early universe (e.g.\ \cite{Rezazadeh_cascading_2024}). 

We explore the observational signatures of an axion-like scalar field that becomes dynamical between big bang nucleosynthesis (BBN) and matter-radiation equality. We refer to this scenario as very Early Dark Energy (vEDE). The presence of vEDE alters the growth of dark matter (DM) density fluctuations, but leaves the CMB minimally altered if the scalar field becomes dynamical sufficiently early. Such vEDE scenarios can leave distinctive imprints on the matter power spectrum that, if observed, would provide additional evidence for the existence of a string axiverse.  

Scalar fields in the early universe have been shown to affect the growth of structure. A period of kination \cite{spokoiny_deflationary_1993-1, joyce_electroweak_1997-1, ferreira_cosmology_1998-2,co_gravitational_2022}, in which the universe is dominated by a fast-rolling scalar field, enhances the growth rate of DM density perturbations by affecting how quickly the expansion rate decreases \cite{redmond_growth_2018,delos_how_2023}. More broadly, scalar fields that undergo a phase of rapidly diluting energy density (RDED), during which the energy density decreases faster than radiation, have been shown to produce a bump in the matter power spectrum \cite{jaber_imprint_2020,de_la_macorra_cosmological_2021}. For example, a RDED phase manifests in bound dark energy (BDE) \cite{de_la_macorra_testing_2018, almaraz_bound_2019}, in which the BDE energy density initially tracks that of Standard Model radiation and then decreases rapidly. Rotating axion fields can also experience an RDED phase that enhances power on scales that enter the horizon while the axion field is dominant \cite{co_gravitational_2022}. 
In contrast, the RDED phase associated with EDE does not lead to an enhancement of the matter power spectrum. When all other cosmological parameters are held fixed, adding EDE suppresses the matter power spectrum for $k \gtrsim 10^{-2} \SI{}{\hmpc} $ \cite{poulin_ups_2023}.\footnote{The EDE models that alleviate the Hubble tension have best-fit values for the spectral tilt and dark matter density that differ from their $\LCDM$ values, and these changes lead to the enhancement of the small-scale matter power spectrum that is often associated with EDE \cite{klypin_clustering_2021,mcdonough_observational_2023-1, goldstein_canonical_2023}.} Therefore, the impact of RDED on the matter power spectrum seems to be more complicated than the enhanced growth rate during an RDED phase would suggest.

We find that an enhancement in the matter power spectrum is only achieved if the scalar field fast-rolls while its energy density dominates the Hubble rate \textit{and} the field becomes dynamical well before matter-radiation equality. Under these conditions, DM density fluctuations on scales that enter the horizon just after the field becomes dynamical experience a sustained period of enhanced growth, which results in power on these scales exceeding that of the standard matter power spectrum. The power on scales on both sides of this enhancement is reduced compared to the standard power spectrum. However, only the scales that are within the horizon while the scalar field significantly affects the Hubble rate are impacted. Additionally, we find that enhancements to the matter power spectrum are not generated by certain shapes of the scalar field potential. Since EDE models that alleviate the Hubble tension employ relatively shallow potentials and scalar fields that become dynamical around the time of matter-radiation equality, these models do not enhance the matter power spectrum when the $\LCDM$ cosmological parameters are held fixed. 

In contrast, scalar fields that become dynamical well before matter-radiation equality can enhance the small-scale matter power spectrum without affecting scales accessible to the CMB. Enhancing the matter power spectrum on small scales (i.e.\ $k \gtrsim \SI{1}{\hmpc}$) could address the discrepancy between predictions of the $\LCDM$ model and the apparent excess of massive high-redshift galaxies observed by the James Webb Space Telescope (JWST) \cite{padmanabhan_alleviating_2023-1, Tkachev_Excess_2024}. Additionally, observations of Milky Way satellites suggest that dwarf galaxies may be more concentrated than predicted by $\LCDM$ \cite{esteban_milky_2023}, hinting at the need for an enhanced structure on small scales. Conversely, Lyman-$\alpha$ measurements from the extended Baryon Oscillation Spectroscopic Survey (eBOSS) \cite{lyke_sloan_2020, chabanier_one-dimensional_2019} demonstrate a preference for a suppression in power, rather than an enhancement, compared to $\LCDM$ on scales $k \simeq \SI{1}{\hmpc}$ \cite{Bird_2023, Fernandez_2024, walther_emulating_2024, ivanov_fundamental_2025, rogers_5_2024, He_Fresh_2025}.

We determine the range of vEDE scenarios that are consistent with observations of the CMB, probes of BAO, uncalibrated Type Ia supernovae, and measurements of the abundance of primordial elements. We find that these observations are consistent with vEDE scenarios that enhance the matter power spectrum by roughly $90\%$ on scales around $\SI{200}{\hmpc}$. This enhancement in the matter power spectrum is a direct result of the scalar field's impact on the pre-recombination expansion rate. Therefore we expect that any mechanism that generates a qualitatively similar expansion history to that of vEDE can enhance the matter power spectrum while maintaining consistency with CMB observations and other probes of the expansion history.

This paper is organized as follows. In Sec.~\ref{sec:Scalar_Field_Model} we introduce the vEDE model with an axion-like potential. Section \ref{sec:Effects_on_the_Growth_of_DM_structures} contains a discussion of the effects that a vEDE cosmology has on the growth of DM perturbations, which culminates in a unique matter power spectrum that we discuss in Sec.~\ref{sec:Effects_on_the_Matter_Power_Spectrum}. We detail the effects of vEDE on the CMB temperature anisotropy spectrum  in Sec.~\ref{sec:Effects_on_the_Temperature_Spectrum}. A discussion of the constraints placed on vEDE from observations of the CMB, BAO, uncalibrated Type Ia supernovae, and BBN abundances is provided in Sec.~\ref{sec:MCMC_results}. In Sec.~\ref{sec:Summary_and_Conclusions}, we summarize our results and discuss how probes of the small-scale matter power spectrum could detect the distinctive features of vEDE. Appendix \ref{sec:appendix_small-scale_suppression} contains a model for the evolution of density perturbations that enter the horizon before vEDE alters the Hubble rate, and Appendix \ref{sec:appendix_supplemental_mcmc_results} contains supplemental results from our analysis of observational constraints on vEDE. Throughout this work, we use natural units ($c=\hbar=k_b=1$).

\section{\MakeLowercase{v}EDE Model} \label{sec:Scalar_Field_Model}
We employ an axion-like scalar field similar to that used in EDE scenarios \cite{karwal_early_2016,poulin_cosmological_2018-1,smith_oscillating_2020}, in which an ultra-light scalar field ($\phi$) rolls down a potential of the form
\begin{equation}
    V_n(\phi) = m^2 f^2 \left[ 1 - \cos(\phi/f) \right]^{n}. \label{eq:V(phi)}
\end{equation}
Here, $f$ is the decay constant of the field, $m$ is the mass of the scalar field, and $n$ is the power-law index that dictates the steepness of the potential. While the $n=1$ axion potential naturally arises in string theory \cite{arvanitaki_string_2010,marsh_axion_2016}, higher powers of $n$ may be generated by higher-order instanton corrections \cite{Abe_natural_inflation_2015,kappl_2016,mcdonough_observational_2023-1,mcdonough_towards_2023}. The field is initially fixed at a value of $\theta_i = \phi_i/f$ via Hubble friction, acting as a fluid with an equation of state $w_\phi \simeq -1$. Then, once the Hubble rate drops below $(d^2V_n/d\phi^2)^{1/2} \propto m$, the field becomes dynamical and begins to oscillate around the bottom of the potential. During this oscillation phase, the energy density of the field ($\rho_\phi$) alternates between being dominated by either kinetic energy or potential energy. If $n > 1$, the oscillations are anharmonic and, as $n$ grows, the oscillation frequency decreases such that $\rho_\phi$ is dominated by kinetic energy for longer periods of time \cite{poulin_early_2019,smith_oscillating_2020}. When the scalar field is dominated by kinetic energy, $\rho_\phi \propto a^{-6}$, where $a$ is scale factor. When the energy density is dominated by potential energy, the field is subject to a brief period of Hubble friction and $\rho_\phi$ is constant. Unless specified otherwise, we fix $n = 8$ to ensure extended periods of fast roll while $\rho_\phi$ is dominant. As we discuss in Sec.~\ref{sec:Effects_on_the_Matter_Power_Spectrum}, we find that scalar fields with $n \lesssim 6$ oscillate too quickly to produce significant enhancements in the matter power spectrum.

\begin{figure}[t]
\centering
  \includegraphics[width=\linewidth]{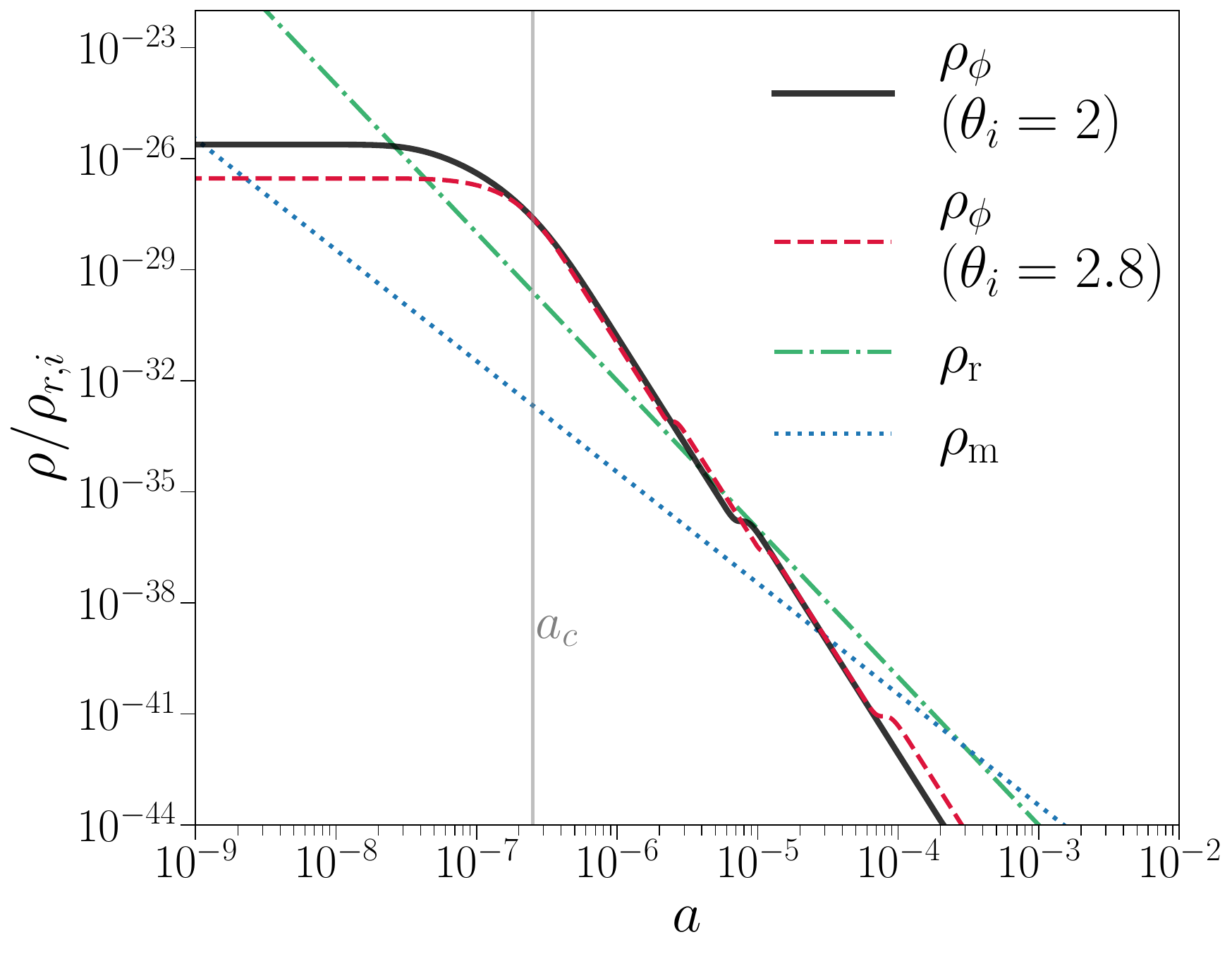}
  \caption{\footnotesize Solid and dashed curves show the energy density of a scalar field with $R_\phi = 99$, $a_c = 10^{-6.6}$, and $n = 8$ but different values of $\theta_i$. The energy density of the scalar field is constant until $a \sim a_c$ (marked by the vertical solid line), after which $\rho_\phi \propto a^{-6}$. Energy densities for standard radiation and non-relativistic matter are shown by the dot-dashed and dotted lines, respectively. All curves are normalized by the radiation energy density at some initial time, $\rho_{r,i}$.}
  \label{fig:rho}
\end{figure}

At fixed $n$ and $\theta_i$, the evolution of the scalar field energy density is determined by $a_c$ and $f_\phi \equiv \rho_\phi(a_c)/\rho_{tot}(a_c)$ \cite{poulin_cosmological_2018-1}, where $a_c = 1/(1+z_c)$ is the critical scale factor at which the scalar field contribution to the total energy density is maximized, and  $\rho_{tot}$ is the total energy density of the universe including $\rho_\phi$. The parameter $f_\phi$, however, is a poor measure of the physical impact of the scalar field energy density when $f_\phi$ approaches unity. For example, increasing $f_\phi$ from $0.9$ to $0.99$ translates to $\rho_\phi(a_c)$ increasing by an order of magnitude. Instead, we parameterize the scalar field contribution to the total energy density with the quantity
\begin{equation}
    R_\phi \equiv \frac{\rho_\phi(a_c)}{\rho_{tot}(a_c) - \rho_\phi(a_c)} = \frac{f_\phi}{1 - f_\phi}, 
    \label{eq:R_phi}
\end{equation}
which better correlates to the impact of $\rho_\phi$. Figure \ref{fig:rho} shows an example of the evolution of $\rho_\phi$ resulting from $R_\phi = 99$, $a_c = 10^{-6.6}$, $n=8$, and different values of $\theta_i$. In this scenario, $\rho_\phi$ comes to dominate the total energy density of the universe. Just before $a_c$, the field becomes dynamical and $\rho_\phi$ begins to decrease as $a^{-6}$. For the case of $\theta_i=2$ (solid line), the energy density becomes dominated by potential energy  near a scale factor of $a\approx 10^{-5}$ and Hubble friction leads to a brief period of constant $\rho_\phi$ before the field fast rolls again. Setting $\theta_i=2.8$ (dashed line) increases the oscillation frequency, and pauses in the evolution of $\rho_\phi$ occur at scale factors of roughly $2.5\times10^{-6}$, $1\times10^{-5}$, and $8\times10^{-5}$.

A fast rolling scalar field is difficult to achieve with potentials other than Eq.~\eqref{eq:V(phi)} if the field is initially fixed by Hubble friction. For an inverse power law, $V(\phi) \propto \phi^{-n}$, the slow-roll parameters $\eta$ and $\epsilon$ are both proportional to $1/\phi^2$. Even if the slow-roll condition of $\eta, \epsilon \ll 1$ is briefly violated, the shape of the runaway potential will force the value of $\phi$ to increase and slow roll will be restored. An inverse power law potential cannot induce fast roll unless the scalar field is initialized in a fast-rolling phase (as in BDE \cite{almaraz_bound_2019}). Similarly, axion kination achieves a $w=1$ phase with an $n=1$ potential following a cosmological history that leaves the axion in a rotating rather than oscillating state \cite{co_axiogenesis_2020,co_gravitational_2022}. 

This work focuses on scenarios in which $a_c$ occurs before matter-radiation equality. We employ a modified version of the Cosmic Linear Anisotropy Solving System ({\tt CLASS}) \cite{blas_cosmic_2011} Boltzmann solver known as {\tt AxiCLASS}\footnote{\url{https://github.com/PoulinV/AxiCLASS}} \cite{poulin_cosmological_2018-1,smith_oscillating_2020} to calculate the background and perturbative dynamics of this vEDE cosmology.

\section{Effects on the Growth of Dark Matter structures} \label{sec:Effects_on_the_Growth_of_DM_structures}
The presence of a scalar field alters the expansion rate and, as a result, some perturbation modes experience a delay in the time of horizon entry. The change in the expansion rate also alters the growth rate of the fractional density perturbation of DM ($\delta_\cdm$) on subhorizon scales. Throughout this work, we use a perturbed Friedmann–Lema\^itre–Robertson–Walker (FLRW) metric in conformal Newtonian gauge given by
\begin{equation}
    ds^2 = a(\tau)^2\left[ -(1+2\Psi)d\tau^2 + \delta_{ij}\left(1 + 2\Phi \right)dx^i dx^j  \right], 
\end{equation}
where $\Phi$ and $\Psi$ are scalar perturbations to the metric and $\tau$ is conformal time. 

Consider a perturbation mode of wave number $k_1$ that enters the cosmological horizon at a scale factor $a_{1}$ when $k_1 = a_{1} H_{\LCDM}(a_{1})$, where $H_{\LCDM}$ is the expansion rate in a $\LCDM$ cosmology. If the expansion rate at $a_1$ increases as the result of vEDE, then that same mode of wave number $k_1$ will be outside the horizon at $a_1$ ($k_1 < a_1 H_{\mathrm{vEDE}}(a_1)$). As $H_{\mathrm{vEDE}}$ decreases, eventually $k_1 = a_2 H_{\mathrm{vEDE}}(a_2)$, where $a_2 > a_1$. Therefore, increasing the expansion rate compared to $\LCDM$ delays horizon entry for a given mode, which will act to suppress the amplitude of that mode.

The rate at which $H(a)$ decreases dictates the subhorizon growth rate of $\delta_{\cdm}$. Upon horizon entry of a perturbation mode, DM particles are initially accelerated by the gravitational potential $\Phi$. If horizon entry occurs during radiation domination, the amplitude of $\Phi$ quickly decays after horizon entry and begins to oscillate around zero. Although there is no longer a net gravitational pull on DM particles, they continue to drift toward initially overdense regions. The subsequent comoving displacement of these massive particles is given by 
\begin{equation}
    \vec{s} = \int \vec{v}\frac{dt}{a} \propto \int \frac{da}{a^3 H(a)},  \label{eq:comoving_distance}
\end{equation}
where $t$ is proper time and $\vec{v}\propto 1/a$ is the physical velocity of DM particles. At linear order, \mbox{$\delta_\cdm = -\vec{\nabla}\cdot\vec{s}$} and so the growth of $\delta_\cdm$ is set by the amount of comoving distance that DM particles with a given  velocity are able to cover when drifting toward initially overdense regions \cite{redmond_growth_2018}. During radiation domination $H(a) \propto a^{-2}$, so Eq.~\eqref{eq:comoving_distance} implies $\vec{s}\propto\ln(a)$ and $\delta_\cdm$ grows logarithmically. If $H(a)$ decreases \textit{slower} than $a^{-2}$, then the expansion rate of the comoving grid is faster than the comoving drift velocity of DM and $\delta_{\cdm}$ will not grow.\footnote{Note that linear growth of $\delta_\cdm$ during matter domination is not due to the DM comoving drift velocity, but rather the fact that $\Phi$ is non-zero and constant during matter domination.} On the other hand, $H(a)$ decreasing \textit{faster} than $a^{-2}$ means the expansion of the comoving grid slows down more quickly than in radiation domination and DM particles of a given velocity are able to travel a larger comoving distance, leading to an increased growth rate of $\delta_\cdm$ \cite{redmond_growth_2018}. We define the deviation from the standard $\LCDM$ expansion rate due to the presence of the scalar field as 
\begin{equation}
    \frac{\Delta H}{H} \equiv \frac{H_{\mathrm{vEDE}}}{H_{\LCDM}} - 1,\label{eq:delta_H}
\end{equation}
and we refer to the time period with $\Delta H/H \gtrsim 10^{-3}$ as the vEDE era. This threshold of $\Delta H/H \gtrsim 10^{-3}$ roughly coincides with the times at which $p = d\ln H/d\ln a$ deviates from the standard values of $p= -2$ during radiation domination or $p = -3/2$ during matter domination. 

\begin{figure}[t]
\centering
    \includegraphics[width=\linewidth]{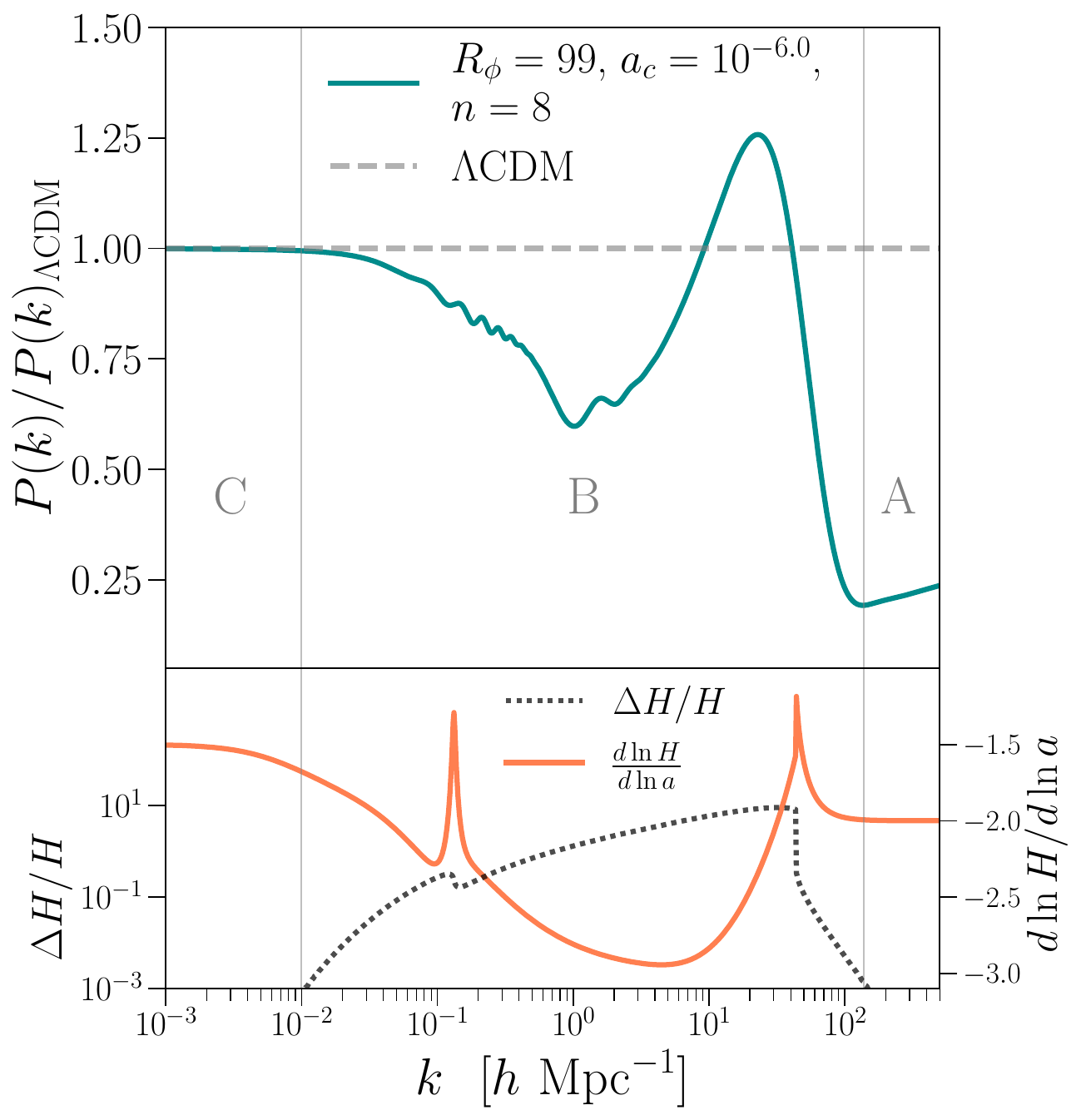}
\caption{\footnotesize The top panel shows how vEDE alters the present-day matter power spectrum for an example vEDE cosmology with $R_\phi=99$, $a_c = 10^{-6}$, $n=8$, and $\theta_i =2$. For both $\LCDM$ and vEDE, we assume \textit{Planck} 2018 TT,TE,EE,lowE best-fit values for $\omega_b$, $\omega_{\cdm}$, $h$, $A_s$, $n_s$, and $\tau_{reio}$ \cite{planck_collaboration_parameters}. The dotted and solid curves in the bottom panel, respectively, show the values of $\Delta H/H$ defined by Eq.~\eqref{eq:delta_H}, and $d\ln H/d\ln a$ at the time of horizon entry for each mode. The evolution of modes in Regimes A, B, and C are detailed in Secs.~\ref{sec:regime_A}, \ref{sec:regime_B}, and \ref{sec:regime_C}, respectively. }
  \label{fig:example_transfer}
\end{figure}

The combination of delayed horizon entry and the altered growth of $\delta_\cdm$ leads to perturbation modes being either suppressed or enhanced compared to $\LCDM$, which culminates in changes to the matter power spectrum such as those shown in top panel of Fig.~\ref{fig:example_transfer}. The dotted curve in the bottom panel of Fig.~\ref{fig:example_transfer} depicts the value of $\Delta H/H$ at the time of horizon entry for each mode, while the solid line shows the corresponding values of $d\ln H/d\ln a$. The example in Fig.~\ref{fig:example_transfer} can be divided into three regimes. Regime A ($k \gtrsim \SI{137}{\hmpc}$) contains modes that enter the horizon before the vEDE era so that the time of horizon entry for these modes is unaffected by the scalar field. Modes in Regime B ($\SI{0.01}{\hmpc} \lesssim k \lesssim \SI{137}{\hmpc}$) enter the horizon during the vEDE era and thus these modes experience both horizon entry delay and an altered subhorizon growth of $\delta_\cdm$. Finally, Regime C ($k \lesssim \SI{0.01}{\hmpc}$) contains modes that enter the horizon after the vEDE era: these modes are unaffected by vEDE. In the following subsections, we detail the possible evolutions of $\delta_\cdm$ in these three regimes.

\begin{figure*}[t]
\centering
    \includegraphics[width=\linewidth]{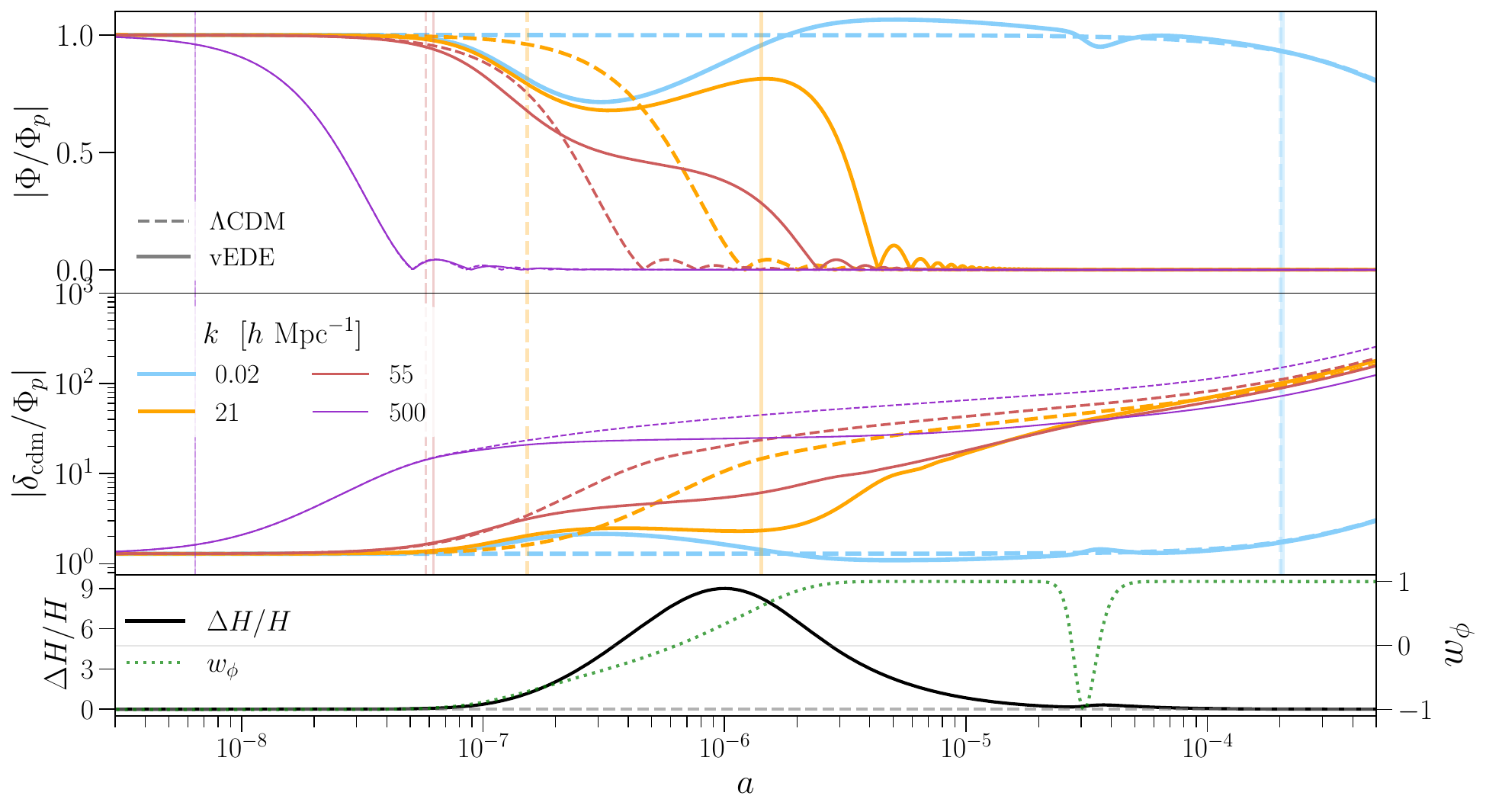}
  \caption{\footnotesize Evolution of gravitational potential ($\Phi$) and DM density perturbation ($\delta_{\cdm}$) for different wavelength modes, normalized by the initial value of $\Phi$ deep in radiation domination ($\Phi_p$). The vertical lines mark horizon entry for each mode. Solid lines result from a vEDE scenario with $R_\phi = 99$, $a_c = 10^{-6}$, $n=8$, and $\theta_i=2$, while dashed lines are $\LCDM$. The solid line in the bottom panel shows the deviation from the standard $\LCDM$ expansion rate defined by Eq.~\eqref{eq:delta_H}, while the dotted line depicts the evolving equation of state of the scalar field. For both $\LCDM$ and vEDE, we assume \textit{Planck} 2018 TT,TE,EE,lowE best-fit values for $\omega_b$, $\omega_\cdm$, $h$, $A_s$, $n_s$, and $\tau_{reio}$ \cite{planck_collaboration_parameters}. The altered evolution of these modes results in the matter power spectrum shown in Fig.~\ref{fig:example_transfer}.  }
  \label{fig:Phi_delta_cdm_H}
\end{figure*}

\subsection{Regime A}  \label{sec:regime_A}
For small-scale modes that enter the horizon before the vEDE era, vEDE only affects the evolution of $\Phi$ after it is oscillating and insignificant (see the top panel of Fig.~\ref{fig:Phi_delta_cdm_H}). DM particles obtain a drift velocity upon horizon entry, and $\delta_{\cdm}$ grows logarithmically while $H(a) \propto a^{-2}$. Once $\rho_\phi$ becomes dominant, $H(a)$ begins to decrease at a slower rate than during radiation domination, causing the growth rate of $\delta_{\cdm}$ to be suppressed. This can be seen for the $k = \SI{500}{\hmpc}$ mode of Fig.~\ref{fig:Phi_delta_cdm_H}. The middle panel shows that $\delta_{\cdm}$ initially grows logarithmically after horizon entry but then the growth of $\delta_{\cdm}$ stops once $\Delta H/H$ exceeds unity.

As seen in Fig.~\ref{fig:Phi_delta_cdm_H}, $\delta_\cdm$ for the \mbox{$k = \SI{500}{\hmpc}$} mode appears to remain constant as $\rho_\phi$ rapidly decreases with $w_\phi = +1$, contrary to the expectation of enhanced growth when $H(a)$ decreases faster than $a^{-2}$. This apparent behavior can be understood by considering Eq.~\eqref{eq:comoving_distance}, which describes the comoving distance traveled by DM particles after horizon entry of a given mode. If a mode enters the horizon at a scale factor of $a_{hor}<a_c$, then Eq.~\eqref{eq:comoving_distance} can be represented as a piecewise integral of the form 
\begin{equation}
    \vec{s}(a) \propto \int_{a_{hor}}^{a_c}\frac{d\tilde{a}}{\tilde{a}^3 H(\tilde{a})} + \int_{a_{c}}^{a}\frac{d\tilde{a}}{\tilde{a}^3 H(\tilde{a})}, \label{eq:piecewise_integral}
\end{equation}
when $a > a_c$. The first integral in Eq.~\eqref{eq:piecewise_integral} is simply a constant whereas the second integral grows with increasing $a$. For a perturbation mode that enters the horizon just before $a_c$, Eq.~\eqref{eq:piecewise_integral} will be dominated by the growing integral. Meanwhile, Eq.~\eqref{eq:piecewise_integral} for a mode that enters the horizon well before $a_c$ will be dominated by the constant term. As a result, the $k = \SI{500}{\hmpc}$ mode in Fig.~\ref{fig:Phi_delta_cdm_H} does not appear to grow when $a \gtrsim a_c$ even though $d\delta_{\cdm}/da$ is nearly constant.

After the vEDE era, $\delta_\cdm$ resumes its standard evolution: subhorizon modes grow logarithmically during radiation domination and then transition to linear growth at matter-radiation equality. Depending on the values of $a_c$ and $R_\phi$, the end of the vEDE era may be close to the time of matter-radiation equality. The vEDE era in Fig.~\ref{fig:Phi_delta_cdm_H} ends around $a\approx 10^{-3.5}$, and so all modes quickly lock on to linear growth in $\delta_\cdm$ once the vEDE era is complete.

All modes that enter the horizon before the vEDE era ultimately end up suppressed compared to $\LCDM$. The level of suppression for these modes depends on how soon the vEDE era begins after horizon entry; a mode that enters just prior to the vEDE era is more suppressed than a mode that enters the horizon well before the vEDE era. We demonstrate this effect in Appendix \ref{sec:appendix_small-scale_suppression} by constructing a toy model of $\delta_\cdm(a)$ that consists of standard logarithmic growth during radiation domination, followed by a period of stunted growth and a subsequent period of enhanced growth. We find that starting the deviation from logarithmic growth closer to horizon entry for a mode leads to a more significant level of suppression (see Fig.~\ref{fig:toy_model}). This effect manifests in Fig.~\ref{fig:example_transfer} for $k \gtrsim \SI{137}{\hmpc}$, with the level of suppression becoming less significant as $k$ increases. 

\vspace{5em}
\subsection{Regime B} \label{sec:regime_B}
Perturbation modes that enter the horizon during the vEDE era experience a delay in the time of horizon entry compared to $\LCDM$. Additionally, the evolution of $\Phi$ for these modes is no longer similar to that of $\LCDM$. 

If a perturbation mode enters the horizon during the $w_\phi < 0$ phase of the vEDE era (e.g.\ $k = \SI{55}{\hmpc}$ in Fig.~\ref{fig:Phi_delta_cdm_H}), then the amplitude of $\Phi$ will begin to decay in a similar manner to the evolution of $\Phi$ during radiation domination. However, as the scalar field becomes dynamical, its equation of state transitions from \mbox{$w_\phi=-1$} to \mbox{$w_\phi= +1$} and thus momentarily crosses a state with $w_\phi=0$ \cite{marsh_axion_2016,poulin_cosmological_2018-1}. This evolution of $w_\phi$ delays the decay of the amplitude of $\Phi$. The non-zero $\Phi$ for \mbox{$k = \SI{55}{\hmpc}$} in Fig.~\ref{fig:Phi_delta_cdm_H} will encourage enhanced growth of $\delta_{\cdm}$. However, $H(a)$ is decreasing at a slower rate than $a^{-2}$ and this Hubble friction ultimately trumps the effects of a non-zero $\Phi$, resulting in a suppressed growth of $\delta_{\cdm}$. To illustrate this further we combine the continuity and Euler equations for DM (e.g. \cite{redmond_growth_2018}),
\begin{equation}
    \delta''_{\cdm} + \frac{3}{2}(1-w_d)\frac{\delta'_\cdm}{a} = \frac{\Phi}{a^2} \left(\frac{k}{aH}\right)^2 - \frac{9}{2}(1-w_d)\frac{\Phi'}{a}-3\Phi'', \label{eq:delta_cdm_growth}
\end{equation}
where a prime denotes differentiation with respect to scale factor, $w_d$ is the equation of state parameter of the dominant energy density of the universe, and \mbox{$H \propto a^{-3(1+w_d)/2}$}. In the limit of constant $\Phi$, Eq.~\eqref{eq:delta_cdm_growth} yields
\begin{equation}
    \delta'_\cdm(a) \propto \begin{cases}
          const., \quad & \, w_d=0 \\
          a^{-1}, \quad & \, w_d=-1/3 \\
          a^{-2}, \quad & \, w_d=-2/3 \\
          a^{-3} \ln(a), \quad & \, w_d=-1 \\
     \end{cases}.
\end{equation}
A nonzero $\Phi$ during matter domination ($w_d = 0$) leads to a constant growth rate of $\delta_\cdm$, as expected. However, that same nonzero $\Phi$ results in a diminishing growth rate in $\delta_\cdm$ when $w_d < 0$. Revisiting the \mbox{$ k = \SI{55}{\hmpc}$} mode in Fig.~\ref{fig:Phi_delta_cdm_H}, $\Phi$ is significant when $w_\phi$ is primarily less than zero and thus the growth rate of $\delta_\cdm$ is suppressed compared to what one would expect with non-negligible values of $\Phi$.

As $H(a)$ begins to decrease rapidly in the $w_\phi = +1$ phase, the amplitude of $\Phi$ decays and oscillates around zero, and DM particles begin to drift. Since $H(a)$ is decreasing more rapidly than it would during radiation domination, $\delta_{\cdm}$ grows at an enhanced rate; Eq.~\eqref{eq:comoving_distance} dictates that $\delta_{\cdm}$ grows linearly when the dominant energy density scales as $\rho \propto a^{-6}$ \cite{redmond_growth_2018}. This is demonstrated for $k = \SI{55}{\hmpc}$ in Fig.~\ref{fig:Phi_delta_cdm_H}. Note that $\delta_\cdm$ for the $k = \SI{55}{\hmpc}$ mode ultimately ends up being suppressed compared to that in $\LCDM$ despite having experienced a boosted growth rate. This is a case in which the combined effects of horizon entry delay and suppressed growth during the $w_\phi <0 $ phase win out over the phase of enhanced growth when $w_\phi = +1$. This overall suppression is seen in Fig.~\ref{fig:example_transfer} where the matter power spectrum is suppressed at $k = \SI{55}{\hmpc}$ compared to $\LCDM$.

If a perturbation mode enters the horizon during the RDED phase of the vEDE era, such as \mbox{$k = \SI{21}{\hmpc}$} in Fig.~\ref{fig:Phi_delta_cdm_H}, the amplitude of $\Phi$ will rapidly decay and oscillate around zero upon horizon entry. The effects that dictate the evolution of $\delta_{\cdm}$ for these modes are simply delayed horizon entry and enhanced growth from the rapid decrease of $H(a)$. The period of enhanced growth continues up until $\Delta H/H$ becomes negligible, at which point $\delta_\cdm$ for all subhorizon modes transitions to logarithmic growth if the universe is dominated by radiation or linear growth if the universe is matter dominated. Modes that enter the horizon just after $a_c$ will benefit from the longest period of augmented growth and can end up with $\delta_{\cdm}$ enhanced compared to $\LCDM$. Indeed, $k = \SI{21}{\hmpc}$ in Fig.~\ref{fig:Phi_delta_cdm_H} experiences a sustained period of increased growth that ultimately leads to the power spectrum in Fig.~\ref{fig:example_transfer} having the most enhancement at $ k \approx \SI{21}{\hmpc}$. 

\subsection{Regime C} \label{sec:regime_C}
Modes that enter the horizon after the vEDE era will experience no delay in horizon entry and their subhorizon growth of $\delta_{\cdm}$ will be  unaltered. Thus, the matter power spectrum on larger scales is not affected  by vEDE (e.g.\ $k \lesssim \SI{0.01}{\hmpc}$ in Figs.~\ref{fig:example_transfer} and \ref{fig:Phi_delta_cdm_H}). However, the changes to $H(a)$ during the vEDE era alter the evolution of $\Phi$ on superhorizon scales. This evolution of $\Phi$ is readily apparent for $k = \SI{0.02}{\hmpc}$ in Fig.~\ref{fig:Phi_delta_cdm_H}. The superhorizon evolution in $\Phi$ plays a key role in keeping acoustic oscillations unaltered on CMB scales (see Sec.~\ref{sec:Effects_on_the_Temperature_Spectrum}).

The superhorizon evolution of $\Phi$ is such that the curvature perturbation of each isolated species is conserved. The curvature perturbation for species $i$ is given by $\zeta_i = \Phi + \delta_i/3(1+w_i)$. During radiation domination, $\Phi$ remains constant on superhorizon scales and thus the quantity $|\delta_\phi/3(1+w_\phi)|$ remains constant at its initial value.\footnote{By default, {\tt AxiCLASS} assumes $\delta_\phi$ is initially zero. However, $\delta_\phi$ quickly matches an attractor solution on superhorizon scales \cite{ballesteros_dark_2010,poulin_cosmological_2018-1}.} When $\rho_\phi$ comes to dominate the energy density of the universe, the quantity $|\delta_\phi/3(1+w_\phi)|$ grows with increasing $a$ on superhorizon scales during the \mbox{$-1 < w_\phi < 0$} phase of the vEDE era due to the growth of $\delta_\phi$. Therefore, keeping $\zeta_\phi$ fixed requires the amplitude of $\Phi$ to decrease when the expansion is dominated by $\rho_\phi$ while $-1 < w_\phi < 0$. When the expansion is dominated by a rapidly decreasing $\rho_\phi$ with $w_\phi \simeq 1$, the amplitude of $\Phi$ will evolve toward $(9/8)\Phi_p$, where $\Phi_p$ is the initial value of $\Phi$ deep in radiation domination \cite{redmond_growth_2018}. 

\begin{figure*}[t]
    \subfigure[\footnotesize fixed $n = 8$.]{
    \label{fig:different_transfers_fixed_n}
    \includegraphics[width=0.49\linewidth]{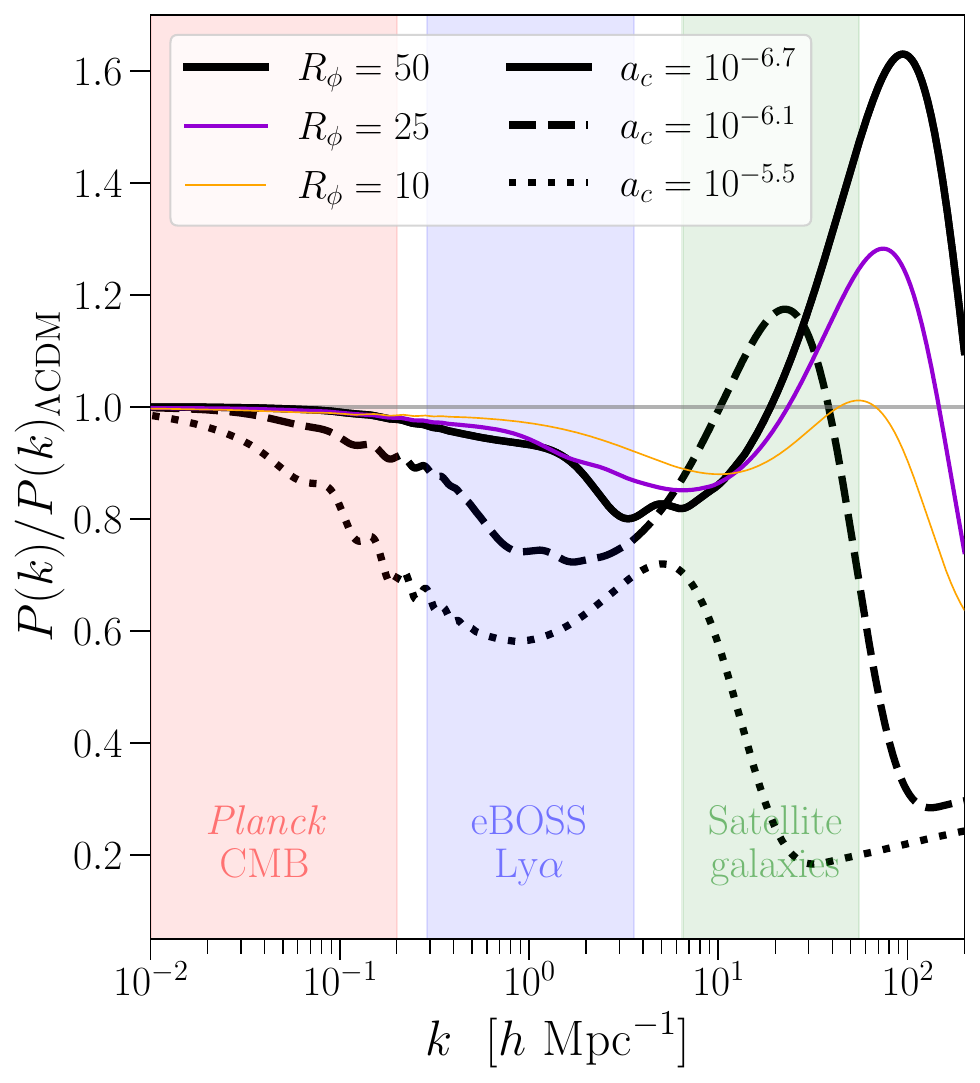}}
    \subfigure[\footnotesize fixed $R_\phi = 19$, $a_c = 10^{-6.5}$.]{
    \label{fig:different_transfers_fixed_fac}
    \includegraphics[width=0.49\linewidth]{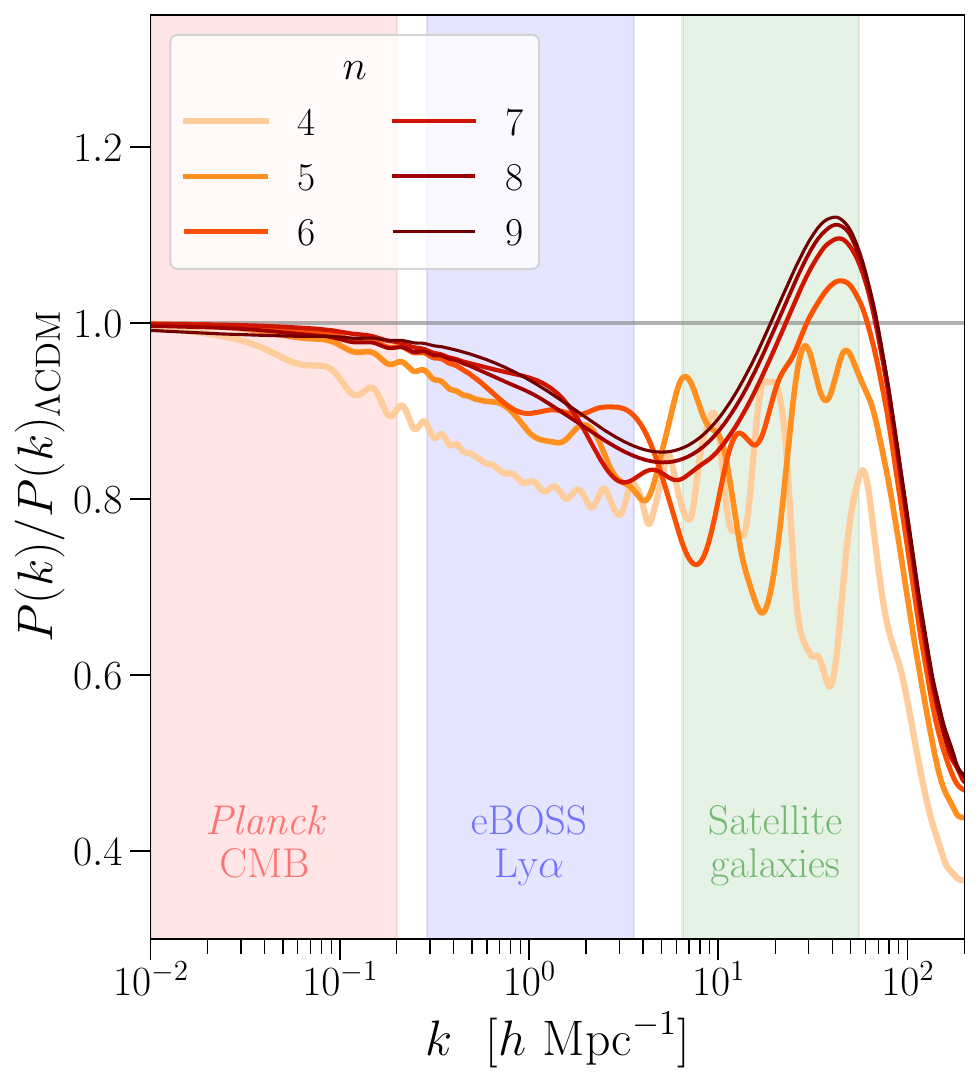}} 
    \caption{\footnotesize Present-day transfer functions resulting from vEDE with fixed $\theta_i=2$ and various values of $R_\phi$, $a_c$, and $n$. Shaded regions show the approximate range of scales to which \textit{Planck} CMB and eBOSS Ly$\alpha$ forest \cite{lyke_sloan_2020,chabanier_one-dimensional_2019} observations are sensitive, as well as scales that can be probed with observations of velocity dispersions in Milky Way satellites \cite{esteban_milky_2023}. For both $\LCDM$ and vEDE, we assume \textit{Planck} 2018 TT,TE,EE,lowE best-fit values for $\omega_b$, $\omega_\cdm$, $h$, $A_s$, $n_s$, and $\tau_{reio}$ \cite{planck_collaboration_parameters}.}
	\label{fig:different_transfers}
\end{figure*}
%

\section{Effects on the Matter Power Spectrum} \label{sec:Effects_on_the_Matter_Power_Spectrum}
The features in the matter power spectrum that result from vEDE depend on the parameters $R_\phi$ and $a_c$. Figure \ref{fig:different_transfers_fixed_n} demonstrates the transfer functions that result from various combinations of $R_\phi$ and $a_c$. The scale at which the peak of the bump is located in the matter power spectrum ($k_p$) roughly corresponds to the mode that enters the horizon just after $a_c$: $k_p \approx a_c H(a_c)$ to within a factor of 0.2 for the range of $R_\phi$ and $a_c$ considered in this work. As discussed in Sec.~\ref{sec:Effects_on_the_Growth_of_DM_structures}, such a mode enters the horizon as the RDED phase of the vEDE era begins and therefore experiences an extended period of enhanced growth. This is demonstrated by the black solid, dashed, and dotted curves in Fig.~\ref{fig:different_transfers_fixed_n}, which all have a fixed value of $R_\phi = 50$. Increasing $a_c$ alters which mode enters the horizon just after $a_c$ and thus moves the bump to larger scales. 

Increasing $a_c$ also  decreases the height of the bump. This effect is a consequence of altering the quantity $a_c/a_{eq}$, where $a_{eq}$ is the scale factor of matter-radiation equality. After matter-radiation equality, $\delta_\cdm$ grows linearly with scale factor. If the universe immediately transitions to being dominated by matter at the end of the vEDE era, then the ratio $a_c/a_{eq}$ sets the duration of the enhanced growth that results from vEDE. If $a_c/a_{eq}$ is too large, the period of enhanced growth is too short to overcome the delayed horizon entry of affected modes, and the matter power spectrum is suppressed for all scales that are affected by vEDE. If $a_c/a_{eq}$ is sufficiently small, the universe returns to a period of radiation domination after the vEDE era, at which point $\delta_\cdm$ will grow logarithmically on subhorizon scales. During radiation domination, $\delta_\cdm(a,k) = \Phi_p A_k\log(B_k\, a/a_{hor})$, where $A_k$ and $B_k$ are constants set by the evolution of each mode during horizon entry \cite{hu_small_1996}. For $\LCDM$, $A_k=9.11$ and $B_k=0.594$ for all modes that enter the horizon during radiation domination. In the context of vEDE, however, $A_k$ and $B_k$ are mode-dependent \cite{redmond_growth_2018}. For scales that enter the horizon during the vEDE era, the altered value of $B_k$ implies that $\delta_\cdm^{\text{vEDE}}/\delta^{\LCDM}_\cdm$ continues to grow between the end of the vEDE era and the start of matter domination. It is not until matter domination that $P(k)/P(k)_{\LCDM}$ becomes constant.

For a fixed value of $a_c$, varying $R_\phi$ alters the height of the bump seen in the matter power spectrum that results from vEDE. This can be seen for the solid curves in Fig.~\ref{fig:different_transfers_fixed_n}, which have varying values of $R_\phi$ and a fixed value of $a_c = 10^{-6.7}$. Decreasing $R_\phi$ reduces the period that $H(a)$ is dominated by a RDED. Therefore, the $k_p$ mode experiences shorter periods of enhanced growth as $R_\phi$ decreases, leading to a reduction in the height of the bump in the matter power spectrum. 

Increasing $R_\phi$ while keeping $n$, $\theta_i$, and $a_c$ fixed also increases the oscillation frequency of the scalar field. Increasing $R_\phi$ corresponds to increasing the amplitude, $(mf)^2$, of the potential described by Eq.~\eqref{eq:V(phi)}. For $n > 1$, oscillations are anharmonic and increasing the amplitude results in an increased oscillation frequency \cite{poulin_cosmological_2018-1}. A larger oscillation frequency makes it more likely that $\rho_\phi$ is still significantly altering the expansion rate when $w_\phi$ briefly returns to $-1$ after the initial RDED period. This leads to a secondary bump in $\Delta H/H$ following the primary enhancement to the Hubble rate at $a_c$ (e.g.\ see \mbox{$a\approx 4\times10^{-5}$} in Fig.~\ref{fig:Phi_delta_cdm_H}). Modes that enter the horizon during the secondary $\Delta H/H$ bump experience a delay in the time of horizon entry. By increasing the oscillation frequency, increasing $R_\phi$ results in the secondary $\Delta H/H$ bump occurring earlier in time and delays horizon entry for smaller scales.

For example, the [$R_\phi = 50$, $a_c=10^{-6.7}$] case in Fig.~\ref{fig:different_transfers_fixed_n} has a non-negligible secondary bump in $\Delta H/H$ that induces extra horizon entry delay for \mbox{$\SI{0.3}{\hmpc} \lesssim k \lesssim \SI{1}{\hmpc}$}, which manifests as extra suppression in the matter power spectrum compared to a similar scenario with no secondary bump in $\Delta H/H$. Decreasing $R_\phi$ to $25$ lessens the suppression effect of this secondary bump, and also decreases the oscillation frequency so that the secondary deviation in $\Delta H/H$ now affects modes in the range of \mbox{$\SI{0.2}{\hmpc} \lesssim k \lesssim \SI{0.5}{\hmpc}$}. Due to the shifting position of the second $\Delta H/H$ bump, Fig.~\ref{fig:different_transfers_fixed_n} shows that $R_\phi$ values of 10, 25, and 50 have similar impacts on CMB scales for $a_c = 10^{-6.7}$.

As discussed above, an enhancement to the matter power spectrum can only be achieved if $R_\phi$ is sufficiently large ($R_\phi \gtrsim 10$), and $a_c$ is significantly smaller than $a_{eq}$. If these criteria are met, vEDE can enhance power on scales $k \gtrsim 10\,h\,\text{Mpc}^{-1}$. Consequently, vEDE cannot explain the possible excess of massive galaxies at redshifts $z \gtrsim 10$ observed by the JWST (e.g.\ \cite{atek_revealing_2022,finkelstein_long_2022,harikane_comprehensive_2023,naidu_schrodingers_2022-1,labbe_population_2023,yan_first_2023}). \textcite{padmanabhan_alleviating_2023-1} demonstrate that an enhancement to the matter power spectrum near $ k\simeq\SI{3}{\hmpc}$ is able to reconcile these observed massive JWST galaxies with astrophysical expectations for the UV luminosity function at high redshift, while \textcite{Tkachev_Excess_2024} find that an enhancement centered at $k\simeq\SI{7}{\hmpc}$ can explain the JWST observations. We find that vEDE is not capable of producing an enhancement on scales $k \lesssim 10\,h\,\text{Mpc}^{-1}$. 

The range of values for $R_\phi$ and $a_c$ that are necessary to produce an enhancement in the matter power spectrum exclude those generally considered in the context of EDE. For EDE, \textit{Planck} CMB observations alone demonstrate a preference for $a_c \approx a_{eq}$ \cite{hill_early_2020,murgia_early_2021,mcdonough_observational_2023-1}, and the combination of CMB observations with probes of the late-time expansion history and SH0ES calibrated supernovae \cite{Riess_large_2019}
exhibits a preference for $a_c \approx a_{eq}$ and $R_\phi \simeq 0.1$  \cite{ smith_oscillating_2020, hill_early_2020,niedermann_resolving_2020, murgia_early_2021, smith_hints_2022,mcdonough_observational_2023-1}. Such a scenario cannot enhance the growth of $\delta_\cdm$. 

Furthermore, EDE studies primarily consider power-law indices of $ n = 3$ and thus do not include extended periods with $w \simeq 1$, which are necessary to support enhanced growth of DM perturbations. As discussed in Sec.~\ref{sec:Scalar_Field_Model}, the value of $n$ dictates the oscillation frequency of the scalar field, with larger values of $n$ corresponding to $\rho_\phi$ being dominated by kinetic energy for longer periods of time. While $\delta_\cdm$ will indeed experience periods of enhanced growth for all scenarios with $n>2$, the length of these periods decreases with shrinking values of $n$. 

This effect is illustrated in Fig.~\ref{fig:different_transfers_fixed_fac}, which shows the transfer functions of various vEDE scenarios with fixed $R_\phi$ and $a_c$ but varying values of $n$. In these scenarios, scales with $k \gtrsim \SI{1}{\hmpc}$ enter the horizon while $\rho_\phi$ is dominant. Even if $\rho_\phi$ is dominant and decreases faster than radiation, small values of $n$ result in high oscillation frequencies of the scalar field and prevent extended periods of enhanced growth. This manifests in Fig.~\ref{fig:different_transfers_fixed_fac}: $n=4$ has rapid oscillations and cannot achieve an enhancement to the matter power spectrum. In contrast, $n \gtrsim 6$ results in a long oscillation period and thus $H(a)$ rapidly decreases for an extended period of time.

For the vEDE scenarios depicted in Fig.~\ref{fig:different_transfers_fixed_fac}, modes of $k \lesssim \SI{1}{\hmpc}$ enter the horizon after $\rho_\phi$ becomes sub-dominant. While $\rho_\phi$ is sub-dominant, the value of $n$ determines if the field continues to oscillate. Scalar field potentials of the form $V(\phi) \propto \phi^\alpha$, with $\alpha >0$, lead to stable attractor behavior where the scalar field energy density scales as a power-law of the form $\rho_\phi \propto a^{-\beta}$ as long as 
\begin{equation}
    \alpha > 2 \left(\frac{6+\gamma}{6-\gamma}\right),
\end{equation}
where $\gamma$ is the scaling index of the dominant energy density of the universe (i.e. $\rho_d \propto a^{-\gamma}$) \cite{liddle_classification_1998}. If $\alpha$ does not satisfy this inequality (and is even) the field oscillates around the value that minimizes its potential. The index $\beta$ is fully determined by the shape of the potential and $\gamma$: $\beta = \alpha \gamma/(\alpha-2)$. 

This attractor behavior allows us to understand the evolution of the matter power spectrum as we change the value of $n$ in Fig.~\ref{fig:different_transfers_fixed_fac}. For the scalar field potential we consider here, $\alpha = 2 n$ once the field gets close to the minimum of the potential. Once $\rho_\phi$ is sub-dominant, we have $\gamma = 4$ for radiation domination. This implies that as long as $\alpha > 10$ (which corresponds to $n>5$) the vEDE field will stop oscillating after it becomes sub-dominant. This behavior is evident in Fig.~\ref{fig:different_transfers_fixed_fac}, where scenarios with $n\leq 5$ are modulated on scales $k \lesssim \SI{1}{\hmpc}$ due to the oscillations in the vEDE field, and those with $n>5$ are smoother due to the presence of the stable power-law attractor. 

The value of $n$ also affects the stability of the homogeneous scalar condensate. An scalar field that oscillates in a potential that has $V(\phi) \propto \phi^n$ near its minimum and flattens away from that minimum experiences a parametric resonance that causes the field to fragment into fluctuations, at which point the equation of state of the scalar field approaches 1/3 \cite{lozanov_equation_2017,lozanov_self-resonance_2018,smith_oscillating_2020}.  For scalar fields that dominate the energy density of the universe, the fragmentation timescale depends on $n$ and the value of the field at which the potential deviates from $V(\phi) \propto \phi^n$, which corresponds to the decay constant $f$ in Eq.~\eqref{eq:V(phi)}. Potentials that generate vEDE and EDE have \mbox{$f \gtrsim 0.01 M_{pl}$}, where $M_{pl}$ is the reduced Planck mass \cite{smith_oscillating_2020}. For such potentials, the number of e-folds that the field oscillates before fragmenting is proportional to $n+1$ for $n>1$ and $n\neq 2$ \cite{lozanov_equation_2017,lozanov_self-resonance_2018}.  Since a scalar field with $n=3$ can coherently oscillate for several e-folds before fragmenting \cite{lozanov_equation_2017,lozanov_self-resonance_2018}, we do not expect the vEDE to fragment before it becomes subdominant. 

\begin{figure}[t]
\centering
    \includegraphics[width=0.9\linewidth]{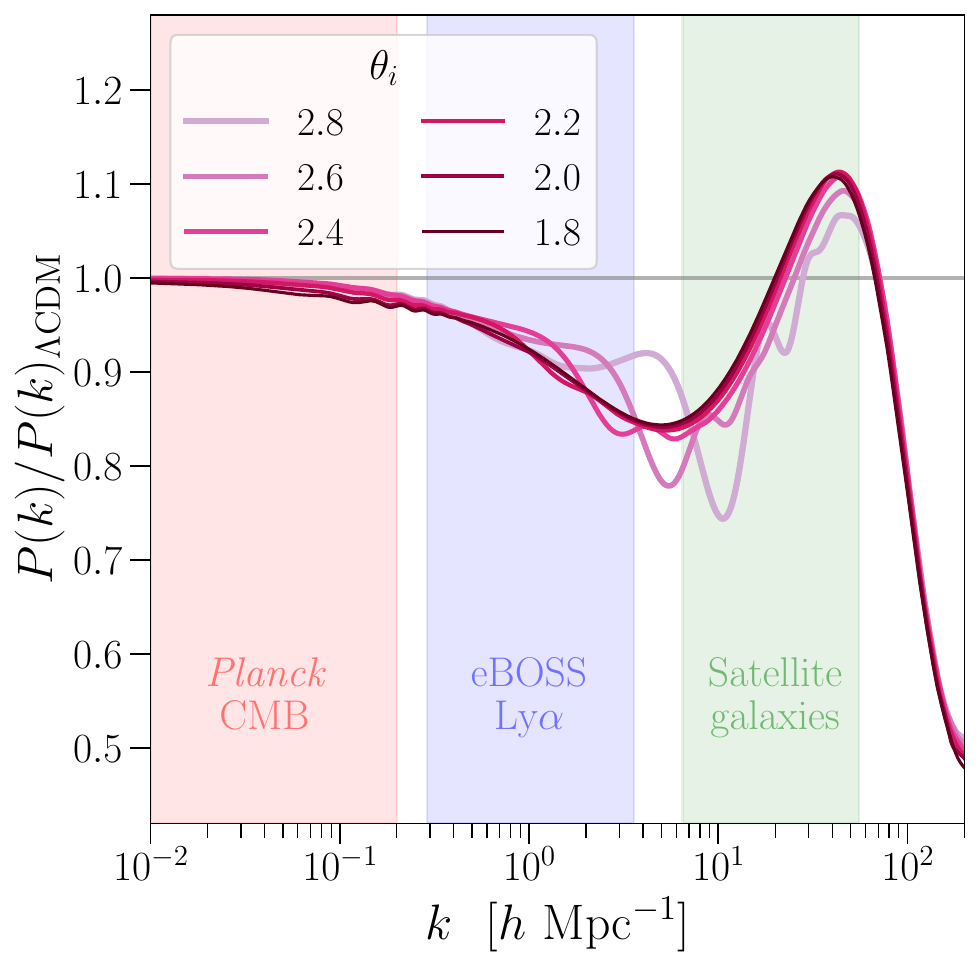}
  \caption{\footnotesize Present-day transfer functions resulting from vEDE with fixed $R_\phi = 19$, $a_c = 10^{-6.5}$, $n=8$, and various values of $\theta_i$. Shaded regions show the approximate range of scales to which \textit{Planck} CMB and eBOSS Ly$\alpha$ forest \cite{lyke_sloan_2020,chabanier_one-dimensional_2019} observations are sensitive, as well as scales that can be probed with observations of velocity dispersions in Milky Way satellites \cite{esteban_milky_2023}. For both $\LCDM$ and vEDE, we assume \textit{Planck} 2018 TT,TE,EE,lowE best-fit values for $\omega_b$, $\omega_\cdm$, $h$, $A_s$, $n_s$, and $\tau_{reio}$ \cite{planck_collaboration_parameters}.}
  \label{fig:different_transfers_theta}
\end{figure}

The matter power spectrum resulting from vEDE is also sensitive to changes in the initial field value, $\theta_i$. For fixed $R_\phi$, $a_c$, and $n$, increasing $\theta_i$ requires a decrease in the  parameters $m$ and $f$ of Eq.~\eqref{eq:V(phi)}, which results in a faster transition of the field from frozen to fast-rolling \cite{smith_oscillating_2020}. Consequently, the initial value of $\rho_\phi$ does not need to be as large in order to achieve the same $\rho_\phi(a_c)$ (see Fig.~\ref{fig:rho}). Keeping $\rho_\phi(a_c)$ fixed, this leads to the vEDE era lasting for shorter periods as $\theta_i$ grows, which ultimately inhibits the height of the bump in the matter power spectrum (see Fig.~\ref{fig:different_transfers_theta}). Larger values of $\theta_i$ also coincide with a higher oscillation frequency \cite{poulin_cosmological_2018-1,smith_oscillating_2020}, which prevents extended periods of $\rho_\phi$ rapidly decreasing and thereby limits periods of enhanced growth. 

Shaded regions in both Figs.~\ref{fig:different_transfers} and \ref{fig:different_transfers_theta} show the approximate range of wavenumbers to which \textit{Planck} CMB observations are sensitive, as well as the approximate range of scales probed by line-of-sight correlations within the Lyman-$\alpha$ forest of quasars from eBOSS \cite{lyke_sloan_2020,chabanier_one-dimensional_2019}. Additionally, measurements of the stellar velocity dispersions and half-light radii of dwarf galaxies orbiting the Milky Way can probe the linear matter power spectrum on scales of \mbox{$\SI{6}{\hmpc} \lesssim k \lesssim \SI{55}{\hmpc}$} \cite{esteban_milky_2023}, which are highlighted in Figs.~\ref{fig:different_transfers} and \ref{fig:different_transfers_theta}.

BDE is another scalar field model that has been shown to produce a bump feature in the matter power spectrum \cite{almaraz_bound_2019, de_la_macorra_testing_2018, de_la_macorra_cosmological_2021}. BDE consists of scalar field energy density that initially tracks that of radiation and then, at a condensation scale of $a_c$, the energy density transitions to diluting faster than radiation. Similar to vEDE, BDE has been found to produce an enhancement in the matter power spectrum at a scale of $k_p \approx a_c H(a_c)$. Unlike that resulting from vEDE, the matter power spectrum resulting from BDE presented in Ref.~\cite{de_la_macorra_cosmological_2021} does not exhibit suppression on either side of the bump. We expect that the evolution of perturbation modes with $k > k_p$ would indeed be qualitatively different between BDE and vEDE since BDE initially tracks radiation whereas vEDE initially behaves as a cosmological constant. However, the mechanism that induces a suppression in the matter power spectrum on scales of $k < k_p$ should be the same for both vEDE and BDE. Reference \cite{de_la_macorra_cosmological_2021} does not state if or how perturbations in the BDE scalar field were calculated, but we obtained matter power spectra that are qualitatively similar to those presented for BDE when we neglected perturbations in the vEDE scalar field. Neglecting the scalar field perturbations while including the scalar field's contribution to the background evolution results in the perturbed stress-energy tensor no longer being conserved, and thus the Bianchi identity is invalid.

\section{Effects on the CMB} \label{sec:Effects_on_the_Temperature_Spectrum}
\begin{figure}[t]
\centering
    \includegraphics[width=\linewidth]{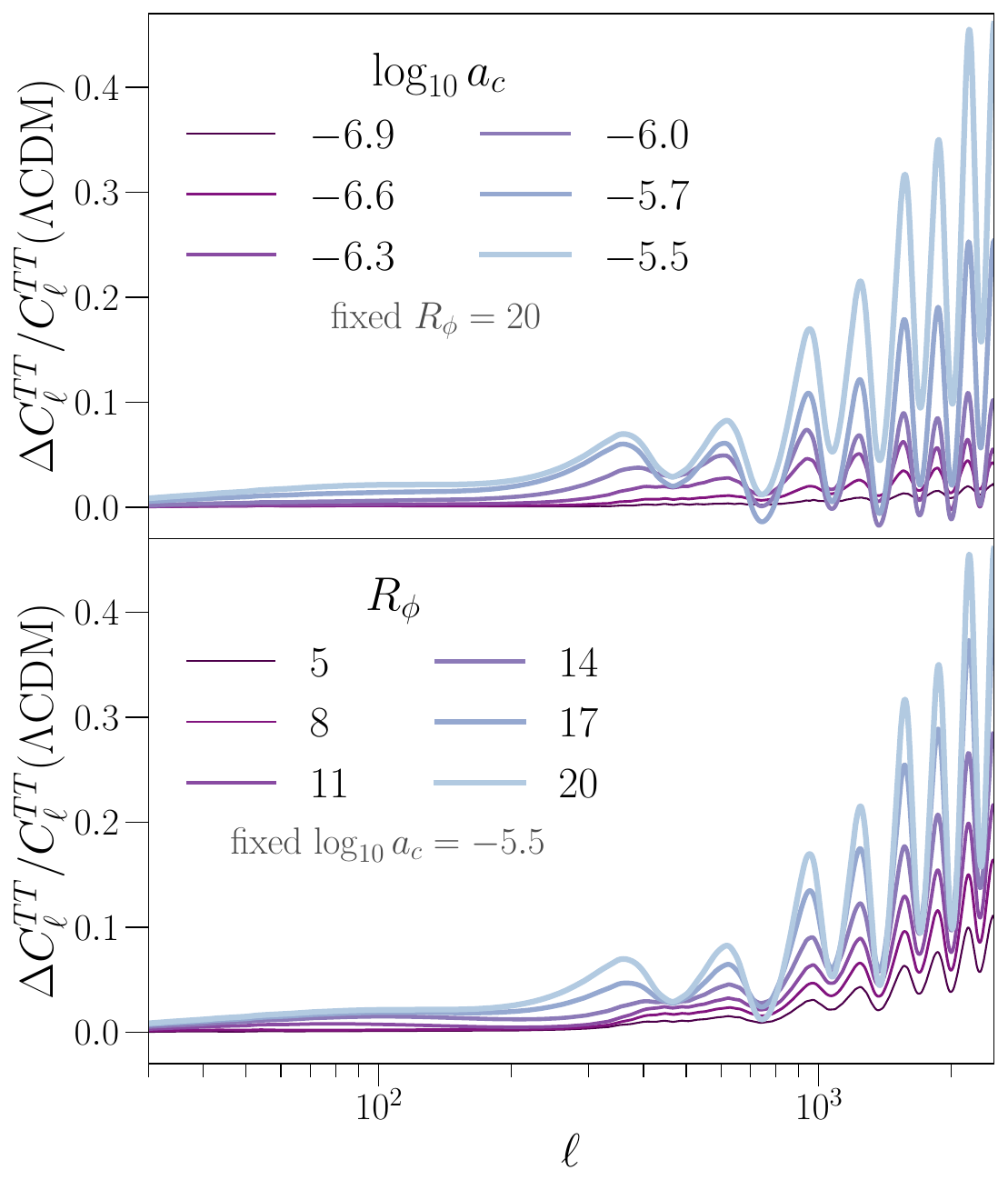}
  \caption{\footnotesize Fractional residuals comparing CMB temperature anisotropy spectra of vEDE with fixed $n=8$ and $\theta_i=2$ to that of $\LCDM$. The top panel is 
  for vEDE with fixed $R_\phi = 20$, and the bottom panel is for vEDE with a fixed $a_c = 10^{-5.5}$. The vEDE induces a scale-dependent enhancement and phase shift on small CMB scales. As $a_c$ and/or $R_\phi$ increases, these effects become more significant. For both $\LCDM$ and vEDE, we assume \textit{Planck} 2018 TT,TE,EE,lowE best-fit values for $\omega_b$, $\omega_\cdm$, $h$, $A_s$, $n_s$, and $\tau_{reio}$ \cite{planck_collaboration_parameters}, and we keep the helium abundance fixed. }
  \label{fig:Cls_residuals}
\end{figure}

As $a_c$ increases, scales that are accessible to CMB measurements begin to experience a suppression in the growth of DM perturbations due to their delayed horizon entry. This suppression is illustrated in Fig.~\ref{fig:different_transfers}, where certain vEDE scenarios exhibit reduced power for $k \lesssim \SI{0.2}{\hmpc}$. Such a suppression impacts the Sachs-Wolfe component of the CMB temperature spectrum, which is the sum of the temperature monopole $\Theta_0$ and the gravitational perturbation $\Psi$. When DM perturbations are suppressed due to vEDE, gravitational potential wells become shallower, resulting in a less negative $\Psi$ in overdense regions ($\Theta_0 > 0)$. Consequently, $\Theta_0 + \Psi$ increases, leading to an enhancement of the CMB temperature anisotropies. Since vEDE scenarios do not uniformly suppress DM perturbations across all CMB scales, this enhancement to the temperature spectrum is scale-dependent with the smallest CMB scales experiencing the most enhancement. This enhancement is more pronounced for larger values of $R_\phi$ and $a_c$, as seen in Fig.~\ref{fig:Cls_residuals}. Decreasing the baryon energy density ($\omega_b$) or increasing the primordial helium abundance ($\YHe$) can partially compensate for this increased small-scale amplitude. Decreasing $\omega_b$ or increasing $\YHe$ results in fewer free electrons at recombination, which results in more photon diffusion and more damping of small-scale anisotropies. 

Even if $a_c$ is sufficiently small such that the primary deviation from the standard Hubble rate has ended before \textit{Planck} scales enter the horizon, vEDE can still enhance small-scale CMB anisotropies through the secondary bump in $\Delta H/H$ for sufficiently large values of $R_\phi$. Decreasing $R_\phi$ pushes the secondary $\Delta H/H$ bump to later times, which affects larger CMB scales. This effect is apparent in the bottom panel of Fig.~\ref{fig:Cls_residuals}; the scenario with $R_\phi = 11$ exhibits enhancement on multipole moments $\ell \gtrsim 350$ due to the primary $\Delta H/H$ bump, but also has a slight enhancement in the temperature spectrum for $\ell \lesssim 150$ induced by the secondary bump in $\Delta H/H$.

Figure \ref{fig:Cls_residuals} also shows that vEDE induces a scale-dependent phase shift in the acoustic peaks. To explain this effect, we employ the Boltzmann equations for photons in the tightly-coupled limit. Neglecting drag effects due to baryons,\footnote{Accounting for baryons only alters the amplitude of oscillations and not the phase \cite{dodelson_scott_modern_2020}.} the acoustic oscillations of photons are governed by the equation 
\begin{equation}
    \left\{\frac{d^2}{d\tau^2} + k^2 c_s^2\right\}[\Theta_0 + \Phi] = \frac{k^2}{3}[\Phi - \Psi], \label{eq:photon_boltzmann}
\end{equation}
where $c_s$ is the sound speed. The solution to Eq.~\eqref{eq:photon_boltzmann} is the sum of the homogeneous solution and the particular solution (PS). By enforcing that both $d\Theta_0/d\tau$ and $d\Phi/d\tau$ vanish as $\tau \rightarrow 0$, the solution to Eq.~\eqref{eq:photon_boltzmann} at the time of recombination ($\tau_*$) becomes
\begin{align*}
    [\Theta_0(k, \tau_*) + \Phi(k, \tau_*)] &=  [\Theta_0(k, 0) + \Phi(k, 0)]\cos(k r_s(\tau_*)) \\
    &\hspace{3.5cm} + \mathrm{PS}(k, \tau_*), \stepcounter{equation}\tag{\theequation}\label{eq:combined_photon_solution}
\end{align*}
where $r_s$ is the sound horizon given by
\begin{equation}
    r_s(\tau) = \int_0^{\tau} c_s(\tilde{\tau})  d\tilde{\tau} 
    \hspace{0.5cm} \text{or} \hspace{0.5cm} 
    r_s(a) = \int_0^{a} \frac{c_s(\tilde{a})}{\tilde{a}^2 H(\tilde{a})} d\tilde{a}, \label{eq:r_s}
\end{equation}
and the PS is
\begin{align*}
    \mathrm{PS}(k,\tau_*) &= \frac{k}{\sqrt{3}}\int_0^{\tau_*} [\Phi(k, \tilde{\tau}) - \Psi(k, \tilde{\tau}) ] \\
    &\hspace{1.8cm} \times \sin\left[k\left(r_s(\tau_*) - r_s(\tilde{\tau})\right) \right] d\tilde{\tau}.  \stepcounter{equation}\tag{\theequation}\label{eq:particular_solution}
\end{align*}
A vEDE scenario will increase $H(a)$ before recombination and thereby decrease the size of the sound horizon at recombination. Therefore, all $k$ modes experience a phase shift compared to $\LCDM$ via the $\cos(k r_s(\tau_*))$ term in Eq.~\eqref{eq:combined_photon_solution}. However, as detailed in Sec.~\ref{sec:Effects_on_the_Growth_of_DM_structures}, modes that enter the horizon after the scalar field energy density becomes negligible also experience superhorizon evolution in $\Phi$ (e.g. $k = \SI{0.02}{\hmpc}$ in Fig.~\ref{fig:Phi_delta_cdm_H}). This superhorizon evolution in $\Phi$ (and thereby evolution in $\Psi$) during the vEDE era results in a change to the PS of Eq.~\eqref{eq:particular_solution}. We numerically determined that, if $\Phi$ and $\Psi$ return to their standard $\LCDM$ values before horizon entry, the change to the PS cancels out the phase shift caused by $\cos(k r_s(\tau_*))$. Thus, large-scale modes that enter the horizon after $\Delta H/H \rightarrow 0$ experience no net phase shift. This finding coincides with the common intuition that perturbation modes are unaffected by changes to the expansion while those modes are outside of the cosmological horizon. In contrast, small-scale modes that enter before or during the vEDE era do not have this cancellation and \textit{do} experience a net phase shift in oscillations. A similar scale-dependent phase shift in the CMB has been observed for the Wess-Zumino Dark Radiation (WZDR) model, which induces a step-like increase in the radiation energy density \cite{aloni_step_2022-2}. 

If the scalar field becomes dynamical early enough, then all scales accessible to \textit{Planck} enter the horizon after the vEDE era, and there is no phase shift in the acoustic peaks in the CMB temperature spectrum. However, at larger values of $a_c$, some of the small-scale \textit{Planck} modes enter the horizon before $\Delta H/H$ is negligible. On these scales, $\Phi$ and $\Psi$ do not return to their $\LCDM$ values before horizon entry, and there is a residual phase shift in the CMB spectrum. Figure \ref{fig:Cls_residuals} illustrates that, at fixed $R_\phi$ (top panel), increasing $a_c$ results in increasingly larger CMB scales entering the horizon when $\Delta H/H \neq 0$ and thus larger and larger scales experience a phase shift compared to $\LCDM$. For fixed $a_c$, increasing $R_\phi$ increases the period over which $\Delta H/H$ is significant, which leads to a phase shift on larger scales, as seen in the bottom panel of Fig.~\ref{fig:Cls_residuals}.

\section{\MakeLowercase{v}EDE Constraints} \label{sec:MCMC_results}
We employ measurements of the CMB, BAO, uncalibrated Type Ia supernovae, and primordial light element abundances to constrain vEDE scenarios. Although we only consider vEDE cases in which the scalar field becomes dynamical after BBN, the constant energy density of the scalar field while $w_\phi = -1$ can still significantly increase the expansion rate during BBN and thereby alter the abundance of primordial elements.\footnote{See Ref.~\cite{Mckeen_early_2024} for constraints from primordial abundances on EDE that becomes dynamical \textit{during} BBN.} Increasing the expansion rate during BBN leads to an increase in both the helium ($\YHe$) and deuterium (D/H) abundance. For example, a vEDE scenario with $R_\phi= 20$, $a_c=10^{-8}$, $n = 8$, and $\theta_i=2$ increases the expansion rate at a temperature of $0.1$ MeV by about $60\%$ compared to $\LCDM$. 

To account for this effect, we employ the BBN code {\tt PArthENoPe-v3.0} \cite{gariazzo_parthenope_2022} to create a lookup table for {\tt AxiCLASS} that reads in the baryon-to-photon ratio and the initial value of $\rho_\phi$ and provides values for abundances of helium and deuterium. These tables were created with $N_{\mathrm{eff}} = 3.044$ and a neutron lifetime of $\tau_n = 879.4$ s. As mentioned in Sec.~\ref{sec:Effects_on_the_Matter_Power_Spectrum}, increasing $\theta_i$ results in a faster transition of the field from frozen to fast-rolling, which yields a smaller initial value of $\rho_\phi$ when $R_\phi$, $a_c$, and $n$ are fixed. Decreasing $n$ while fixing $R_\phi$, $a_c$, and $\theta_i$ has the same effect. Therefore, either increasing $\theta_i$ or decreasing $n$ reduces the initial value of $\rho_\phi$ for fixed $R_\phi$ and $a_c$ and thereby lessens the impact of vEDE on primordial abundances. 

We perform a Markov Chain Monte Carlo (MCMC) analysis to quantify the range of allowed vEDE cosmologies. We employ {\tt MontePython-v3}\footnote{\url{https://github.com/brinckmann/montepython_public}}\cite{audren_conservative_2013, brinckmann_montepython_2018-1} with a Metropolis-Hastings algorithm and assume flat priors on the base six cosmological parameters \mbox{\{$\omega_b$, $\omega_{\cdm}$, $h$, $A_s$, $n_s$, $\tau_{reio}$\}}. We consider all chains with a Gelman-Rubin \cite{gelman_inference_1992} criterion of $|R-1| < 0.01$ as converged, and we perform post-processing of chains with {\tt GetDist} \cite{lewis_getdist_2019}, removing the first $30\%$ of samples as burn-in.

The following datasets are employed in our analyses:

\begin{itemize}
    \item \textbf{Planck}: \textit{Planck} 2018 high-$\ell$ and low-$\ell$ TT, TE, EE as well as CMB lensing likelihoods \cite{planck_collaboration_likelihoods}.
    %
    %
    \item \textbf{+BAO/SN}: BAO measurements at $z = 0.106$ from 6dFGS \cite{beutler_6df_2011-1},
    %
    redshift space distortions for the SDSS DR7 Main Galaxy Sample (MGS)  \cite{howlett_clustering_2015},
    %
    BAO likelihood for BOSS DR12 LOWZ and luminous red galaxy (LRG) samples at $z = 0.38, 0.51$ \cite{alam_clustering_2017-1}, as well as the 
    %
    LRG sample at $z=0.7$ \cite{bautista_completed_2020,gil-marin_completed_2020}, 
    %
    quasar (QSO) sample at $z=1.48$ \cite{hou_completed_2020,neveux_completed_2020},
    %
    and the BAO from the auto- and cross-correlation of the Lyman-$\alpha$ absorption and QSOs at an effective $z = 2.33$ \cite{bourboux_completed_2020}, 
    from eBOSS DR16 \cite{eboss_collaboration_completed_2021-2}. 
    We also include the Pantheon+ likelihood \cite{brout_pantheon_2022} based on uncalibrated Type Ia supernovae \cite{scolnic_pantheon_2022}.
    %
    %
    \item \textbf{+D/H}: We employ limits on the primordial deuterium abundance from \textcite{cooke_one_2018}. The reported bound of $(\text{D/H}) = (2.527\pm0.030)\times 10^{-5}$ only includes measurement uncertainty. We incorporate additional uncertainty from nuclear reaction rates by translating the bounds on the baryon-to-photon ratio reported by \textcite{cooke_one_2018}, which include both measurement uncertainty and uncertainty associated with reaction rates, to bounds on D/H. This translation is done by calculating D/H for a range of \mbox{$5.8 \leq 10^{10}\eta \leq 6.88$} using {\tt PArthENoPe}, which yields \mbox{$\text{D/H}\propto \eta^{-1.65}$}. This fit lends a new fractional uncertainty for D/H ($\sigma_{\mathrm{DH}}/\text{DH} = 1.65 \times \sigma_\eta/\eta)$, resulting in \mbox{$(\text{D/H}) = (2.527\pm0.068)\times 10^{-5}$}. We create a Gaussian likelihood function for {\tt MontePython} with a mean of $\mu_{\text{DH}} = 2.527\times10^{-5}$ and standard deviation $\sigma_{\text{DH}} = 6.83 \times 10^{-7}$. 
    %
    %
    \item \textbf{+SPT}: Third generation South Pole Telescope 2018 (SPT-3G) TT, TE, and EE data \cite{chown_maps_2018,dutcher_measurements_2021, balkenhol_measurement_2022} adapted to the {\tt clik} format.\footnote{\url{https://github.com/SouthPoleTelescope/spt3g_y1_dist}}
\end{itemize}

\begin{table}[t]
\centering
    \caption{\footnotesize Prior ranges for vEDE parameters used in different MCMC analyses. }
\begin{tabular*}{\linewidth}{|l@{\extracolsep{\fill}}cc|}
    \hline
    \hline
     & Analysis $A$ & Analysis $B$\\
    \hline
    $R_\phi$ 
    & [0, 15] 
    & [0, 100]  \\
    $\log_{10}(a_c)$ 
    & [-8, -6] 
    & [-7.4, -6.8] \\
    \hline
    \hline
\end{tabular*}
    \label{tab:priors}
\end{table}

To sufficiently sample the parameter space of $R_\phi$ and $\log_{10}(a_c)$ preferred by observations, we perform two separate analyses (see Table \ref{tab:priors}). Analysis $A$ samples the range of $R_\phi = [0,15]$ and $\log_{10}(a_c) = [-8, -6]$. At the lower bound of $\log_{10}(a_c) = -8$, we expect scenarios with $R_\phi \gtrsim 0.1$ to be ruled out by bounds on the deuterium abundance at $2\sigma$. The upper bound of $\log_{10}(a_c) = -6$ is chosen to ensure that the scalar field energy density is subdominant by the time of matter-radiation equality. 

We fix $n=8$ to obtain the extended periods of $w_\phi \simeq 1$ required to generate an enhancement to the matter power spectrum. As discussed in Sec.~\ref{sec:Effects_on_the_Matter_Power_Spectrum}, we expect qualitatively similar results for scenarios with $n \gtrsim 6$. We do not marginalize over $\theta_i$ in this work. As with Bayesian analyses of EDE, our analysis of vEDE is susceptible to prior volume effects \cite{Smith_2021, Gomez_Valent_2022, Herold_2022}. In the limit of $R_\phi \rightarrow 0$, the other vEDE parameters become unconstrained and thereby inflate the posterior density near $R_\phi \approx 0$. The impact of these volume effects is reduced by keeping $\theta_i$ and $n$ fixed.

We consider $\theta_i$ fixed at either 2, 2.5, or 3; we do not analyze $\theta_i$ values less than $1.8$. While it is possible to achieve a period of kination with $\theta_i < 1.8$, such scenarios require very large values of the decay constant $f$. In order for the scalar field to overcome Hubble friction and begin rolling, $\dd^2V(\phi)/\dd\phi^2$ must be greater than $3H^2$. Enforcing this rolling condition when $\rho_\phi \geq \rho_{r}$ sets a minimum required decay constant of
\begin{equation}
    \frac{f}{m_{pl}} \geq \left[  \frac{ \mathcal{F}(n,\theta_i) }{ 16\pi \left(1 - \cos(\theta_i) \right)^n } \right]^{1/2}, \label{eq:minimum_alpha}
\end{equation}
where $m_{pl}$ is the Planck mass and \mbox{$\mathcal{F} = (m^{-2}) \dd^2V(\phi)/\dd\phi^2$}. The right-hand side of Eq.~\eqref{eq:minimum_alpha} increases rapidly for decreasing $\theta_i$. Therefore, simultaneously demanding a small value of $\theta_i$ and vEDE domination necessitates $f\gg m_{pl}$. This requirement makes it difficult for \texttt{AxiCLASS} to numerically solve for $f$ and $m$ when given fixed values for $R_\phi$, $a_c$, and small $\theta_i$; small changes in $f$ correspond to large changes in $R_\phi$ when $a_c$ and $\theta_i$ are fixed. We avoid this problem by only considering $\theta_i = 2$, $2.5$, or $3$, but we note that the constraints on $a_c$ and $R_\phi$ in Appendix \ref{sec:appendix_supplemental_mcmc_results} should be considered only in the context of fixed $n$ and $\theta_i$. A profile likelihood analysis would yield more robust constraints on $R_\phi$ \cite{Gomez_Valent_2022}.

\begin{figure}[t]
\centering
    \includegraphics[width=\linewidth]{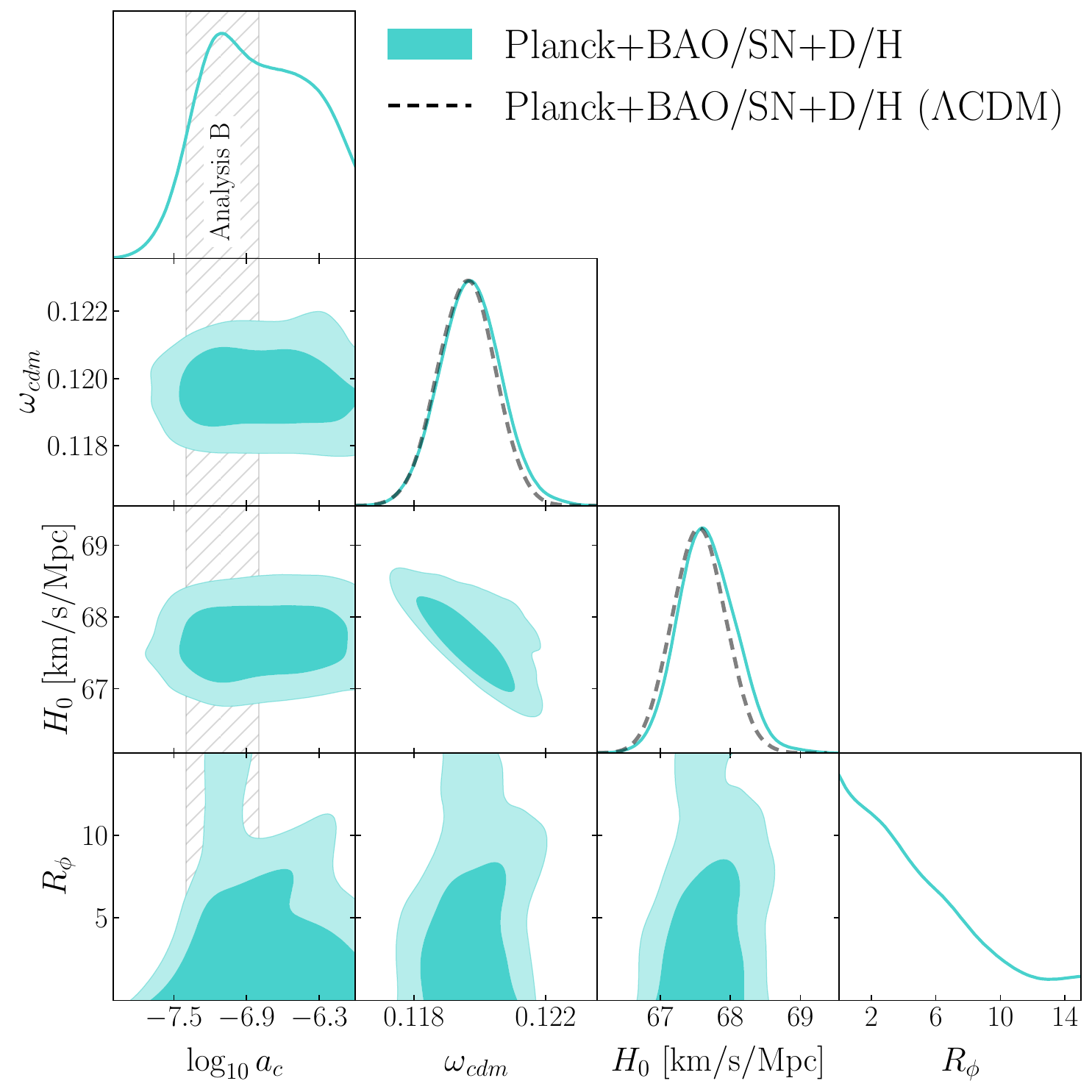}
  \caption{\footnotesize 1D and 2D posterior distributions of select parameters for Analysis $A$. We fix $n=8$ and $\theta_i = 2$. The dashed line shows the 1D posteriors for $\LCDM$ constrained by Planck+BAO/SN+D/H. The hatched region depicts the range of $a_c$ probed by Analysis $B$.}
  \label{fig:AnalysisA_triangle}
\end{figure}

Figure \ref{fig:AnalysisA_triangle} shows the posterior distributions of select parameters in Analysis $A$ with 68\% and 95\% confidence level (C.L.) contours for Planck+BAO/SN+D/H. We keep $\theta_i$ fixed at 2. The dashed line in the 1D posteriors represents a $\LCDM$ model constrained by Planck+BAO/SN+D/H. Contours for \{$\omega_b$, $\omega_{\cdm}$, $h$, $A_s$, $n_s$, $\tau_{reio}$\} can be found in Fig.~\ref{fig:AnalysisA_full_triangle} of Appendix \ref{sec:appendix_supplemental_mcmc_results}. We report the numerical bounds for each parameter of Analysis $A$ in Table \ref{tab:Analysis_A_posteriors} of Appendix \ref{sec:appendix_supplemental_mcmc_results}.

For $\log_{10}a_c \lesssim -7.4$, large values of $R_\phi$ correspond to a significant enhancement to the expansion rate during BBN and thus an increase in the helium abundance. However, a vEDE scenario with $R_\phi = 99$ and $\log_{10}a_c = -7.5$ results in $\YHe = 0.254$, which is well within the reported \textit{Planck} 2018 TT,TE,EE+lowE+lensing+BAO bound of $\YHe = 0.243^{+0.023}_{-0.024}$ (95\% C.L.) \cite{planck_collaboration_parameters}. Therefore, Planck+BAO/SN allows $R_\phi$ values up to 100 for $\log_{10}a_c \lesssim -7$. Compared to the helium abundance, the abundance of deuterium is much more sensitive to changes in the expansion rate during BBN. The $R_\phi$ vs.\ $\log_{10}a_c$ contour in Fig.~\ref{fig:AnalysisA_triangle} demonstrates that the addition of bounds on the deuterium abundance significantly constrains $R_\phi$ for $\log_{10}a_c \lesssim -7.4$. As $a_c$ decreases, the effect that vEDE has on the expansion rate during BBN increases and thus D/H measurements become more constraining. Note that these numerical constraints are specific to $n=8$ and $\theta_i = 2$. Increasing $n$ or decreasing $\theta_i$ causes the value of $\rho_\phi$ during BBN to increase and constraints from D/H will become more stringent. 

In the regime of $\log_{10}a_c \gtrsim -7.4$, D/H bounds hold little constraining power over vEDE and two interesting features emerge in the $R_\phi$ vs.\ $\log_{10}a_c$ plane of Fig.~\ref{fig:AnalysisA_triangle}: Planck+BAO/SN+D/H allows for slightly larger $R_\phi$ at $\log_{10}a_c \approx -6.3$ as well as significantly large values of $R_\phi$ at $\log_{10}a_c \approx -7$. Both of these features derive from the presence of the secondary bump in $\Delta H/H$.

A vEDE scenario with [$R_\phi = 10$, $\log_{10}a_c = -6.8$] has a non-negligible secondary bump in $\Delta H/H$ when the smallest scales that are accessible to \textit{Planck} are entering the horizon. This leads to an enhancement in the CMB temperature spectrum on small scales while leaving large CMB scales unaffected. In comparison, a vEDE scenario with [$R_\phi = 10$, $\log_{10}a_c = -6.3$] results in the \textit{primary} $\Delta H/H$ bump occurring when small CMB scales are entering the horizon, and the \textit{secondary} $\Delta H/H$ bump occurs when large CMB scales are entering the horizon. Thus, both large and small CMB scales experience an enhancement in the temperature spectrum, which can be compensated for by increasing $\omega_{\cdm}$. 

For $\log_{10}a_c \approx -7$, a scenario with $R_\phi = 10$ experiences a non-negligible secondary bump in $\Delta H/H$ as the smallest scales accessible to \textit{Planck} are entering the horizon and thereby induces a scale-dependent enhancement to the CMB temperature spectrum. If $R_\phi \lesssim 6$, then the amplitude of the secondary bump in $\Delta H/H$ is negligible and CMB scales are unaltered. As discussed in Sec.~\ref{sec:Effects_on_the_Matter_Power_Spectrum}, increasing $R_\phi$ not only increases the amplitude of the secondary bump in $\Delta H/H$, but it also increases the oscillation frequency of the scalar field and leads to the secondary bump in $\Delta H/H$ occurring earlier in time. If $R_\phi \gtrsim 30$, the non-negligible secondary bump in $\Delta H/H$ occurs before scales accessible to \textit{Planck} enter the horizon and effects to the CMB are minimized. In other words, for a vEDE scenario with $\log_{10}a_c \approx -7$, \textit{Planck} observations will disfavor $6 \lesssim R_\phi \lesssim 30$ compared to either $R_\phi \lesssim 6$ or $R_\phi \gtrsim 30$.

\begin{figure}[t]
\centering
    \includegraphics[width=\linewidth]{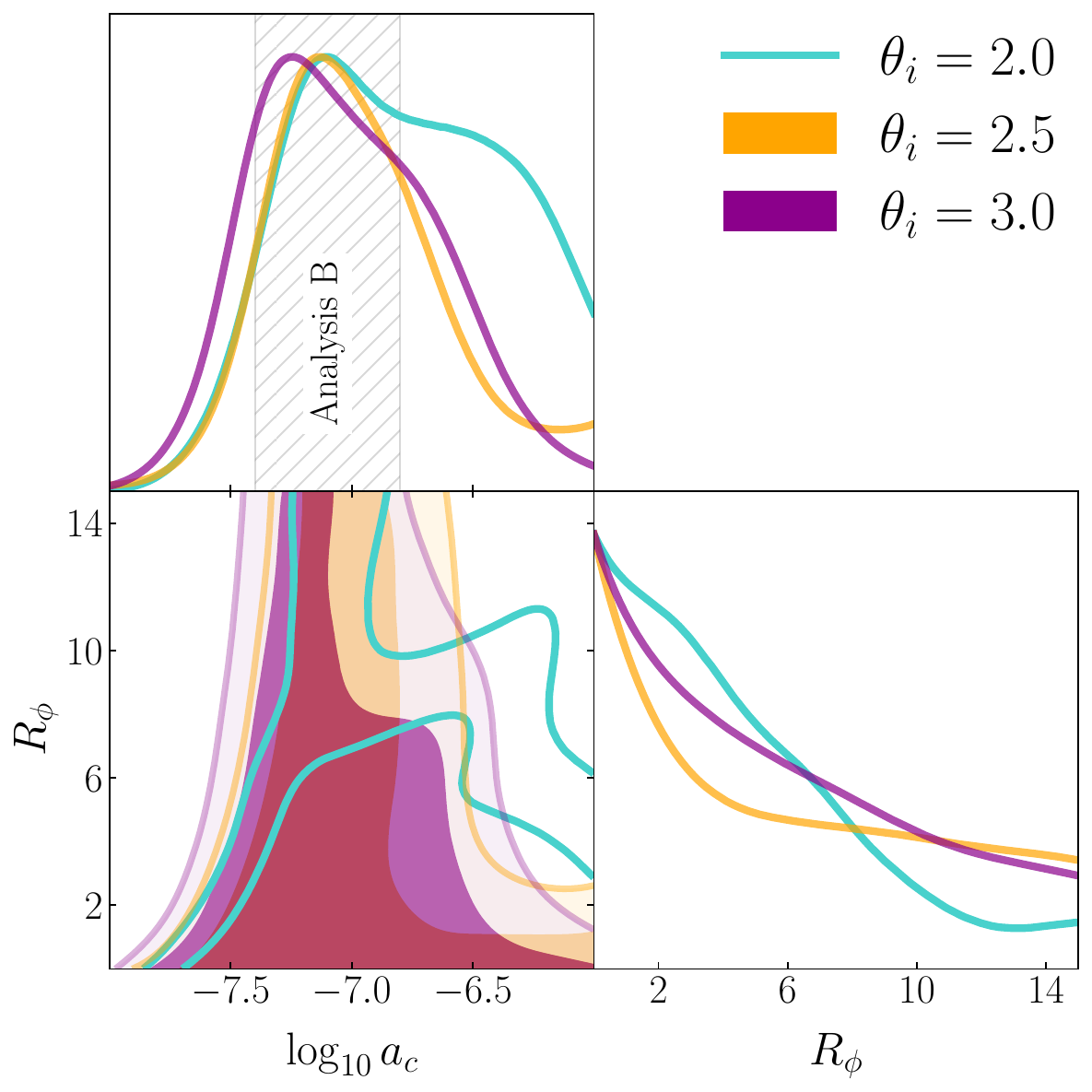}
  \caption{\footnotesize 1D and 2D posterior distributions of the vEDE parameters for Analysis $A$ with different values of $\theta_i$; posteriors for other parameters are shown in Appendix \ref{sec:appendix_supplemental_mcmc_results}. Note that the $1\sigma$ contour for $\theta_i=2$ only extends to $R_\phi \simeq 7$, while the $1\sigma$ contours for $\theta_i=2.5$ and $\theta_i=3$ reach the edge of the prior on $R_\phi$ ($R_\phi=15$). }
  \label{fig:AnalysisA_thetas}
\end{figure}

While the posteriors of Fig.~\ref{fig:AnalysisA_triangle} are specific to $\theta_i = 2$, changing the value of $\theta_i$ results in qualitatively similar posteriors. This is demonstrated by Fig.~\ref{fig:AnalysisA_thetas}, which shows posteriors for cases with $\theta_i$ fixed at either $2$, $2.5$, or $3$. It is clear that large values of $R_\phi$ are still permitted by Planck+BAO/SN+D/H for a range of $\log_{10}a_c$ when using $\theta_i = 2.5$ or $\theta_i = 3$. Neither $\theta_i = 2.5$ or $\theta_i = 3$ have a defined $2\sigma$ limit on $R_\phi$ within the prior range of $R_\phi = [0, 15]$. Additionally, the $1\sigma$ contour reaches the upper bound of this $R_\phi$ prior for both $\theta_i = 2.5$ and $\theta_i = 3$. In contrast, the $1\sigma$ contour for $\theta_i=2$ only reaches a maximum of $R_\phi \simeq 7$, and there is a $2\sigma$ limit of $R_\phi < 11.7$ over this range of $a_c$ values. 

Setting $\theta_i$ to 2.5 or 3 generates a weaker constraint on $R_\phi$ because increasing $\theta_i$ increases the scalar field's oscillation frequency. As a result, the secondary $\Delta H/H$ bump occurs before Planck-accessible scales enter the horizon. Increasing $\theta_i$ from 2.5 to 3 raises the oscillation frequency such that there is a tertiary non-negligible $\Delta H/H$ bump when Planck-accessible scales are entering the horizon. Therefore, the $1\sigma$ contour for $\theta_i=3$ only reaches $R_\phi \simeq 7$ at $\log_{10}a_c = -7$. Scenarios with $\log_{10}a_c \simeq -7.25$ and $\theta_i=3$, however, see the tertiary $\Delta H/H$ bump occurring before Planck scales enter the horizon. As a result, $R_\phi = 15$ lies within the $1\sigma$ contour when $\theta_i = 3$ and $\log_{10}a_c \simeq -7.25$.

Compared to the $\theta_i = 2$ case, larger $\theta_i$ cases are more constrained in $R_\phi$ for $\log_{10}a_c \gtrsim -6.4$. For this range of $\log_{10}a_c$, the larger oscillation frequency of the scalar field yielded by larger $\theta_i$ means the primary and secondary bumps in $\Delta H/H$ are no longer affecting small and large CMB scales simultaneously. Therefore, such vEDE scenarios with $\log_{10}a_c \gtrsim -6.4$ and $\theta_i =2.5$ or $\theta_i=3$ become difficult to accommodate with adjustments to $\omega_{\cdm}$. The $\theta_i = 2.5$ and $\theta_i = 3$ cases demonstrate less stringent constraints on $R_\phi$ for $\log_{10}a_c \lesssim -7.5$. Increasing $\theta_i$ leads to a smaller value of $\rho_\phi$ during BBN and thus the deuterium abundance is less altered compared to the case with $\theta_i=2$.

\begin{figure}[t]
\centering
    \includegraphics[width=\linewidth]{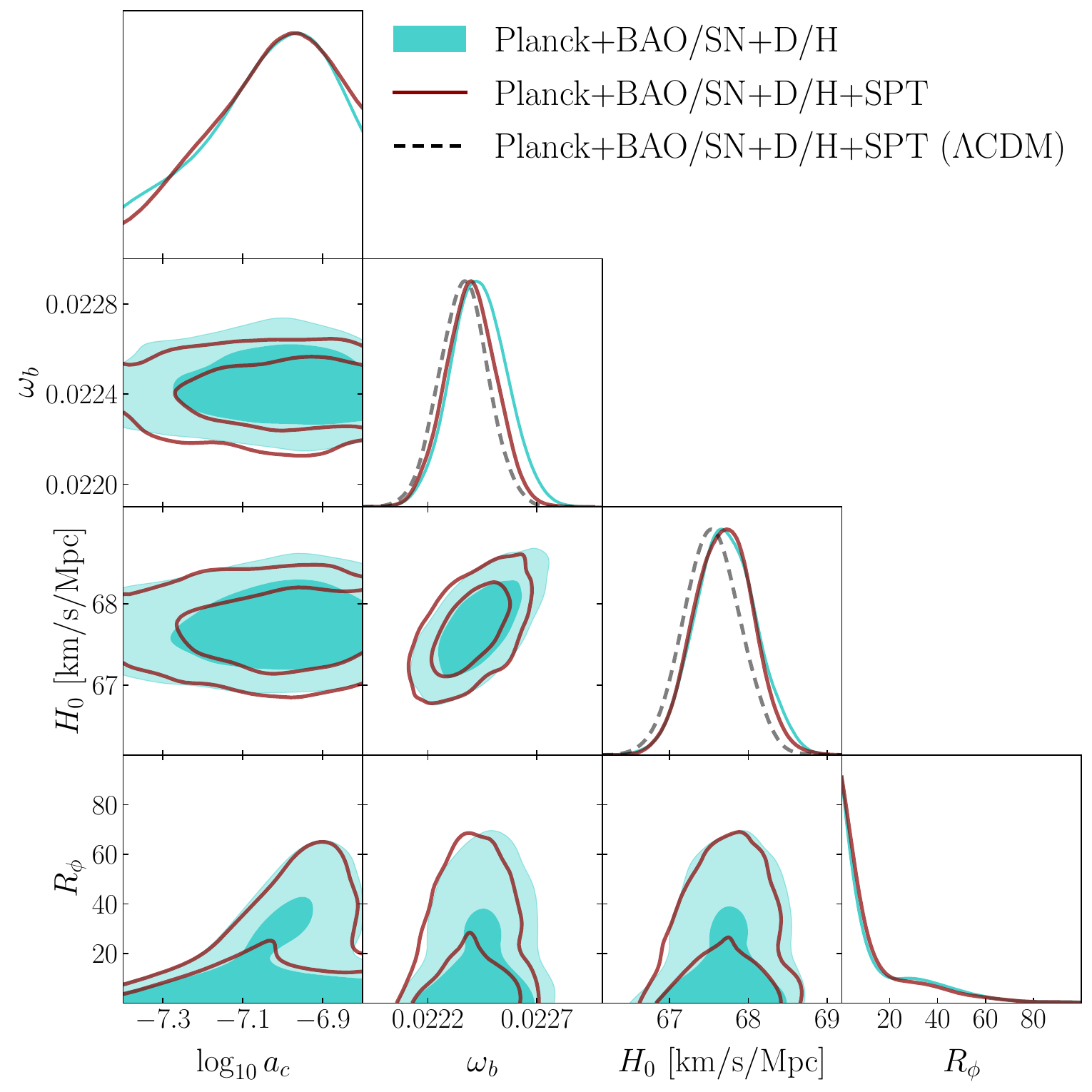}
  \caption{\footnotesize 1D and 2D posterior distributions of select parameters for Analysis $B$. We fix $n=8$ and $\theta_i = 2$. The dashed line shows the 1D posteriors for $\LCDM$ constrained by Planck+BAO/SN+D/H+SPT.}
  \label{fig:AnalysisB_triangle}
\end{figure}

Figure \ref{fig:AnalysisA_thetas} confirms the intuition established by Fig.~\ref{fig:different_transfers_theta} that, aside from changes in the oscillation frequency, increasing $\theta_i$ does not significantly alter the effects of vEDE on the CMB and light element abundances. Regardless of the value of $\theta_i$, Fig.~\ref{fig:AnalysisA_thetas} demonstrates that $\log_{10}a_c \simeq -7$ is a particularly interesting region of parameter space for vEDE.
We perform a separate MCMC analysis (Analysis $B$) to further investigate constraints for vEDE scenarios with $\log_{10}a_c \simeq -7$. If $R_\phi \gtrsim 10$, any vEDE case with $\theta_i=2$ and $\log_{10}a_c \lesssim -7.4$ will be ruled out at $2\sigma$ by bounds on the deuterium abundance. If $R_\phi = 10$, the secondary bump in $\Delta H/H$ only affects horizon entry of scales observable by \textit{Planck} when $\log_{10}a_c \gtrsim -6.8$ and $\theta_i=2$. To probe the region of vEDE parameter space that minimally alters both D/H and the CMB, we choose priors of $\log_{10}a_c = [-7.4, -6.8]$ and $R_\phi = [0, 100].$ While these priors are motivated by cases with $\theta_i = 2$, Fig.~\ref{fig:AnalysisA_thetas} demonstrates that the range of $\log_{10}a_c = [-7.4, -6.8]$ also encompasses $a_c$ values for which $R_\phi$ is least constrained when $\theta_i$ is 2.5 or 3. The resulting posterior distributions for select parameters of this analysis with $\theta_i=2$ are presented in Fig.~\ref{fig:AnalysisB_triangle}. Contours for \{$\omega_b$, $\omega_{\cdm}$, $h$, $A_s$, $n_s$, $\tau_{reio}$\} can be found in Fig.~\ref{fig:AnalysisB_full_triangle} of Appendix \ref{sec:appendix_supplemental_mcmc_results} and numerical posteriors are reported in Table \ref{tab:Analysis_B_posteriors} of Appendix \ref{sec:appendix_supplemental_mcmc_results}.

\begin{figure*}[t]
\centering
    \includegraphics[width=\linewidth]{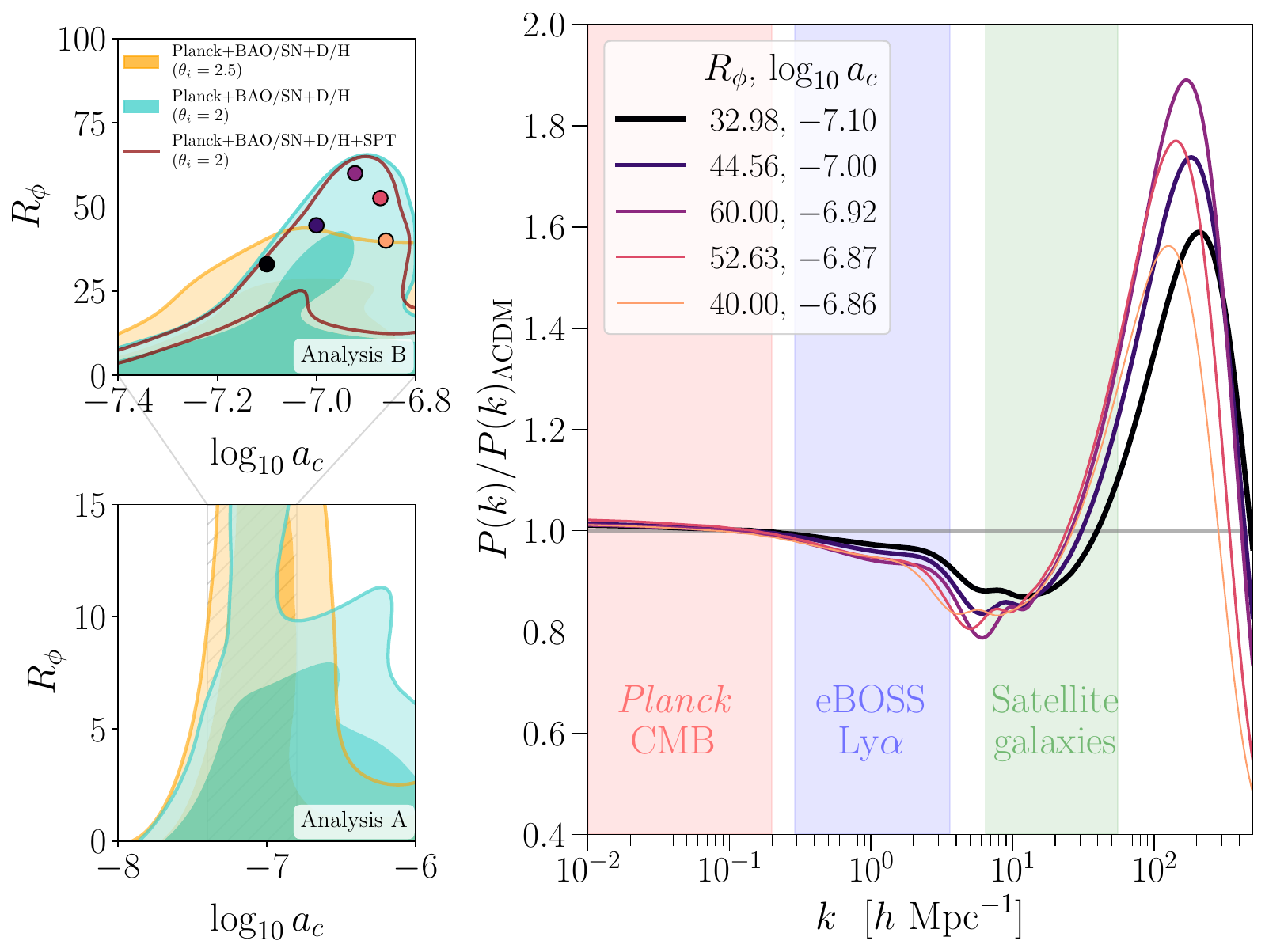}
  \caption{\footnotesize (\textit{Bottom Left}) $1\sigma$ and $2\sigma$ contours for vEDE parameter posteriors from Analysis $A$, constrained by Planck+BAO/SN+D/H. The hatched region depicts the range of $a_c$ probed by Analysis $B$. (\textit{Top Left}) $1\sigma$ and $2\sigma$ contours for vEDE parameter posteriors from Analysis $B$. Colored dots mark example vEDE cosmologies with $\theta_i=2$ that have a $\Delta \chi^2 < 5$ compared to $\LCDM$ when constrained by Planck+BAO/SN+D/H+SPT. (\textit{Right}) Transfer functions of those vEDE cosmologies marked by dots in the top left panel. Shaded regions show the approximate range of scales to which \textit{Planck} CMB and eBOSS Ly$\alpha$ forest \cite{lyke_sloan_2020,chabanier_one-dimensional_2019} observations are sensitive, as well as scales that can be probed with observations of velocity dispersions in Milky Way satellites \cite{esteban_milky_2023}. For the transfer functions, we assume the best-fit values of $\omega_b$, $\omega_\cdm$, $h$, $A_s$, $n_s$, and $\tau_{reio}$ for each respective combination of $R_\phi$ and $a_c$ and best-fit values for $\LCDM$}.
  \label{fig:2sig_allowed_transfers}
\end{figure*}

At fixed $\log_{10}a_c=-7$ and $\theta_i=2$, values of $R_\phi\lesssim 15$ are within the $1\sigma$ contour of $R_\phi$ vs.~$\log_{10}a_c$ for Planck+BAO/SN+D/H in Fig.~\ref{fig:AnalysisB_triangle}. Additionally, a limited range of $23 \lesssim R_\phi \lesssim 40$ is contained within the $1\sigma$ contour at $\log_{10}a_c=-7$. This $1\sigma$ allowed region of parameter space at larger values of $R_\phi$ corresponds to vEDE scenarios in which the secondary $\Delta H/H$ bump slightly suppresses the smallest CMB scales, leading to a $\lesssim 2\%$ enhancement in the damping tail of the CMB temperature spectrum compared to that of $\LCDM$. Since SPT data constrain the small-scale tail of the CMB spectrum, the $1\sigma$ contour for Planck+BAO/SN+D/H+SPT in Fig.~\ref{fig:AnalysisB_triangle} no longer contains this lobe of parameter space. Otherwise, the inclusion of SPT data does not significantly alter the range of allowed $R_\phi$ values for $\log_{10}a_c = [-7.4, -6.8]$. 

Figure \ref{fig:AnalysisB_triangle} demonstrates that the $2\sigma$ limit of Planck+BAO/SN+D/H+SPT permits vEDE scenarios with a maximum of $R_\phi \approx 60$ at $\log_{10}a_c \approx -6.9$. Such a vEDE scenario can result in a $\sim 90\%$ enhancement to the matter power spectrum compared to $\LCDM$. Note that Analysis $B$ focuses on scenarios in which the secondary $\Delta H/H$ bump does \textit{not} significantly influence the CMB. Changes in $n$ or $\theta_i$ can alter which scales are affected by the secondary (or tertiary) $\Delta H/H$ bump and thus the numerical results presented in Fig.~\ref{fig:AnalysisB_triangle} are sensitive to changes in $n$ or $\theta_i$.

Figures \ref{fig:AnalysisA_triangle} and \ref{fig:AnalysisB_triangle} show that the inclusion of vEDE does not increase the value of $H_0$ preferred by Planck+BAO/SN+D/H data. As discussed in Sec.~\ref{sec:Effects_on_the_Temperature_Spectrum}, vEDE scenarios decrease the sound horizon at recombination. However, if $a_c$ is small enough that the scalar field energy density is insignificant by the time CMB scales enter the horizon, vEDE does not cause a shift to the acoustic peaks of the CMB. Therefore, even though the value of $r_s$ is reduced compared to that for $\LCDM$, an increase in $H_0$ is not required by observations of the CMB and BAO. 

Figure \ref{fig:2sig_allowed_transfers} depicts the $R_\phi$ vs. $\log_{10}a_c$ contours from both Analysis $A$ and Analysis $B$ in the bottom left and top left panels, respectively. We include contours for $\theta_i$ fixed at either 2 or 2.5, demonstrating that large values of $R_\phi$ are allowed for both $\theta_i=2$ and $\theta_i = 2.5$.
The right panel of Fig.~\ref{fig:2sig_allowed_transfers} depicts the transfer functions resulting from select vEDE scenarios marked by dots in the top left panel of Fig.~\ref{fig:2sig_allowed_transfers}. These vEDE scenarios only slightly worsen the fit to Planck+BAO/SN+D/H+SPT compared to $\LCDM$  ($\Delta \chi^2<5$). The best-fit for each of these scenarios was identified by the minimum $\chi^2$ in respective MCMC analyses that fixed $R_\phi$, $a_c$, $\theta_i$, and $n$. A dedicated minimization would likely identify an even better fit of vEDE to Planck+BAO/SN+D/H+SPT data \cite{Karwal:2024qpt}. 

The specific vEDE scenarios shown in Fig.~\ref{fig:2sig_allowed_transfers} correspond to $\theta_i = 2$, a mass of $m \simeq 10^{-20}$ eV, and a decay constant of $f \simeq 0.8 \, m_{pl}$. Each transfer function assumes the best-fit values of $\omega_b$, $\omega_\cdm$, $h$, $A_s$, $n_s$, and $\tau_{reio}$ for each respective combination of $R_\phi$ and $a_c$. These best-fit values are within $1\sigma$ of the \textit{Planck} 2018 reported TT,TE,EE+lowE+lensing+BAO values \cite{planck_collaboration_parameters}. Scenarios with $R_\phi \approx 40$ and $\log_{10}a_c \approx -7$ are allowed at $2\sigma$ by Planck+BAO/SN+D/H when $\theta_i=2.5$, as shown in Fig.~\ref{fig:2sig_allowed_transfers}. These vEDE scenarios generate qualitatively similar enhancements to the matter power spectrum as those resulting from $\theta_i=2.$

Figure \ref{fig:2sig_allowed_transfers} demonstrates that current observations permit vEDE scenarios that result in significant enhancements to the matter power spectrum on scales near \mbox{$k\approx\SI{200}{\hmpc}$} compared to that resulting from $\LCDM$. Furthermore, the transfer functions depicted in Fig.~\ref{fig:2sig_allowed_transfers} induce a red-tilted suppression in power on scales probed by eBOSS Lyman-$\alpha$ while minimally altering \textit{Planck} CMB scales. The slight excess in power seen in Fig.~\ref{fig:2sig_allowed_transfers} for $k\lesssim\SI{0.07}{\hmpc}$ results from the vEDE best-fit values of $n_s$ being slightly smaller than the best-fit $n_s$ value for $\LCDM$. The allowed transfer functions in Fig.~\ref{fig:2sig_allowed_transfers} do not significantly alter the matter fluctuation parameter \mbox{$S_8 \equiv \sigma_8 \sqrt{\Omega_m/0.3}$}, where $\sigma_8$ is the standard deviation of density fluctuations within spheres with radii of $\SI{8}{\mpch}$, and $\Omega_m$ is the relative contribution of matter to the total present day energy density. For example, the \mbox{[$R_\phi = 60$, $\log_{10}a_c = -6.92$]} scenario of Fig.~\ref{fig:2sig_allowed_transfers} results in $S_8 = 0.830$, whereas $S_8 = 0.838$ for the Planck+BAO/SN+D/H+SPT best-fit $\LCDM$ model.

\section{Summary and Discussion} \label{sec:Summary_and_Conclusions}

This work examines the effects of vEDE generated by a fast-rolling axion-like scalar field that becomes dynamical between BBN and matter-radiation equality. If this scalar field comes to dominate the energy density of the universe, it significantly alters the rate at which $H(a)$ decreases. Perturbation modes experience a delay in the time of horizon entry as well an altered subhorizon growth rate of DM density perturbations if those modes enter the horizon while the scalar field significantly alters the Hubble rate. These effects produce a distinctive matter power spectrum that can exhibit an enhancement compared to the matter power spectrum generated by a standard $\LCDM$ expansion history. The power on scales on both sides of this enhancement is reduced compared to standard. However, only scales that are within the horizon while the scalar field significantly affects the Hubble rate are altered.

We find that an enhancement in the matter power spectrum is only achieved if vEDE becomes dynamical well before matter-radiation equality. If the scalar field begins to fast-roll not long before matter-radiation equality, the period of enhanced growth due to vEDE will be short-lived because the fractional density perturbation in DM for all subhorizon modes locks on to linear growth when the universe transitions to matter domination. Additionally, enhancements to the matter power spectrum are difficult to produce when the potential power-law index, $n$, is less than 5. Decreasing $n$ increases the oscillation frequency of the scalar field, which prevents extended periods of enhanced growth for DM density perturbations. EDE studies focus on axion-like scalar fields that become dynamical around the time of matter-radiation equality and potentials with $n = 3$; these models cannot generate an enhancement in the matter power spectrum on scales $k \gtrsim \SI{0.01}{\hmpc}$ while the $\Lambda$CDM cosmological parameters are kept fixed. Instead, enhancements to the matter power spectrum that result from EDE are driven by changes in cosmological parameters such as $\omega_\cdm$ and $n_s$. 

If the scalar field's energy density is still significant when scales with $k\lesssim\SI{0.2}{\hmpc}$ enter the horizon, vEDE induces a scale-dependent enhancement and phase shift to the CMB spectrum. If the scalar field becomes dynamical soon after BBN, the scalar field energy density can increase the expansion rate during BBN and thus amplify the abundance of primordial elements like deuterium and helium. We employ observations of the CMB from \textit{Planck} and SPT-3G, as well as BAO observations, uncalibrated Type Ia supernovae, and measurements of the abundance of primordial deuterium (D/H), to assess the range of vEDE scenarios that are consistent with current observations. The numerical constraints presented here are specific to vEDE scenarios with $n=8$ and fixed $\theta_i$. Fixing $\theta_i$ and $n$ can help mitigate prior volume effects as $R_\phi \rightarrow 0$. However, our Bayesian analysis results are consequently limited by not marginalizing over $\theta_i$ or $n$. Even so, we expect changing $\theta_i$ or varying $n$ within $n \geq 6$ to produce qualitatively similar results. 

Once $n$ and $\theta_i$ are chosen, a vEDE cosmology can be described by two parameters: the scale factor at which the vEDE contribution to the total energy density is maximized ($a_c$), and the ratio of the scalar field energy density to the energy density of everything else at $a_c$ ($R_\phi$). We find that the combination of \textit{Planck} CMB observations, BAO measurements, uncalibrated Type Ia supernovae, and the abundance of D/H permit vEDE scenarios with $R_\phi$ values approaching $60$ at the $2\sigma$ level. The addition of SPT data minimally alters these results. Such vEDE cosmologies with $R_\phi \approx 60$ and $\log_{10}a_c \approx -6.9$ can enhance the matter power spectrum on scales $\SI{30}{\hmpc} \lesssim k \lesssim \SI{500}{\hmpc}$, with up to a $\sim90\%$ increase in $P(k)$ at $k\sim\SI{200}{\hmpc}$ compared to $\LCDM$. 

Probes of the small-scale matter power spectrum could detect these allowed enhancements generated by vEDE. The concentrations of Milky Way satellites depend on the formation times of their DM halos. Consequently, observations of satellite concentrations can probe the matter power spectrum on scales \mbox{$\SI{6}{\hmpc} \lesssim k \lesssim \SI{55}{\hmpc}$} \cite{esteban_milky_2023}. Strong lensing of quasars also probes this range of scales in the matter power spectrum \cite{gilman_primordial_2022}; however, these constraints are currently weaker than those derived from concentrations of Milky Way satellites \cite{esteban_milky_2023}. Interestingly, measurements of the stellar velocity dispersion and half-light radius of dwarf galaxies orbiting the Milky Way demonstrate a slight preference for an enhancement in the matter power spectrum compared to $\LCDM$ for $\SI{15}{\hmpc} \lesssim k \lesssim \SI{55}{\hmpc}$ \cite{esteban_milky_2023,dekker_constraints_2024}, albeit only at the $1\sigma$ level. 

Conversely, the abundance of these satellites is sensitive to suppressions in the matter power spectrum. The current census of satellite galaxies rules out any suppression to the matter power spectrum that reduces $P(k)$ by more than $30\%$ compared to $\LCDM$ on scales $k\sim \SI{30}{\hmpc}$ (95\% C.L.) \cite{nadler_constraints_2019,nadler_milky_2021}. We find that those vEDE scenarios allowed by observations of the CMB, measurements of the primordial deuterium abundance, and probes of the the late-time expansion history do not result in this level of suppression in power at $k\sim \SI{30}{\hmpc}$. 
The Press-Schechter formalism \cite{press-schechter} can accurately predict the impact of an enhanced matter power spectrum on the halo mass function \cite{gomez-navarro_impact_2023}. 
For the allowed vEDE scenarios that nearly double the amplitude of the matter power spectrum on scales around $\SI{200}{\hmpc}$ (Fig.~\ref{fig:2sig_allowed_transfers}), Press-Schechter halo mass functions predict up to a $\sim 60\%$ increase in the abundance of DM halos with masses \mbox{$ 10^6 M_\odot \lesssim M \lesssim 10^9 M_\odot$}. These scenarios only cause a depletion of up to $\sim 13\%$ in the abundance of \mbox{$10^9 M_\odot \lesssim M \lesssim 10^{11} M_\odot$} halos. Upcoming galaxy surveys with the Vera C.\ Rubin Observatory \cite{ivezic_lsst_2019} will further improve constraints on suppressions to the small-scale matter power spectrum via halo counts \cite{nadler_forecasts_2024}.  

Observations of perturbations to stellar streams due to DM subhalos can also probe the halo mass function in the range of \mbox{$ 10^6 M_\odot \lesssim M \lesssim 10^9 M_\odot$}. Current observational sensitivity places $2\sigma$ upper limits on the abundance of subhalos determined by stellar streams that are consistent with allowed vEDE scenarios \cite{banik_evidence_2021,banik_novel_2021}. Observations from Gaia \cite{bonaca_stellar_2024} and the Nancy Grace Roman Space Telescope \cite{aganze_prospects_2024} will improve these constraints from stellar streams.

Hydrodynamic simulations can be used to infer the amplitude ($\Delta^2_{\mathrm{lin}}$) and tilt ($n_{\mathrm{lin}}$) of the $z=3$ linear matter power spectrum at a wavenumber\footnote{ Converting from velocity space to position space requires a factor that depends on the expansion rate. For standard late-time cosmologies, a wavenumber of $\SI{0.009}{\second\per\kilo\meter}$ in velocity space corresponds to $k\simeq \SI{1}{\hmpc}$.} of $k\simeq \SI{1}{\hmpc}$ that generates the observed line-of-sight correlations in the Lyman-$\alpha$ forest \cite{mcdonald_linear_2005,chabanier_one-dimensional_2019, pedersen_compressing_2023, Bird_2023}. The original analysis of the eBOSS Lyman-$\alpha$ spectrum \cite{chabanier_one-dimensional_2019} found $\Delta^2_{\mathrm{lin}} = 0.31 \pm 0.02$ and $n_{\mathrm{lin}} = -2.339 \pm 0.006$, which are inconsistent with the matter power spectrum that follows from the \textit{Planck} best-fit cosmology \cite{rogers_5_2024}. More recent simulations have eliminated the disagreement between \textit{Planck} and eBOSS values of $n_\mathrm{lin}$, but the disagreement in $\Delta^2_{\mathrm{lin}}$ between \textit{Planck} and eBOSS persists \cite{Fernandez_2024, walther_emulating_2024, ivanov_fundamental_2025}. A cosmological model that results in a suppression on Lyman-$\alpha$ scales in the matter power spectrum could alleviate this tension \cite{rogers_5_2024, he_self-interacting_2024, He_Fresh_2025}. Allowed vEDE scenarios result in a suppression in power on scales probed by eBOSS Lyman-$\alpha$ while minimally altering scales accessible to \textit{Planck} (see Fig.~\ref{fig:2sig_allowed_transfers}). In contrast, EDE models that alleviate the Hubble tension worsen this tension between \textit{Planck} and eBOSS Lyman-$\alpha$ \cite{goldstein_canonical_2023}.

This work demonstrates that vEDE can induce distinctive features in the matter power spectrum. Current limits from observations of the CMB, BAO, uncalibrated Type Ia supernovae, and primordial element abundances permit vEDE scenarios that result in significant enhancements to the small-scale matter power spectrum compared to that resulting from $\LCDM$. We expect that similar enhancements are possible with alternative models; mechanisms that generate a period of kination around a scale factor of roughly $10^{-7}$ (e.g.\ \cite{co_gravitational_2022}) will generate qualitatively similar enhancements to the matter power spectrum as those in Fig.~\ref{fig:2sig_allowed_transfers} produced by vEDE.
These enhancements in the matter power spectrum lie on the cusp of our current observational capabilities. The dynamics of Milky Way satellites can probe enhancements to the small-scale matter power spectrum, and the Vera C.\ Rubin Observatory will increase the number of observed dwarf galaxies \cite{Mutlu-Pakdil_resolved_2021}. Meanwhile, vEDE results in a suppression in power for $\SI{0.3}{\hmpc} \lesssim k\lesssim \SI{30}{\hmpc}$ in the matter power spectrum, which is preferred by measurements of the matter power spectrum derived from eBOSS Lyman-$\alpha$. The census of satellite galaxies can serve as a probe of such suppressions generated by vEDE, and our ability to constrain the population of DM subhalos will improve dramatically with the Vera C.\ Rubin Observatory. Finally,  constraints on the small-scale matter power spectrum from perturbations to stellar streams are expected to improve with observations from Gaia and the Nancy Grace Roman Space Telescope. The detection of unique vEDE signatures in the matter power spectrum would provide additional evidence for a collection of cosmological scalar fields, which is a generic prediction of string theory.

\section{Acknowledgments} \label{sec:Acknowledgements}

We thank Marc Kamionkowski, Vera Gluscevic, Jessie Shelton, and Mustafa Amin for useful discussions. Computational work for this was in part performed on Firebird, a cluster supported by Swarthmore College and Lafayette College. ACS and ALE received support from NSF Grant PHY-2310719. TLS is supported by NSF Grants AST-2009377 and AST-2308173. This research was supported in part by NSF grant PHY-2309135 to the Kavli Institute for Theoretical Physics (KITP).

\vspace{2em}
\appendix

\section{Suppression in Regime A} \label{sec:appendix_small-scale_suppression}
\begin{figure}[t]
\centering
    \includegraphics[width=\linewidth]{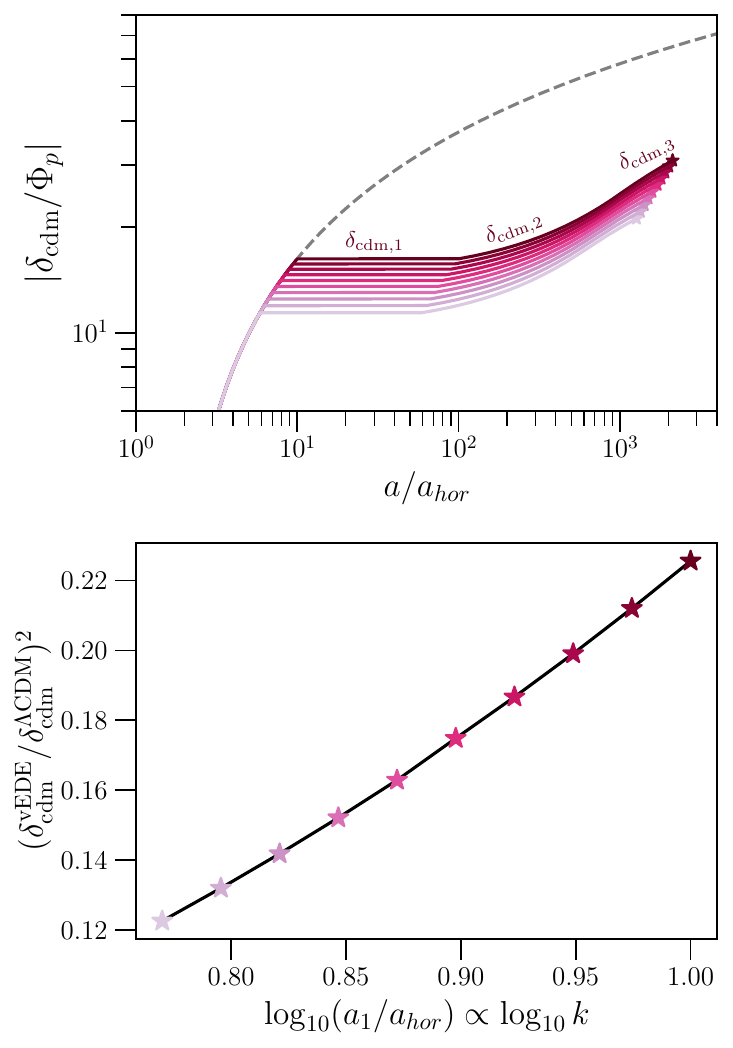}
  \caption{\footnotesize (\textit{Top}) Piecewise function for $\delta_\cdm$ described by Eqs.~\eqref{eq:delta1}, \eqref{eq:delta2}, and \eqref{eq:delta3}. Different modes are represented by different values of $\tilde{a}_1 = a_1/a_{hor}$. Each mode has the same values of $\tilde{a}_2/\tilde{a}_1 = \tilde{a}_3/\tilde{a}_2 = 10$ and $\tilde{a}_4/\tilde{a}_3 = 10^{1/3}$.     (\textit{Bottom}) Value of $\delta_\cdm(\tilde{a}_4)$ for different modes resulting from our piecewise model compared to $\delta_\cdm(\tilde{a}_4)$ from Eq.~\eqref{eq:delta_log}. This mirrors the scale-dependent suppression seen in the transfer function resulting from vEDE. }
  \label{fig:toy_model}
\end{figure}

By definition, Regime A contains modes that enter the horizon before vEDE alters the expansion rate. One might assume that all perturbation modes in Regime A would experience identical evolution upon horizon entry. Instead, there is a $k$-dependent level of suppression for these modes, with those that enter the horizon just before the vEDE era experiencing the most suppression (see Fig.~\ref{fig:example_transfer}). Here we construct a toy model to understand the effect that vEDE has on the matter power spectrum in Regime A. The subhorizon evolution of $\delta_\cdm$ during radiation domination is
\begin{equation}
    \delta_\cdm(\tilde{a}) = \Phi_p \left[A \log(B\tilde{a})\right], \label{eq:delta_log}
\end{equation}
where $A=9.11$, $B=0.594$, $\Phi_p$ is the initial value of $\Phi$ deep in radiation domination, $\tilde{a}\equiv a/a_{hor}$, and $a_{hor}$ is the scale factor at the time of horizon entry for a given mode \cite{hu_small_1996}. This evolution is depicted in the top panel of Fig.~\ref{fig:toy_model} by the dashed line. We insert a pause in this logarithmic growth starting at $\tilde{a}_1$ and ending at $\tilde{a}_2$ in order to mimic the suppressed subhorizon growth experienced by small-scale modes when $H(a)$ decreases slower than $a^{-2}$. It follows that the evolution of $\delta_\cdm$ during the pause is simply
\begin{equation}
    \delta_{\cdm,1}(\tilde{a}) = \Phi_p \left[A\log(B\tilde{a}_1)\right], \hspace{0.5cm} \tilde{a}_1 \leq \tilde{a} < \tilde{a}_2. \label{eq:delta1}
\end{equation}
After $a_c$, vEDE facilitates enhanced growth on subhorizon scales: $\tilde{a}\delta'_\cdm(\tilde{a})$ increases when $H(a)$ decreases faster than $a^{-2}$ as a result of vEDE, whereas $\tilde{a}\delta'_\cdm(\tilde{a})$ is constant during radiation domination (a prime denotes differentiation with respect to $\tilde{a}$). If $\delta_{\cdm,2}$ is the DM density perturbation during the period of enhanced growth that extends from $\tilde{a}_2$ to $\tilde{a}_3$, we set 
\begin{equation}
    \frac{d\log_{10}(\tilde{a}\delta'_{\cdm,2})}{d \log_{10}(\tilde{a})} = s \label{eq:delta2_condition}, 
\end{equation}
where $s$ is a positive constant. Under the requirement that $\delta_{\cdm,2}(\tilde{a}_2) = \delta_{\cdm,1}$, the solution to Eq.~\eqref{eq:delta2_condition} is
\begin{align*}
    \delta_{\cdm,2}(\tilde{a}) &=  \delta_{\cdm,1} + \frac{C }{s} \left [ \left(\tilde{a}\right)^s  - \left(\tilde{a}_2\right)^s \right ], \hspace{0.5cm} \tilde{a}_2 \leq \tilde{a} < \tilde{a}_3. \stepcounter{equation}\tag{\theequation}\label{eq:delta2}
\end{align*}
We find that $C = (5.72\Phi_p)/(645)^s$ and $s=0.79$ for a vEDE scenario with $R_\phi = 99$ and $a_c = 10^{-6.3}$. 

Once the vEDE era is complete, subhorizon modes will once again grow logarithmically if the universe is dominated by radiation. We model this return to logarithmic growth by enforcing that $\tilde{a}\delta'_\cdm$ remain constant from $\tilde{a}_3$ to $\tilde{a}_4$, after which point matter-radiation equality occurs and all modes experience the same linear growth in $\delta_\cdm$. In other words, we require that 
\begin{equation}
    \tilde{a}\delta'_{\cdm,3}(\tilde{a}) = C(\tilde{a}_3)^s, \label{eq:delta3_condition}
\end{equation}
where the right-hand side comes from evaluating $a\delta'_{\cdm,2}$ at $\tilde{a}_3$. With the condition that $\delta_{\cdm,3} = \delta_{\cdm,2}$ at $\tilde{a}_3$, the solution to Eq.~\eqref{eq:delta3_condition} is
\begin{align*}
    \delta_{\cdm,3}(\tilde{a}) &= \delta_{\cdm,2}\left(\tilde{a}_3\right) + C \left(\tilde{a}_3\right)^s \log\left(\frac{\tilde{a}}{\tilde{a}_3}\right), \\
    &\hspace{5.1cm}  \tilde{a}_3 \leq \tilde{a} < \tilde{a}_4.      \stepcounter{equation}\tag{\theequation}\label{eq:delta3}
\end{align*}
We stitch Eqs.~\eqref{eq:delta_log}, \eqref{eq:delta1}, \eqref{eq:delta2}, and \eqref{eq:delta3} together in a piecewise fashion to model the evolution of $\delta_\cdm$ for a given $k$ mode. Every small-scale mode that enters the horizon before the vEDE era experiences the same subsequent periods of suppressed, enhanced, and logarithmic growth. Therefore, the values of $\tilde{a}_2/\tilde{a}_1$, $\tilde{a}_3/\tilde{a}_2$, and $\tilde{a}_4/\tilde{a}_3$ should be the same for all modes that enter before the vEDE era. However, different $k$ modes enter the horizon at different times before the vEDE era begins and thus $k \propto \tilde{a}_1 = a_1/a_{hor}$. The top panel of Fig.~\ref{fig:toy_model} shows the piecewise evolution of $\delta_\cdm$ for different modes (i.e. different values of $\tilde{a}_1$), each with fixed values of $\tilde{a}_2/\tilde{a}_1 = \tilde{a}_3/\tilde{a}_2 = 10$ and $\tilde{a}_4/\tilde{a}_3 = 10^{1/3}$. Here it can be seen that, compared to the standard growth described by Eq.~\eqref{eq:delta_log}, modes that experience this interruption in logarithmic growth are ultimately suppressed. Furthermore, modes that enter the horizon closer to the start of the vEDE era (i.e. smaller $\tilde{a}_1$) exhibit more suppression than those modes that enter well before the vEDE era. 

The scale-dependent suppression seen in the top panel of Fig.~\ref{fig:toy_model} manifests in the matter power spectrum. This is demonstrated by the bottom panel of Fig.~\ref{fig:toy_model}, which depicts the value of $\delta_\cdm(\tilde{a}_4)$ for different modes resulting from our piecewise model compared to that resulting from Eq.~\eqref{eq:delta_log}. We find $P(k)$ decreases nearly linearly with decreasing $\log_{10}(k)$, mirroring the scale-dependent suppression observed in the matter power spectrum that results from vEDE (see Regime A of Fig.~\ref{fig:example_transfer}).

\onecolumngrid
\clearpage
\section{Supplemental MCMC Results} \label{sec:appendix_supplemental_mcmc_results}

\begin{table}[b]
\centering
    \caption{\footnotesize Posteriors for vEDE and cosmological parameters in Analysis $A$, corresponding to those shown in Figs.~\ref{fig:AnalysisA_full_triangle} and \ref{fig:AnalysisA_full_triangle_thetas}. Uncertainties are reported at 68\% C.L. and upper limits are given at 95\% C.L.}
 \begin{tabularx}{\linewidth}{|l|>{\centering\arraybackslash}X|>{\centering\arraybackslash}X|>{\centering\arraybackslash}X|>{\centering\arraybackslash}X|}
    \hline
    \hline
     \multirow{2}{*}{} & \multicolumn{4}{c|}{Planck+BAO/SN+D/H} \\
    \cline{2-5}
      & \multicolumn{1}{c}{$\Lambda$CDM} & \multicolumn{1}{c}{vEDE ($\theta_i = 2$)} & \multicolumn{1}{c}{vEDE ($\theta_i=2.5$)} & \multicolumn{1}{c|}{vEDE ($\theta_i = 3$)} \\
    \hline
    $R_\phi$ 
    & ...
    & $< 11.7$  
    & ...
    & ... \\
    $\log_{10}(a_c)$ 
    & ...
    & $-6.81^{+0.44}_{-0.49} $ 
    & $ -6.98^{+0.28}_{-0.40} $ 
    & $-7.02^{+0.33}_{-0.44}   $\\
    $\omega_b$ 
    & $0.02239\pm 0.00012 $
    & $0.02242\pm 0.00014$ 
    & $0.02241\pm 0.00013 $
    & $0.02241\pm 0.00013 $\\
    $\omega_{\cdm}$ 
    & $0.11960\pm 0.00087 $
    & $0.11972\pm 0.00094$ 
    & $0.11954\pm 0.00086 $
    & $0.11954\pm 0.00085 $\\
    $H_0 [\,\SI[per-mode=fraction]{}{\kilo\meter\per\second\per\mega\parsec}] $ 
    & $67.56\pm 0.38 $
    & $67.67^{+0.38}_{-0.44}$ 
    & $67.64\pm 0.38 $
    & $67.64\pm 0.38 $\\
    $10^{-9}A_s$ 
    & $2.113\pm 0.030 $
    & $2.114\pm 0.030$ 
    & $2.113\pm 0.030 $
    & $2.112\pm 0.030 $\\
    $n_s$ 
    & $0.9660\pm 0.0036 $
    & $0.9663\pm 0.0038$     
    & $0.9658\pm 0.0037 $
    & $0.9657\pm 0.0037 $\\
    $\tau_{reio}$ 
    & $0.0576\pm 0.0072 $
    & $0.0578\pm 0.0071$     
    & $0.0577\pm 0.0072 $
    & $0.0576\pm 0.0071 $\\
    $S_{8}$ 
    & $0.8293\pm 0.0098 $
    & $0.827^{+0.011}_{-0.0095}$
    & $0.827\pm 0.010 $
    & $0.827\pm 0.010 $\\
    \hline
    \hline
\end{tabularx}
    \label{tab:Analysis_A_posteriors}
\end{table}

\begin{table}[b]
\centering
    \caption{\footnotesize Posteriors for vEDE and cosmological parameters in Analysis $B$, corresponding to those shown in Fig.~\ref{fig:AnalysisB_full_triangle}.  Uncertainties are reported at 68\% C.L. and upper limits are given at 95\% C.L.}
    \setlength{\tabcolsep}{6pt} 
\begin{tabularx}{\linewidth}{|l|>{\centering\arraybackslash}p{0.15\linewidth}>{\centering\arraybackslash}p{0.15\linewidth}>{\centering\arraybackslash}X|>{\centering\arraybackslash}p{0.15\linewidth}>{\centering\arraybackslash}X|}
    \hline
    \hline
     \multirow{2}{*}{} & \multicolumn{3}{c|}{Planck+BAO/SN+D/H} & \multicolumn{2}{c|}{Planck+BAO/SN+D/H+SPT} \\
    \cline{2-6}
     & \multicolumn{1}{c}{$\Lambda$CDM} & \multicolumn{1}{c}{vEDE ($\theta_i=2$)} & \multicolumn{1}{c|}{vEDE ($\theta_i=2.5$)}& \multicolumn{1}{c}{$\Lambda$CDM} & \multicolumn{1}{c|}{vEDE ($\theta_i=2$)}\\
    \hline
    $R_\phi$ 
    & ...
    & $< 54.8$  
    & $ < 39.4 $
    & ...
    &  $< 52.5$        \\
    $\log_{10}(a_c)$ 
    & ...
    & $-7.05^{+0.22}_{-0.088} $
    & $ -7.07^{+0.21}_{-0.13}  $
    & ...
    & $-7.04^{+0.22}_{-0.082} $    \\
    $\omega_b$ 
    & $0.02239\pm 0.00012 $
    & $0.02243\pm 0.00013 $
    & $0.02243\pm 0.00013  $
    & $0.02236\pm 0.00011 $
    &   $0.02240\pm 0.00012 $    \\
    $\omega_{\cdm}$ 
    & $0.11960\pm 0.00087 $
    & $0.11955\pm 0.00088 $ 
    & $0.11957\pm 0.00086  $
    & $0.11955\pm 0.00084 $
    &   $0.11952\pm 0.00085 $   \\
    $H_0 [\,\SI[per-mode=fraction]{}{\kilo\meter\per\second\per\mega\parsec}] $ 
    & $67.56\pm 0.38 $
    & $67.71\pm 0.39 $ 
    & $67.69\pm 0.39  $
    & $67.54\pm 0.36 $
    &   $67.68\pm 0.37 $   \\
    $10^{-9}A_s$ 
    & $2.113\pm 0.030 $
    & $2.114^{+0.028}_{-0.032} $ 
    & $ 2.114^{+0.028}_{-0.032} $
    & $2.106\pm 0.028 $
    &   $2.104^{+0.026}_{-0.031} $    \\
    $n_s$ 
    & $0.9660\pm 0.0036 $
    & $0.9657\pm 0.0037 $ 
    & $0.9659\pm 0.0038  $
    & $0.9661\pm 0.0035 $
    &    $0.9652^{+0.0039}_{-0.0035} $    \\
    $\tau_{reio}$ 
    & $0.0576\pm 0.0072 $
    & $0.0582^{+0.0067}_{-0.0076} $ 
    & $0.0580^{+0.0067}_{-0.0076}  $
    & $ 0.0557\pm 0.0069 $
    &    $0.0557^{+0.0064}_{-0.0075} $         \\
    $S_{8}$ 
    & $0.8293\pm 0.0098 $
    & $0.825\pm 0.011$
    & $ 0.826\pm 0.010 $
    & $0.8278\pm 0.0096 $
    &   $0.8229\pm 0.0099 $           \\
    \hline
    \hline
\end{tabularx}
    \label{tab:Analysis_B_posteriors}
\end{table}

\begin{figure*}[b]
\centering
    \includegraphics[width=\linewidth]{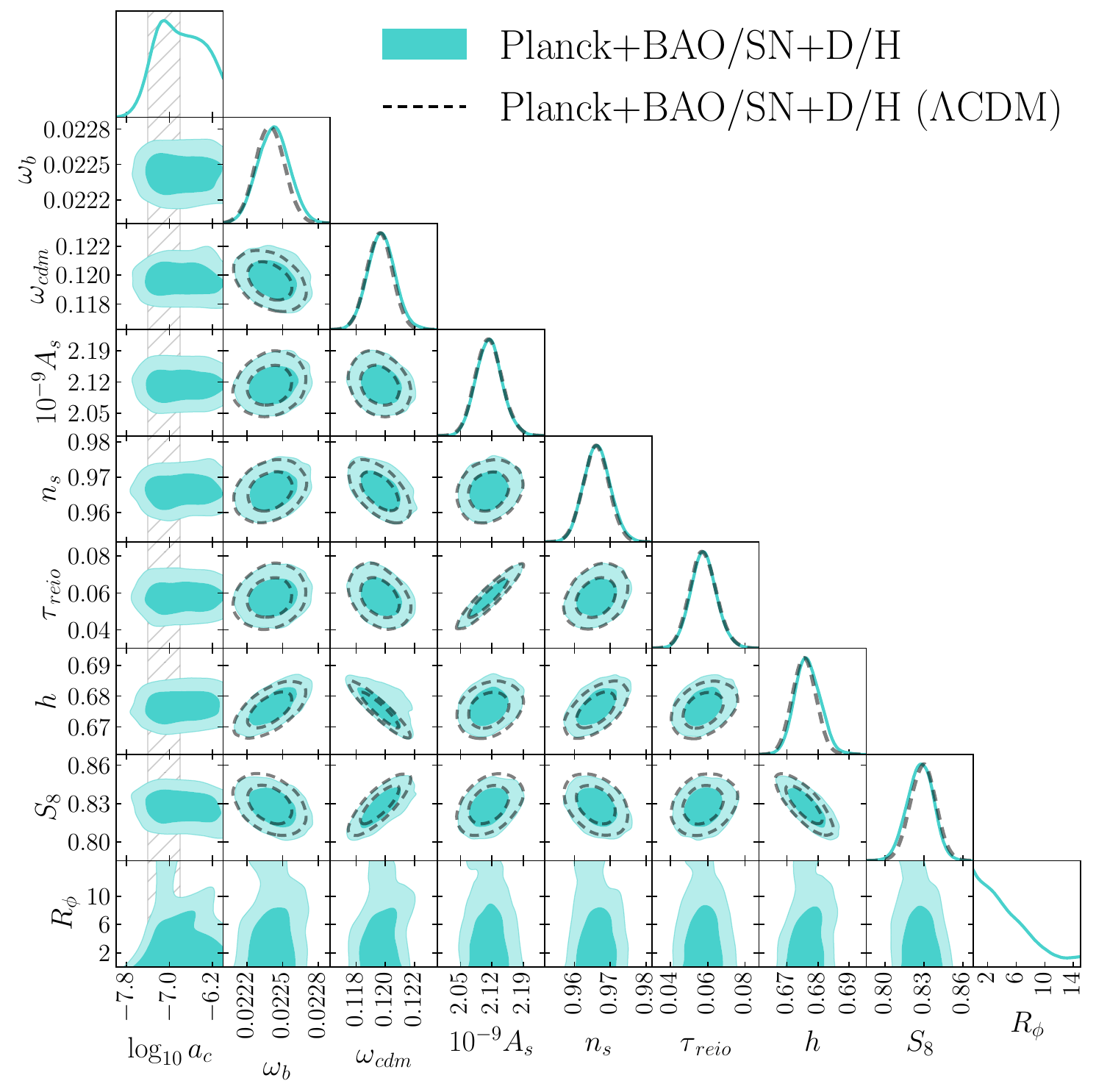}
  \caption{\footnotesize  1D and 2D posterior distributions of select parameters for Analysis $A$. We fix $n=8$ and $\theta_i = 2$. The dashed line shows the 1D posteriors for $\LCDM$ constrained by Planck+BAO/SN+D/H. The hatched region depicts the range of $a_c$ probed by Analysis $B$.}
  \label{fig:AnalysisA_full_triangle}
\end{figure*}

\begin{figure*}[b]
\centering
    \includegraphics[width=\linewidth]{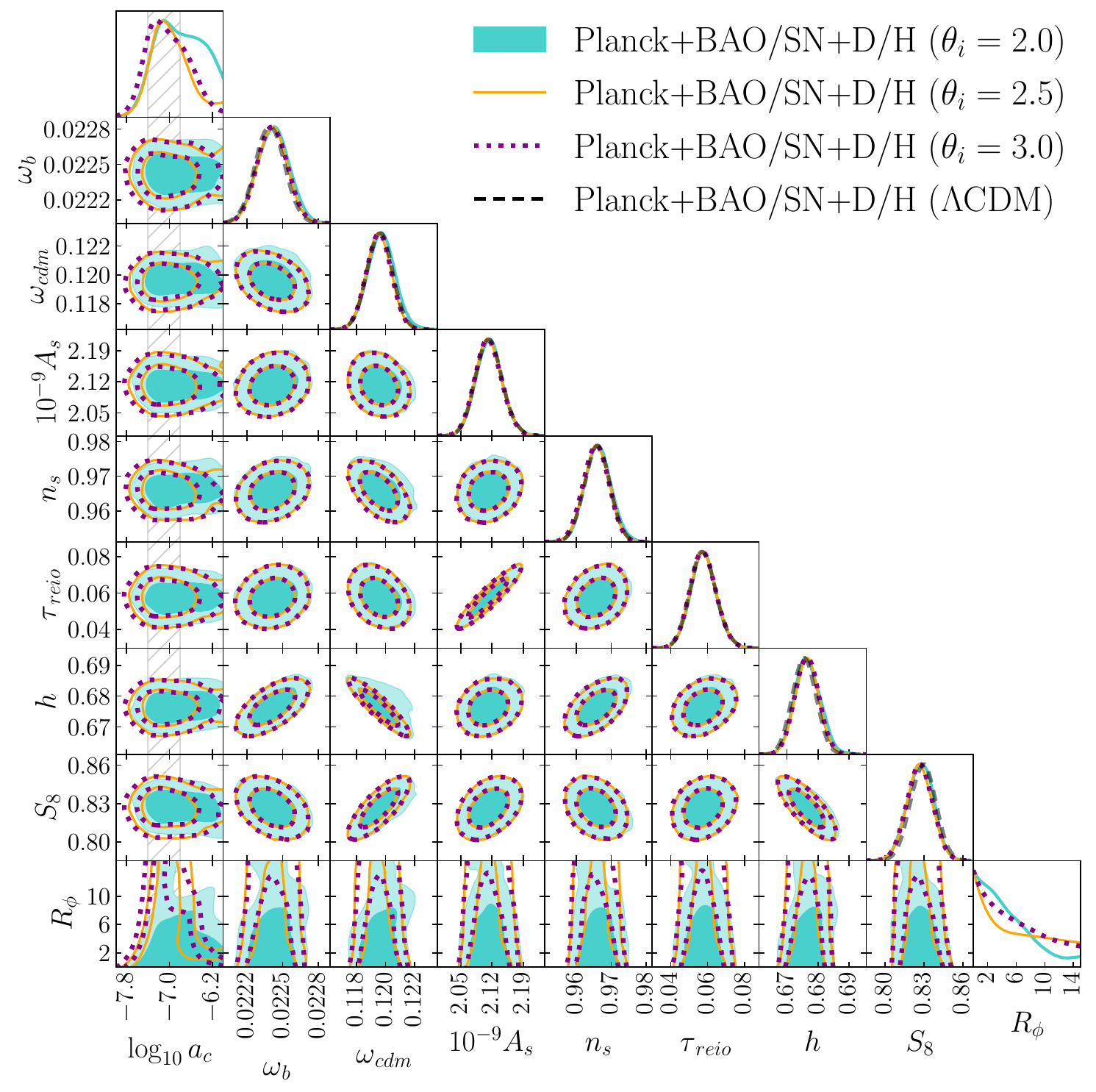}
  \caption{\footnotesize The same as Fig.~\ref{fig:AnalysisA_full_triangle}, but with additional posteriors for fixing $\theta_i = 2.5$ or $\theta_i=3.0$.}
  \label{fig:AnalysisA_full_triangle_thetas}
\end{figure*}

\begin{figure*}[b]
\centering
    \includegraphics[width=\linewidth]{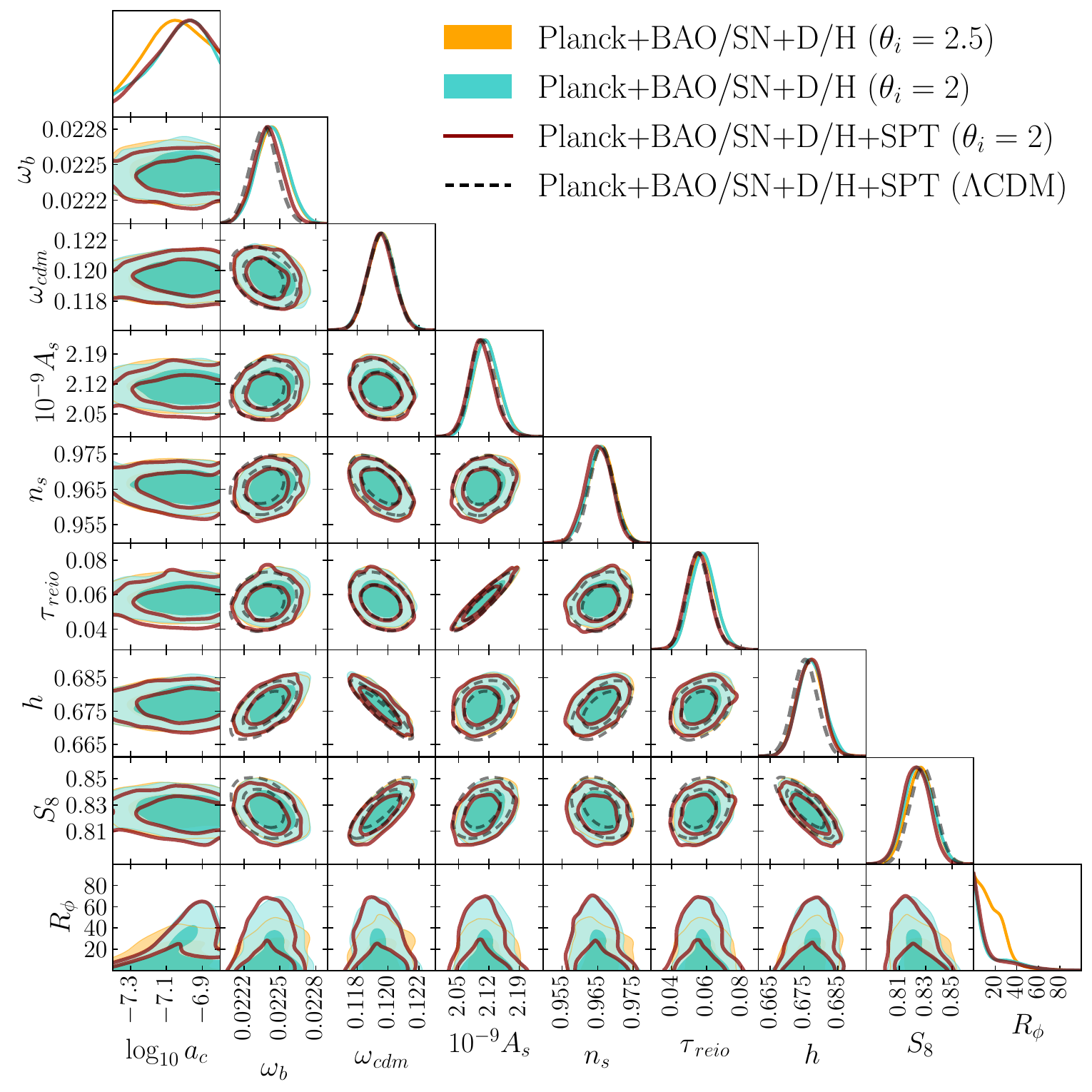}
  \caption{\footnotesize  1D and 2D posterior distributions of select parameters for Analysis $B$. We fix $n=8$ and show contours for $\theta_i$ fixed at either 2 or 2.5. The dashed line shows the 1D posteriors for $\LCDM$ constrained by Planck+BAO/SN+D/H+SPT.}
  \label{fig:AnalysisB_full_triangle}
\end{figure*}

\twocolumngrid

\clearpage
\bibliography{Signatures_of_very_Early_Dark_Energy_in_the_Matter_Power_Spectrum}{}

\begin{thebibliography}{115}%
\makeatletter
\providecommand \@ifxundefined [1]{%
 \@ifx{#1\undefined}
}%
\providecommand \@ifnum [1]{%
 \ifnum #1\expandafter \@firstoftwo
 \else \expandafter \@secondoftwo
 \fi
}%
\providecommand \@ifx [1]{%
 \ifx #1\expandafter \@firstoftwo
 \else \expandafter \@secondoftwo
 \fi
}%
\providecommand \natexlab [1]{#1}%
\providecommand \enquote  [1]{``#1''}%
\providecommand \bibnamefont  [1]{#1}%
\providecommand \bibfnamefont [1]{#1}%
\providecommand \citenamefont [1]{#1}%
\providecommand \href@noop [0]{\@secondoftwo}%
\providecommand \href [0]{\begingroup \@sanitize@url \@href}%
\providecommand \@href[1]{\@@startlink{#1}\@@href}%
\providecommand \@@href[1]{\endgroup#1\@@endlink}%
\providecommand \@sanitize@url [0]{\catcode `\\12\catcode `\$12\catcode `\&12\catcode `\#12\catcode `\^12\catcode `\_12\catcode `\%12\relax}%
\providecommand \@@startlink[1]{}%
\providecommand \@@endlink[0]{}%
\providecommand \url  [0]{\begingroup\@sanitize@url \@url }%
\providecommand \@url [1]{\endgroup\@href {#1}{\urlprefix }}%
\providecommand \urlprefix  [0]{URL }%
\providecommand \Eprint [0]{\href }%
\providecommand \doibase [0]{https://doi.org/}%
\providecommand \selectlanguage [0]{\@gobble}%
\providecommand \bibinfo  [0]{\@secondoftwo}%
\providecommand \bibfield  [0]{\@secondoftwo}%
\providecommand \translation [1]{[#1]}%
\providecommand \BibitemOpen [0]{}%
\providecommand \bibitemStop [0]{}%
\providecommand \bibitemNoStop [0]{.\EOS\space}%
\providecommand \EOS [0]{\spacefactor3000\relax}%
\providecommand \BibitemShut  [1]{\csname bibitem#1\endcsname}%
\let\auto@bib@innerbib\@empty
\bibitem [{\citenamefont {Svrcek}(2006)}]{svrcek_cosmological_2006}%
  \BibitemOpen
  \bibfield  {author} {\bibinfo {author} {\bibfnamefont {P.}~\bibnamefont {Svrcek}},\ }\bibfield  {journal} {\bibinfo  {journal} {arXiv e-prints}\ }\href {https://doi.org/10.48550/arXiv.hep-th/0607086} {10.48550/arXiv.hep-th/0607086} (\bibinfo {year} {2006}),\ \Eprint {https://arxiv.org/abs/hep-th/0607086} {arXiv:hep-th/0607086} \BibitemShut {NoStop}%
\bibitem [{\citenamefont {Svrcek}\ and\ \citenamefont {Witten}(2006)}]{svrcek_axions_2006}%
  \BibitemOpen
  \bibfield  {author} {\bibinfo {author} {\bibfnamefont {P.}~\bibnamefont {Svrcek}}\ and\ \bibinfo {author} {\bibfnamefont {E.}~\bibnamefont {Witten}},\ }\href {https://doi.org/10.1088/1126-6708/2006/06/051} {\bibfield  {journal} {\bibinfo  {journal} {J. High Energy Phys.}\ }\textbf {\bibinfo {volume} {2006}}\bibfield  {number} {\bibinfo  {number} { (06)},\ \bibinfo {pages} {051}},\ }\Eprint {https://arxiv.org/abs/hep-th/0605206} {arXiv:hep-th/0605206} \BibitemShut {NoStop}%
\bibitem [{\citenamefont {Arvanitaki}\ \emph {et~al.}(2010)\citenamefont {Arvanitaki}, \citenamefont {Dimopoulos}, \citenamefont {Dubovsky}, \citenamefont {Kaloper},\ and\ \citenamefont {{March-Russell}}}]{arvanitaki_string_2010}%
  \BibitemOpen
  \bibfield  {author} {\bibinfo {author} {\bibfnamefont {A.}~\bibnamefont {Arvanitaki}}, \bibinfo {author} {\bibfnamefont {S.}~\bibnamefont {Dimopoulos}}, \bibinfo {author} {\bibfnamefont {S.}~\bibnamefont {Dubovsky}}, \bibinfo {author} {\bibfnamefont {N.}~\bibnamefont {Kaloper}},\ and\ \bibinfo {author} {\bibfnamefont {J.}~\bibnamefont {{March-Russell}}},\ }\href {https://doi.org/10.1103/PhysRevD.81.123530} {\bibfield  {journal} {\bibinfo  {journal} {Phys. Rev. D}\ }\textbf {\bibinfo {volume} {81}},\ \bibinfo {pages} {123530} (\bibinfo {year} {2010})},\ \Eprint {https://arxiv.org/abs/0905.4720} {arXiv:0905.4720} \BibitemShut {NoStop}%
\bibitem [{\citenamefont {Marsh}(2016)}]{marsh_axion_2016}%
  \BibitemOpen
  \bibfield  {author} {\bibinfo {author} {\bibfnamefont {D.~J.~E.}\ \bibnamefont {Marsh}},\ }\href {https://doi.org/10.1016/j.physrep.2016.06.005} {\bibfield  {journal} {\bibinfo  {journal} {Physics Reports}\ }\textbf {\bibinfo {volume} {643}},\ \bibinfo {pages} {1} (\bibinfo {year} {2016})},\ \Eprint {https://arxiv.org/abs/1510.07633} {arXiv:1510.07633} \BibitemShut {NoStop}%
\bibitem [{\citenamefont {Dodelson}\ \emph {et~al.}(2000)\citenamefont {Dodelson}, \citenamefont {Kaplinghat},\ and\ \citenamefont {Stewart}}]{dodelson_solving_2000}%
  \BibitemOpen
  \bibfield  {author} {\bibinfo {author} {\bibfnamefont {S.}~\bibnamefont {Dodelson}}, \bibinfo {author} {\bibfnamefont {M.}~\bibnamefont {Kaplinghat}},\ and\ \bibinfo {author} {\bibfnamefont {E.}~\bibnamefont {Stewart}},\ }\href {https://doi.org/10.1103/PhysRevLett.85.5276} {\bibfield  {journal} {\bibinfo  {journal} {Phys. Rev. Lett.}\ }\textbf {\bibinfo {volume} {85}},\ \bibinfo {pages} {5276} (\bibinfo {year} {2000})},\ \Eprint {https://arxiv.org/abs/astro-ph/0002360} {arXiv:astro-ph/0002360} \BibitemShut {NoStop}%
\bibitem [{\citenamefont {Griest}(2002)}]{griest_toward_2002}%
  \BibitemOpen
  \bibfield  {author} {\bibinfo {author} {\bibfnamefont {K.}~\bibnamefont {Griest}},\ }\href {https://doi.org/10.1103/PhysRevD.66.123501} {\bibfield  {journal} {\bibinfo  {journal} {Phys. Rev. D}\ }\textbf {\bibinfo {volume} {66}},\ \bibinfo {pages} {123501} (\bibinfo {year} {2002})},\ \Eprint {https://arxiv.org/abs/astro-ph/0202052} {arXiv:astro-ph/0202052} \BibitemShut {NoStop}%
\bibitem [{\citenamefont {Kamionkowski}\ \emph {et~al.}(2014)\citenamefont {Kamionkowski}, \citenamefont {Pradler},\ and\ \citenamefont {Walker}}]{kamionkowski_dark_2014}%
  \BibitemOpen
  \bibfield  {author} {\bibinfo {author} {\bibfnamefont {M.}~\bibnamefont {Kamionkowski}}, \bibinfo {author} {\bibfnamefont {J.}~\bibnamefont {Pradler}},\ and\ \bibinfo {author} {\bibfnamefont {D.~G.~E.}\ \bibnamefont {Walker}},\ }\href {https://doi.org/10.1103/PhysRevLett.113.251302} {\bibfield  {journal} {\bibinfo  {journal} {Phys. Rev. Lett.}\ }\textbf {\bibinfo {volume} {113}},\ \bibinfo {pages} {251302} (\bibinfo {year} {2014})},\ \Eprint {https://arxiv.org/abs/1409.0549} {arXiv:1409.0549} \BibitemShut {NoStop}%
\bibitem [{\citenamefont {Karwal}\ and\ \citenamefont {Kamionkowski}(2016)}]{karwal_early_2016}%
  \BibitemOpen
  \bibfield  {author} {\bibinfo {author} {\bibfnamefont {T.}~\bibnamefont {Karwal}}\ and\ \bibinfo {author} {\bibfnamefont {M.}~\bibnamefont {Kamionkowski}},\ }\href {https://doi.org/10.1103/PhysRevD.94.103523} {\bibfield  {journal} {\bibinfo  {journal} {Phys. Rev. D}\ }\textbf {\bibinfo {volume} {94}},\ \bibinfo {pages} {103523} (\bibinfo {year} {2016})},\ \Eprint {https://arxiv.org/abs/1608.01309} {arXiv:1608.01309} \BibitemShut {NoStop}%
\bibitem [{\citenamefont {Yuan}\ \emph {et~al.}(2019)\citenamefont {Yuan}, \citenamefont {Riess}, \citenamefont {Macri}, \citenamefont {Casertano},\ and\ \citenamefont {Scolnic}}]{yuan_consistent_2019}%
  \BibitemOpen
  \bibfield  {author} {\bibinfo {author} {\bibfnamefont {W.}~\bibnamefont {Yuan}}, \bibinfo {author} {\bibfnamefont {A.~G.}\ \bibnamefont {Riess}}, \bibinfo {author} {\bibfnamefont {L.~M.}\ \bibnamefont {Macri}}, \bibinfo {author} {\bibfnamefont {S.}~\bibnamefont {Casertano}},\ and\ \bibinfo {author} {\bibfnamefont {D.}~\bibnamefont {Scolnic}},\ }\href {https://doi.org/10.3847/1538-4357/ab4bc9} {\bibfield  {journal} {\bibinfo  {journal} {ApJ}\ }\textbf {\bibinfo {volume} {886}},\ \bibinfo {pages} {61} (\bibinfo {year} {2019})},\ \Eprint {https://arxiv.org/abs/1908.00993} {arXiv:1908.00993} \BibitemShut {NoStop}%
\bibitem [{\citenamefont {Wong}\ \emph {et~al.}(2020)\citenamefont {Wong}, \citenamefont {Suyu}, \citenamefont {Chen}, \citenamefont {Rusu}, \citenamefont {Millon}, \citenamefont {Sluse}, \citenamefont {Bonvin}, \citenamefont {Fassnacht}, \citenamefont {Taubenberger}, \citenamefont {Auger} \emph {et~al.}}]{wong_h0licow_2020}%
  \BibitemOpen
  \bibfield  {author} {\bibinfo {author} {\bibfnamefont {K.~C.}\ \bibnamefont {Wong}}, \bibinfo {author} {\bibfnamefont {S.~H.}\ \bibnamefont {Suyu}}, \bibinfo {author} {\bibfnamefont {G.~C.-F.}\ \bibnamefont {Chen}}, \bibinfo {author} {\bibfnamefont {C.~E.}\ \bibnamefont {Rusu}}, \bibinfo {author} {\bibfnamefont {M.}~\bibnamefont {Millon}}, \bibinfo {author} {\bibfnamefont {D.}~\bibnamefont {Sluse}}, \bibinfo {author} {\bibfnamefont {V.}~\bibnamefont {Bonvin}}, \bibinfo {author} {\bibfnamefont {C.~D.}\ \bibnamefont {Fassnacht}}, \bibinfo {author} {\bibfnamefont {S.}~\bibnamefont {Taubenberger}}, \bibinfo {author} {\bibfnamefont {M.~W.}\ \bibnamefont {Auger}}, \emph {et~al.},\ }\href {https://doi.org/10.1093/mnras/stz3094} {\bibfield  {journal} {\bibinfo  {journal} {Monthly Notices of the Royal Astronomical Society}\ }\textbf {\bibinfo {volume} {498}},\ \bibinfo {pages} {1420} (\bibinfo {year} {2020})},\ \Eprint {https://arxiv.org/abs/1907.04869} {arXiv:1907.04869} \BibitemShut {NoStop}%
\bibitem [{\citenamefont {Blakeslee}\ \emph {et~al.}(2021)\citenamefont {Blakeslee}, \citenamefont {Jensen}, \citenamefont {Ma}, \citenamefont {Milne},\ and\ \citenamefont {Greene}}]{blakeslee_hubble_2021}%
  \BibitemOpen
  \bibfield  {author} {\bibinfo {author} {\bibfnamefont {J.~P.}\ \bibnamefont {Blakeslee}}, \bibinfo {author} {\bibfnamefont {J.~B.}\ \bibnamefont {Jensen}}, \bibinfo {author} {\bibfnamefont {C.-P.}\ \bibnamefont {Ma}}, \bibinfo {author} {\bibfnamefont {P.~A.}\ \bibnamefont {Milne}},\ and\ \bibinfo {author} {\bibfnamefont {J.~E.}\ \bibnamefont {Greene}},\ }\href {https://doi.org/10.3847/1538-4357/abe86a} {\bibfield  {journal} {\bibinfo  {journal} {ApJ}\ }\textbf {\bibinfo {volume} {911}},\ \bibinfo {pages} {65} (\bibinfo {year} {2021})},\ \Eprint {https://arxiv.org/abs/2101.02221} {arXiv:2101.02221} \BibitemShut {NoStop}%
\bibitem [{\citenamefont {Riess}\ \emph {et~al.}(2022)\citenamefont {Riess}, \citenamefont {Yuan}, \citenamefont {Macri}, \citenamefont {Scolnic}, \citenamefont {Brout}, \citenamefont {Casertano}, \citenamefont {Jones}, \citenamefont {Murakami}, \citenamefont {Breuval}, \citenamefont {Brink} \emph {et~al.}}]{riess_comprehensive_2022}%
  \BibitemOpen
  \bibfield  {author} {\bibinfo {author} {\bibfnamefont {A.~G.}\ \bibnamefont {Riess}}, \bibinfo {author} {\bibfnamefont {W.}~\bibnamefont {Yuan}}, \bibinfo {author} {\bibfnamefont {L.~M.}\ \bibnamefont {Macri}}, \bibinfo {author} {\bibfnamefont {D.}~\bibnamefont {Scolnic}}, \bibinfo {author} {\bibfnamefont {D.}~\bibnamefont {Brout}}, \bibinfo {author} {\bibfnamefont {S.}~\bibnamefont {Casertano}}, \bibinfo {author} {\bibfnamefont {D.~O.}\ \bibnamefont {Jones}}, \bibinfo {author} {\bibfnamefont {Y.}~\bibnamefont {Murakami}}, \bibinfo {author} {\bibfnamefont {L.}~\bibnamefont {Breuval}}, \bibinfo {author} {\bibfnamefont {T.~G.}\ \bibnamefont {Brink}}, \emph {et~al.},\ }\href {https://doi.org/10.3847/2041-8213/ac5c5b} {\bibfield  {journal} {\bibinfo  {journal} {The Astrophysical Journal Letters}\ }\textbf {\bibinfo {volume} {934}},\ \bibinfo {pages} {L7} (\bibinfo {year} {2022})},\ \Eprint {https://arxiv.org/abs/2112.04510} {arXiv:2112.04510} \BibitemShut {NoStop}%
\bibitem [{\citenamefont {{Planck Collaboration}}\ \emph {et~al.}(2020{\natexlab{a}})\citenamefont {{Planck Collaboration}}, \citenamefont {Aghanim} \emph {et~al.}}]{planck_collaboration_parameters}%
  \BibitemOpen
  \bibfield  {author} {\bibinfo {author} {\bibnamefont {{Planck Collaboration}}}, \bibinfo {author} {\bibfnamefont {N.}~\bibnamefont {Aghanim}}, \emph {et~al.},\ }\href {https://doi.org/10.1051/0004-6361/201833910} {\bibfield  {journal} {\bibinfo  {journal} {A\&A}\ }\textbf {\bibinfo {volume} {641}},\ \bibinfo {pages} {A6} (\bibinfo {year} {2020}{\natexlab{a}})},\ \Eprint {https://arxiv.org/abs/1807.06209} {arXiv:1807.06209} \BibitemShut {NoStop}%
\bibitem [{\citenamefont {Poulin}\ \emph {et~al.}(2018)\citenamefont {Poulin}, \citenamefont {Smith}, \citenamefont {Grin}, \citenamefont {Karwal},\ and\ \citenamefont {Kamionkowski}}]{poulin_cosmological_2018-1}%
  \BibitemOpen
  \bibfield  {author} {\bibinfo {author} {\bibfnamefont {V.}~\bibnamefont {Poulin}}, \bibinfo {author} {\bibfnamefont {T.~L.}\ \bibnamefont {Smith}}, \bibinfo {author} {\bibfnamefont {D.}~\bibnamefont {Grin}}, \bibinfo {author} {\bibfnamefont {T.}~\bibnamefont {Karwal}},\ and\ \bibinfo {author} {\bibfnamefont {M.}~\bibnamefont {Kamionkowski}},\ }\href {https://doi.org/10.1103/PhysRevD.98.083525} {\bibfield  {journal} {\bibinfo  {journal} {Phys. Rev. D}\ }\textbf {\bibinfo {volume} {98}},\ \bibinfo {pages} {083525} (\bibinfo {year} {2018})},\ \Eprint {https://arxiv.org/abs/1806.10608} {arXiv:1806.10608} \BibitemShut {NoStop}%
\bibitem [{\citenamefont {Poulin}\ \emph {et~al.}(2019)\citenamefont {Poulin}, \citenamefont {Smith}, \citenamefont {Karwal},\ and\ \citenamefont {Kamionkowski}}]{poulin_early_2019}%
  \BibitemOpen
  \bibfield  {author} {\bibinfo {author} {\bibfnamefont {V.}~\bibnamefont {Poulin}}, \bibinfo {author} {\bibfnamefont {T.~L.}\ \bibnamefont {Smith}}, \bibinfo {author} {\bibfnamefont {T.}~\bibnamefont {Karwal}},\ and\ \bibinfo {author} {\bibfnamefont {M.}~\bibnamefont {Kamionkowski}},\ }\href {https://doi.org/10.1103/PhysRevLett.122.221301} {\bibfield  {journal} {\bibinfo  {journal} {Phys. Rev. Lett.}\ }\textbf {\bibinfo {volume} {122}},\ \bibinfo {pages} {221301} (\bibinfo {year} {2019})},\ \Eprint {https://arxiv.org/abs/1811.04083} {arXiv:1811.04083} \BibitemShut {NoStop}%
\bibitem [{\citenamefont {Smith}\ \emph {et~al.}(2020)\citenamefont {Smith}, \citenamefont {Poulin},\ and\ \citenamefont {Amin}}]{smith_oscillating_2020}%
  \BibitemOpen
  \bibfield  {author} {\bibinfo {author} {\bibfnamefont {T.~L.}\ \bibnamefont {Smith}}, \bibinfo {author} {\bibfnamefont {V.}~\bibnamefont {Poulin}},\ and\ \bibinfo {author} {\bibfnamefont {M.~A.}\ \bibnamefont {Amin}},\ }\href {https://doi.org/10.1103/PhysRevD.101.063523} {\bibfield  {journal} {\bibinfo  {journal} {Phys. Rev. D}\ }\textbf {\bibinfo {volume} {101}},\ \bibinfo {pages} {063523} (\bibinfo {year} {2020})},\ \Eprint {https://arxiv.org/abs/1908.06995} {arXiv:1908.06995} \BibitemShut {NoStop}%
\bibitem [{\citenamefont {Murgia}\ \emph {et~al.}(2021)\citenamefont {Murgia}, \citenamefont {Abell{\'a}n},\ and\ \citenamefont {Poulin}}]{murgia_early_2021}%
  \BibitemOpen
  \bibfield  {author} {\bibinfo {author} {\bibfnamefont {R.}~\bibnamefont {Murgia}}, \bibinfo {author} {\bibfnamefont {G.~F.}\ \bibnamefont {Abell{\'a}n}},\ and\ \bibinfo {author} {\bibfnamefont {V.}~\bibnamefont {Poulin}},\ }\href {https://doi.org/10.1103/PhysRevD.103.063502} {\bibfield  {journal} {\bibinfo  {journal} {Phys. Rev. D}\ }\textbf {\bibinfo {volume} {103}},\ \bibinfo {pages} {063502} (\bibinfo {year} {2021})},\ \Eprint {https://arxiv.org/abs/2009.10733} {arXiv:2009.10733} \BibitemShut {NoStop}%
\bibitem [{\citenamefont {Smith}\ \emph {et~al.}(2022)\citenamefont {Smith}, \citenamefont {Lucca}, \citenamefont {Poulin}, \citenamefont {Abellan}, \citenamefont {Balkenhol}, \citenamefont {Benabed}, \citenamefont {Galli},\ and\ \citenamefont {Murgia}}]{smith_hints_2022}%
  \BibitemOpen
  \bibfield  {author} {\bibinfo {author} {\bibfnamefont {T.~L.}\ \bibnamefont {Smith}}, \bibinfo {author} {\bibfnamefont {M.}~\bibnamefont {Lucca}}, \bibinfo {author} {\bibfnamefont {V.}~\bibnamefont {Poulin}}, \bibinfo {author} {\bibfnamefont {G.~F.}\ \bibnamefont {Abellan}}, \bibinfo {author} {\bibfnamefont {L.}~\bibnamefont {Balkenhol}}, \bibinfo {author} {\bibfnamefont {K.}~\bibnamefont {Benabed}}, \bibinfo {author} {\bibfnamefont {S.}~\bibnamefont {Galli}},\ and\ \bibinfo {author} {\bibfnamefont {R.}~\bibnamefont {Murgia}},\ }\href {https://doi.org/10.1103/PhysRevD.106.043526} {\bibfield  {journal} {\bibinfo  {journal} {Phys. Rev. D}\ }\textbf {\bibinfo {volume} {106}},\ \bibinfo {pages} {043526} (\bibinfo {year} {2022})},\ \Eprint {https://arxiv.org/abs/2202.09379} {arXiv:2202.09379} \BibitemShut {NoStop}%
\bibitem [{\citenamefont {{Co}}\ \emph {et~al.}(2024)\citenamefont {{Co}}, \citenamefont {{Fernandez}}, \citenamefont {{Ghalsasi}}, \citenamefont {{Harigaya}},\ and\ \citenamefont {{Shelton}}}]{co_axion_2024}%
  \BibitemOpen
  \bibfield  {author} {\bibinfo {author} {\bibfnamefont {R.~T.}\ \bibnamefont {{Co}}}, \bibinfo {author} {\bibfnamefont {N.}~\bibnamefont {{Fernandez}}}, \bibinfo {author} {\bibfnamefont {A.}~\bibnamefont {{Ghalsasi}}}, \bibinfo {author} {\bibfnamefont {K.}~\bibnamefont {{Harigaya}}},\ and\ \bibinfo {author} {\bibfnamefont {J.}~\bibnamefont {{Shelton}}},\ }\href {https://doi.org/10.48550/arXiv.2405.12268} {\bibfield  {journal} {\bibinfo  {journal} {arXiv e-prints}\ ,\ \bibinfo {eid} {arXiv:2405.12268}} (\bibinfo {year} {2024})},\ \Eprint {https://arxiv.org/abs/2405.12268} {arXiv:2405.12268 [hep-ph]} \BibitemShut {NoStop}%
\bibitem [{\citenamefont {{DESI Collaboration}}\ \emph {et~al.}(2024)\citenamefont {{DESI Collaboration}}, \citenamefont {Adame} \emph {et~al.}}]{desi_collaboration_desi_2024-1}%
  \BibitemOpen
  \bibfield  {author} {\bibinfo {author} {\bibnamefont {{DESI Collaboration}}}, \bibinfo {author} {\bibfnamefont {A.~G.}\ \bibnamefont {Adame}}, \emph {et~al.},\ }\bibfield  {journal} {\bibinfo  {journal} {arXiv e-prints}\ }\href {https://doi.org/10.48550/arXiv.2404.03002} {10.48550/arXiv.2404.03002} (\bibinfo {year} {2024}),\ \Eprint {https://arxiv.org/abs/2404.03002} {arXiv:2404.03002} \BibitemShut {NoStop}%
\bibitem [{\citenamefont {Cort{\^e}s}\ and\ \citenamefont {Liddle}(2024)}]{cortes_interpreting_2024-1}%
  \BibitemOpen
  \bibfield  {author} {\bibinfo {author} {\bibfnamefont {M.}~\bibnamefont {Cort{\^e}s}}\ and\ \bibinfo {author} {\bibfnamefont {A.~R.}\ \bibnamefont {Liddle}},\ }\bibfield  {journal} {\bibinfo  {journal} {arXiv e-prints}\ }\href {https://doi.org/10.48550/arXiv.2404.08056} {10.48550/arXiv.2404.08056} (\bibinfo {year} {2024}),\ \Eprint {https://arxiv.org/abs/2404.08056} {arXiv:2404.08056} \BibitemShut {NoStop}%
\bibitem [{\citenamefont {Colg{\'a}in}\ \emph {et~al.}(2024)\citenamefont {Colg{\'a}in}, \citenamefont {Dainotti}, \citenamefont {Capozziello}, \citenamefont {Pourojaghi}, \citenamefont {{Sheikh-Jabbari}},\ and\ \citenamefont {Stojkovic}}]{colgain_does_2024-1}%
  \BibitemOpen
  \bibfield  {author} {\bibinfo {author} {\bibfnamefont {E.~{\'O}.}\ \bibnamefont {Colg{\'a}in}}, \bibinfo {author} {\bibfnamefont {M.~G.}\ \bibnamefont {Dainotti}}, \bibinfo {author} {\bibfnamefont {S.}~\bibnamefont {Capozziello}}, \bibinfo {author} {\bibfnamefont {S.}~\bibnamefont {Pourojaghi}}, \bibinfo {author} {\bibfnamefont {M.~M.}\ \bibnamefont {{Sheikh-Jabbari}}},\ and\ \bibinfo {author} {\bibfnamefont {D.}~\bibnamefont {Stojkovic}},\ }\bibfield  {journal} {\bibinfo  {journal} {arXiv e-prints}\ }\href {https://doi.org/10.48550/arXiv.2404.08633} {10.48550/arXiv.2404.08633} (\bibinfo {year} {2024}),\ \Eprint {https://arxiv.org/abs/2404.08633} {arXiv:2404.08633} \BibitemShut {NoStop}%
\bibitem [{\citenamefont {Wolf}\ and\ \citenamefont {Ferreira}(2023)}]{wolf_underdetermination_2023}%
  \BibitemOpen
  \bibfield  {author} {\bibinfo {author} {\bibfnamefont {W.~J.}\ \bibnamefont {Wolf}}\ and\ \bibinfo {author} {\bibfnamefont {P.~G.}\ \bibnamefont {Ferreira}},\ }\href {https://doi.org/10.1103/PhysRevD.108.103519} {\bibfield  {journal} {\bibinfo  {journal} {Phys. Rev. D}\ }\textbf {\bibinfo {volume} {108}},\ \bibinfo {pages} {103519} (\bibinfo {year} {2023})},\ \Eprint {https://arxiv.org/abs/2310.07482} {arXiv:2310.07482} \BibitemShut {NoStop}%
\bibitem [{\citenamefont {Wolf}\ \emph {et~al.}(2024)\citenamefont {Wolf}, \citenamefont {{Garc{\'i}a-Garc{\'i}a}}, \citenamefont {Bartlett},\ and\ \citenamefont {Ferreira}}]{wolf_scant_2024}%
  \BibitemOpen
  \bibfield  {author} {\bibinfo {author} {\bibfnamefont {W.~J.}\ \bibnamefont {Wolf}}, \bibinfo {author} {\bibfnamefont {C.}~\bibnamefont {{Garc{\'i}a-Garc{\'i}a}}}, \bibinfo {author} {\bibfnamefont {D.~J.}\ \bibnamefont {Bartlett}},\ and\ \bibinfo {author} {\bibfnamefont {P.~G.}\ \bibnamefont {Ferreira}},\ }\bibfield  {journal} {\bibinfo  {journal} {arXiv e-prints}\ }\href {https://doi.org/10.48550/arXiv.2408.17318} {10.48550/arXiv.2408.17318} (\bibinfo {year} {2024}),\ \Eprint {https://arxiv.org/abs/2408.17318} {arXiv:2408.17318} \BibitemShut {NoStop}%
\bibitem [{\citenamefont {Colg{\'a}in}\ and\ \citenamefont {{Sheikh-Jabbari}}(2024)}]{colgain_desi_2024}%
  \BibitemOpen
  \bibfield  {author} {\bibinfo {author} {\bibfnamefont {E.~{\'O}.}\ \bibnamefont {Colg{\'a}in}}\ and\ \bibinfo {author} {\bibfnamefont {M.~M.}\ \bibnamefont {{Sheikh-Jabbari}}},\ }\bibfield  {journal} {\bibinfo  {journal} {arXiv e-prints}\ }\href {https://doi.org/10.48550/arXiv.2412.12905} {10.48550/arXiv.2412.12905} (\bibinfo {year} {2024}),\ \Eprint {https://arxiv.org/abs/2412.12905} {arXiv:2412.12905} \BibitemShut {NoStop}%
\bibitem [{\citenamefont {Berghaus}\ \emph {et~al.}(2024)\citenamefont {Berghaus}, \citenamefont {Kable},\ and\ \citenamefont {Miranda}}]{berghaus_quantifying_2024}%
  \BibitemOpen
  \bibfield  {author} {\bibinfo {author} {\bibfnamefont {K.~V.}\ \bibnamefont {Berghaus}}, \bibinfo {author} {\bibfnamefont {J.~A.}\ \bibnamefont {Kable}},\ and\ \bibinfo {author} {\bibfnamefont {V.}~\bibnamefont {Miranda}},\ }\bibfield  {journal} {\bibinfo  {journal} {arXiv e-prints}\ }\href {https://doi.org/10.48550/arXiv.2404.14341} {10.48550/arXiv.2404.14341} (\bibinfo {year} {2024}),\ \Eprint {https://arxiv.org/abs/2404.14341} {arXiv:2404.14341} \BibitemShut {NoStop}%
\bibitem [{\citenamefont {{Rezazadeh}}\ \emph {et~al.}(2024)\citenamefont {{Rezazadeh}}, \citenamefont {{Ashoorioon}},\ and\ \citenamefont {{Grin}}}]{Rezazadeh_cascading_2024}%
  \BibitemOpen
  \bibfield  {author} {\bibinfo {author} {\bibfnamefont {K.}~\bibnamefont {{Rezazadeh}}}, \bibinfo {author} {\bibfnamefont {A.}~\bibnamefont {{Ashoorioon}}},\ and\ \bibinfo {author} {\bibfnamefont {D.}~\bibnamefont {{Grin}}},\ }\href {https://doi.org/10.3847/1538-4357/ad7b16} {\bibfield  {journal} {\bibinfo  {journal} {\apj}\ }\textbf {\bibinfo {volume} {975}},\ \bibinfo {eid} {137} (\bibinfo {year} {2024})},\ \Eprint {https://arxiv.org/abs/2208.07631} {arXiv:2208.07631 [astro-ph.CO]} \BibitemShut {NoStop}%
\bibitem [{\citenamefont {Spokoiny}(1993)}]{spokoiny_deflationary_1993-1}%
  \BibitemOpen
  \bibfield  {author} {\bibinfo {author} {\bibfnamefont {B.}~\bibnamefont {Spokoiny}},\ }\href {https://doi.org/10.1016/0370-2693(93)90155-B} {\bibfield  {journal} {\bibinfo  {journal} {Physics Letters B}\ }\textbf {\bibinfo {volume} {315}},\ \bibinfo {pages} {40} (\bibinfo {year} {1993})},\ \Eprint {https://arxiv.org/abs/gr-qc/9306008} {arXiv:gr-qc/9306008} \BibitemShut {NoStop}%
\bibitem [{\citenamefont {Joyce}(1997)}]{joyce_electroweak_1997-1}%
  \BibitemOpen
  \bibfield  {author} {\bibinfo {author} {\bibfnamefont {M.}~\bibnamefont {Joyce}},\ }\href {https://doi.org/10.1103/PhysRevD.55.1875} {\bibfield  {journal} {\bibinfo  {journal} {Phys. Rev. D}\ }\textbf {\bibinfo {volume} {55}},\ \bibinfo {pages} {1875} (\bibinfo {year} {1997})},\ \Eprint {https://arxiv.org/abs/hep-ph/9606223} {arXiv:hep-ph/9606223} \BibitemShut {NoStop}%
\bibitem [{\citenamefont {Ferreira}\ and\ \citenamefont {Joyce}(1998)}]{ferreira_cosmology_1998-2}%
  \BibitemOpen
  \bibfield  {author} {\bibinfo {author} {\bibfnamefont {P.~G.}\ \bibnamefont {Ferreira}}\ and\ \bibinfo {author} {\bibfnamefont {M.}~\bibnamefont {Joyce}},\ }\href {https://doi.org/10.1103/PhysRevD.58.023503} {\bibfield  {journal} {\bibinfo  {journal} {Phys. Rev. D}\ }\textbf {\bibinfo {volume} {58}},\ \bibinfo {pages} {023503} (\bibinfo {year} {1998})},\ \Eprint {https://arxiv.org/abs/astro-ph/9711102} {arXiv:astro-ph/9711102} \BibitemShut {NoStop}%
\bibitem [{\citenamefont {Co}\ \emph {et~al.}(2022)\citenamefont {Co}, \citenamefont {Dunsky}, \citenamefont {Fernandez}, \citenamefont {Ghalsasi}, \citenamefont {Hall}, \citenamefont {Harigaya},\ and\ \citenamefont {Shelton}}]{co_gravitational_2022}%
  \BibitemOpen
  \bibfield  {author} {\bibinfo {author} {\bibfnamefont {R.~T.}\ \bibnamefont {Co}}, \bibinfo {author} {\bibfnamefont {D.}~\bibnamefont {Dunsky}}, \bibinfo {author} {\bibfnamefont {N.}~\bibnamefont {Fernandez}}, \bibinfo {author} {\bibfnamefont {A.}~\bibnamefont {Ghalsasi}}, \bibinfo {author} {\bibfnamefont {L.~J.}\ \bibnamefont {Hall}}, \bibinfo {author} {\bibfnamefont {K.}~\bibnamefont {Harigaya}},\ and\ \bibinfo {author} {\bibfnamefont {J.}~\bibnamefont {Shelton}},\ }\href {https://doi.org/10.1007/JHEP09(2022)116} {\bibfield  {journal} {\bibinfo  {journal} {J. High Energ. Phys.}\ }\textbf {\bibinfo {volume} {2022}}\bibfield  {number} {\bibinfo  {number} { (9)},\ \bibinfo {pages} {116}},\ }\Eprint {https://arxiv.org/abs/2108.09299} {arXiv:2108.09299} \BibitemShut {NoStop}%
\bibitem [{\citenamefont {Redmond}\ \emph {et~al.}(2018)\citenamefont {Redmond}, \citenamefont {Trezza},\ and\ \citenamefont {Erickcek}}]{redmond_growth_2018}%
  \BibitemOpen
  \bibfield  {author} {\bibinfo {author} {\bibfnamefont {K.}~\bibnamefont {Redmond}}, \bibinfo {author} {\bibfnamefont {A.}~\bibnamefont {Trezza}},\ and\ \bibinfo {author} {\bibfnamefont {A.~L.}\ \bibnamefont {Erickcek}},\ }\href {https://doi.org/10.1103/PhysRevD.98.063504} {\bibfield  {journal} {\bibinfo  {journal} {Phys. Rev. D}\ }\textbf {\bibinfo {volume} {98}},\ \bibinfo {pages} {063504} (\bibinfo {year} {2018})},\ \Eprint {https://arxiv.org/abs/1807.01327} {arXiv:1807.01327} \BibitemShut {NoStop}%
\bibitem [{\citenamefont {{Delos}}\ \emph {et~al.}(2023)\citenamefont {{Delos}}, \citenamefont {{Redmond}},\ and\ \citenamefont {{Erickcek}}}]{delos_how_2023}%
  \BibitemOpen
  \bibfield  {author} {\bibinfo {author} {\bibfnamefont {M.~S.}\ \bibnamefont {{Delos}}}, \bibinfo {author} {\bibfnamefont {K.}~\bibnamefont {{Redmond}}},\ and\ \bibinfo {author} {\bibfnamefont {A.~L.}\ \bibnamefont {{Erickcek}}},\ }\href {https://doi.org/10.1103/PhysRevD.108.023528} {\bibfield  {journal} {\bibinfo  {journal} {\prd}\ }\textbf {\bibinfo {volume} {108}},\ \bibinfo {eid} {023528} (\bibinfo {year} {2023})},\ \Eprint {https://arxiv.org/abs/2304.12336} {arXiv:2304.12336} \BibitemShut {NoStop}%
\bibitem [{\citenamefont {Jaber}\ \emph {et~al.}(2020)\citenamefont {Jaber}, \citenamefont {Almaraz},\ and\ \citenamefont {{de la Macorra}}}]{jaber_imprint_2020}%
  \BibitemOpen
  \bibfield  {author} {\bibinfo {author} {\bibfnamefont {M.}~\bibnamefont {Jaber}}, \bibinfo {author} {\bibfnamefont {E.}~\bibnamefont {Almaraz}},\ and\ \bibinfo {author} {\bibfnamefont {A.}~\bibnamefont {{de la Macorra}}},\ }\href {https://doi.org/10.1016/j.astropartphys.2019.102388} {\bibfield  {journal} {\bibinfo  {journal} {Astroparticle Physics}\ }\textbf {\bibinfo {volume} {115}},\ \bibinfo {pages} {102388} (\bibinfo {year} {2020})},\ \Eprint {https://arxiv.org/abs/1906.09522} {arXiv:1906.09522} \BibitemShut {NoStop}%
\bibitem [{\citenamefont {{de la Macorra}}\ \emph {et~al.}(2021)\citenamefont {{de la Macorra}}, \citenamefont {{Gomez-Navarro}}, \citenamefont {Aviles}, \citenamefont {Jaber}, \citenamefont {Mastache},\ and\ \citenamefont {Almaraz}}]{de_la_macorra_cosmological_2021}%
  \BibitemOpen
  \bibfield  {author} {\bibinfo {author} {\bibfnamefont {A.}~\bibnamefont {{de la Macorra}}}, \bibinfo {author} {\bibfnamefont {D.~V.}\ \bibnamefont {{Gomez-Navarro}}}, \bibinfo {author} {\bibfnamefont {A.}~\bibnamefont {Aviles}}, \bibinfo {author} {\bibfnamefont {M.}~\bibnamefont {Jaber}}, \bibinfo {author} {\bibfnamefont {J.}~\bibnamefont {Mastache}},\ and\ \bibinfo {author} {\bibfnamefont {E.}~\bibnamefont {Almaraz}},\ }\href {https://doi.org/10.1103/PhysRevD.104.023529} {\bibfield  {journal} {\bibinfo  {journal} {Phys. Rev. D}\ }\textbf {\bibinfo {volume} {104}},\ \bibinfo {pages} {023529} (\bibinfo {year} {2021})},\ \Eprint {https://arxiv.org/abs/2009.12673} {arXiv:2009.12673} \BibitemShut {NoStop}%
\bibitem [{\citenamefont {{de la Macorra}}\ and\ \citenamefont {Almaraz}(2018)}]{de_la_macorra_testing_2018}%
  \BibitemOpen
  \bibfield  {author} {\bibinfo {author} {\bibfnamefont {A.}~\bibnamefont {{de la Macorra}}}\ and\ \bibinfo {author} {\bibfnamefont {E.}~\bibnamefont {Almaraz}},\ }\href {https://doi.org/10.1103/PhysRevLett.121.161303} {\bibfield  {journal} {\bibinfo  {journal} {Phys. Rev. Lett.}\ }\textbf {\bibinfo {volume} {121}},\ \bibinfo {pages} {161303} (\bibinfo {year} {2018})},\ \Eprint {https://arxiv.org/abs/1805.01510} {arXiv:1805.01510} \BibitemShut {NoStop}%
\bibitem [{\citenamefont {Almaraz}\ and\ \citenamefont {{de la Macorra}}(2019)}]{almaraz_bound_2019}%
  \BibitemOpen
  \bibfield  {author} {\bibinfo {author} {\bibfnamefont {E.}~\bibnamefont {Almaraz}}\ and\ \bibinfo {author} {\bibfnamefont {A.}~\bibnamefont {{de la Macorra}}},\ }\href {https://doi.org/10.1103/PhysRevD.99.103504} {\bibfield  {journal} {\bibinfo  {journal} {Phys. Rev. D}\ }\textbf {\bibinfo {volume} {99}},\ \bibinfo {pages} {103504} (\bibinfo {year} {2019})},\ \Eprint {https://arxiv.org/abs/1812.01133} {arXiv:1812.01133} \BibitemShut {NoStop}%
\bibitem [{\citenamefont {Poulin}\ \emph {et~al.}(2023)\citenamefont {Poulin}, \citenamefont {Smith},\ and\ \citenamefont {Karwal}}]{poulin_ups_2023}%
  \BibitemOpen
  \bibfield  {author} {\bibinfo {author} {\bibfnamefont {V.}~\bibnamefont {Poulin}}, \bibinfo {author} {\bibfnamefont {T.~L.}\ \bibnamefont {Smith}},\ and\ \bibinfo {author} {\bibfnamefont {T.}~\bibnamefont {Karwal}},\ }\bibfield  {journal} {\bibinfo  {journal} {arXiv e-prints}\ }\href {https://doi.org/10.48550/arXiv.2302.09032} {10.48550/arXiv.2302.09032} (\bibinfo {year} {2023}),\ \Eprint {https://arxiv.org/abs/2302.09032} {arXiv:2302.09032} \BibitemShut {NoStop}%
\bibitem [{\citenamefont {Klypin}\ \emph {et~al.}(2021)\citenamefont {Klypin}, \citenamefont {Poulin}, \citenamefont {Prada}, \citenamefont {Primack}, \citenamefont {Kamionkowski}, \citenamefont {{Avila-Reese}}, \citenamefont {{Rodriguez-Puebla}}, \citenamefont {Behroozi}, \citenamefont {Hellinger},\ and\ \citenamefont {Smith}}]{klypin_clustering_2021}%
  \BibitemOpen
  \bibfield  {author} {\bibinfo {author} {\bibfnamefont {A.}~\bibnamefont {Klypin}}, \bibinfo {author} {\bibfnamefont {V.}~\bibnamefont {Poulin}}, \bibinfo {author} {\bibfnamefont {F.}~\bibnamefont {Prada}}, \bibinfo {author} {\bibfnamefont {J.}~\bibnamefont {Primack}}, \bibinfo {author} {\bibfnamefont {M.}~\bibnamefont {Kamionkowski}}, \bibinfo {author} {\bibfnamefont {V.}~\bibnamefont {{Avila-Reese}}}, \bibinfo {author} {\bibfnamefont {A.}~\bibnamefont {{Rodriguez-Puebla}}}, \bibinfo {author} {\bibfnamefont {P.}~\bibnamefont {Behroozi}}, \bibinfo {author} {\bibfnamefont {D.}~\bibnamefont {Hellinger}},\ and\ \bibinfo {author} {\bibfnamefont {T.~L.}\ \bibnamefont {Smith}},\ }\href {https://doi.org/10.1093/mnras/stab769} {\bibfield  {journal} {\bibinfo  {journal} {Mon. Not. R. Astron. Soc.}\ }\textbf {\bibinfo {volume} {504}},\ \bibinfo {pages} {769} (\bibinfo {year} {2021})},\ \Eprint {https://arxiv.org/abs/2006.14910} {arXiv:2006.14910} \BibitemShut {NoStop}%
\bibitem [{\citenamefont {McDonough}\ \emph {et~al.}(2023)\citenamefont {McDonough}, \citenamefont {Hill}, \citenamefont {Ivanov}, \citenamefont {La~Posta},\ and\ \citenamefont {Toomey}}]{mcdonough_observational_2023-1}%
  \BibitemOpen
  \bibfield  {author} {\bibinfo {author} {\bibfnamefont {E.}~\bibnamefont {McDonough}}, \bibinfo {author} {\bibfnamefont {J.~C.}\ \bibnamefont {Hill}}, \bibinfo {author} {\bibfnamefont {M.~M.}\ \bibnamefont {Ivanov}}, \bibinfo {author} {\bibfnamefont {A.}~\bibnamefont {La~Posta}},\ and\ \bibinfo {author} {\bibfnamefont {M.~W.}\ \bibnamefont {Toomey}},\ }\bibfield  {journal} {\bibinfo  {journal} {arXiv e-prints}\ }\href {https://doi.org/10.48550/arXiv.2310.19899} {10.48550/arXiv.2310.19899} (\bibinfo {year} {2023}),\ \Eprint {https://arxiv.org/abs/2310.19899} {arXiv:2310.19899} \BibitemShut {NoStop}%
\bibitem [{\citenamefont {Goldstein}\ \emph {et~al.}(2023)\citenamefont {Goldstein}, \citenamefont {Hill}, \citenamefont {Ir{\v s}i{\v c}},\ and\ \citenamefont {Sherwin}}]{goldstein_canonical_2023}%
  \BibitemOpen
  \bibfield  {author} {\bibinfo {author} {\bibfnamefont {S.}~\bibnamefont {Goldstein}}, \bibinfo {author} {\bibfnamefont {J.~C.}\ \bibnamefont {Hill}}, \bibinfo {author} {\bibfnamefont {V.}~\bibnamefont {Ir{\v s}i{\v c}}},\ and\ \bibinfo {author} {\bibfnamefont {B.~D.}\ \bibnamefont {Sherwin}},\ }\href {https://doi.org/10.1103/PhysRevLett.131.201001} {\bibfield  {journal} {\bibinfo  {journal} {Physical Review Letters}\ }\textbf {\bibinfo {volume} {131}},\ \bibinfo {pages} {201001} (\bibinfo {year} {2023})}\BibitemShut {NoStop}%
\bibitem [{\citenamefont {Padmanabhan}\ and\ \citenamefont {Loeb}(2023)}]{padmanabhan_alleviating_2023-1}%
  \BibitemOpen
  \bibfield  {author} {\bibinfo {author} {\bibfnamefont {H.}~\bibnamefont {Padmanabhan}}\ and\ \bibinfo {author} {\bibfnamefont {A.}~\bibnamefont {Loeb}},\ }\href {https://doi.org/10.3847/2041-8213/acea7a} {\bibfield  {journal} {\bibinfo  {journal} {The Astrophysical Journal Letters}\ }\textbf {\bibinfo {volume} {953}},\ \bibinfo {eid} {L4} (\bibinfo {year} {2023})},\ \Eprint {https://arxiv.org/abs/2306.04684} {arXiv:2306.04684} \BibitemShut {NoStop}%
\bibitem [{\citenamefont {{Tkachev}}\ \emph {et~al.}(2024)\citenamefont {{Tkachev}}, \citenamefont {{Pilipenko}}, \citenamefont {{Mikheeva}},\ and\ \citenamefont {{Lukash}}}]{Tkachev_Excess_2024}%
  \BibitemOpen
  \bibfield  {author} {\bibinfo {author} {\bibfnamefont {M.~V.}\ \bibnamefont {{Tkachev}}}, \bibinfo {author} {\bibfnamefont {S.~V.}\ \bibnamefont {{Pilipenko}}}, \bibinfo {author} {\bibfnamefont {E.~V.}\ \bibnamefont {{Mikheeva}}},\ and\ \bibinfo {author} {\bibfnamefont {V.~N.}\ \bibnamefont {{Lukash}}},\ }\href {https://doi.org/10.1093/mnras/stad3279} {\bibfield  {journal} {\bibinfo  {journal} {Monthly Notices of the Royal Astronomical Society}\ }\textbf {\bibinfo {volume} {527}},\ \bibinfo {pages} {1381} (\bibinfo {year} {2024})},\ \Eprint {https://arxiv.org/abs/2307.13774} {arXiv:2307.13774 [astro-ph.CO]} \BibitemShut {NoStop}%
\bibitem [{\citenamefont {Esteban}\ \emph {et~al.}(2023)\citenamefont {Esteban}, \citenamefont {Peter},\ and\ \citenamefont {Kim}}]{esteban_milky_2023}%
  \BibitemOpen
  \bibfield  {author} {\bibinfo {author} {\bibfnamefont {I.}~\bibnamefont {Esteban}}, \bibinfo {author} {\bibfnamefont {A.~H.~G.}\ \bibnamefont {Peter}},\ and\ \bibinfo {author} {\bibfnamefont {S.~Y.}\ \bibnamefont {Kim}},\ }\bibfield  {journal} {\bibinfo  {journal} {arXiv e-prints}\ }\href {https://doi.org/10.48550/arXiv.2306.04674} {10.48550/arXiv.2306.04674} (\bibinfo {year} {2023}),\ \Eprint {https://arxiv.org/abs/2306.04674} {arXiv:2306.04674} \BibitemShut {NoStop}%
\bibitem [{\citenamefont {Lyke}\ \emph {et~al.}(2020)\citenamefont {Lyke}, \citenamefont {Higley}, \citenamefont {McLane}, \citenamefont {Schurhammer}, \citenamefont {Myers}, \citenamefont {Ross}, \citenamefont {Dawson}, \citenamefont {Chabanier}, \citenamefont {Martini}, \citenamefont {Busca} \emph {et~al.}}]{lyke_sloan_2020}%
  \BibitemOpen
  \bibfield  {author} {\bibinfo {author} {\bibfnamefont {B.~W.}\ \bibnamefont {Lyke}}, \bibinfo {author} {\bibfnamefont {A.~N.}\ \bibnamefont {Higley}}, \bibinfo {author} {\bibfnamefont {J.~N.}\ \bibnamefont {McLane}}, \bibinfo {author} {\bibfnamefont {D.~P.}\ \bibnamefont {Schurhammer}}, \bibinfo {author} {\bibfnamefont {A.~D.}\ \bibnamefont {Myers}}, \bibinfo {author} {\bibfnamefont {A.~J.}\ \bibnamefont {Ross}}, \bibinfo {author} {\bibfnamefont {K.}~\bibnamefont {Dawson}}, \bibinfo {author} {\bibfnamefont {S.}~\bibnamefont {Chabanier}}, \bibinfo {author} {\bibfnamefont {P.}~\bibnamefont {Martini}}, \bibinfo {author} {\bibfnamefont {N.~G.}\ \bibnamefont {Busca}}, \emph {et~al.},\ }\href {https://doi.org/10.3847/1538-4365/aba623} {\bibfield  {journal} {\bibinfo  {journal} {ApJS}\ }\textbf {\bibinfo {volume} {250}},\ \bibinfo {pages} {8} (\bibinfo {year} {2020})},\ \Eprint {https://arxiv.org/abs/2007.09001} {arXiv:2007.09001} \BibitemShut {NoStop}%
\bibitem [{\citenamefont {Chabanier}\ \emph {et~al.}(2019)\citenamefont {Chabanier}, \citenamefont {{Palanque-Delabrouille}}, \citenamefont {Y{\`e}che}, \citenamefont {Goff}, \citenamefont {Armengaud}, \citenamefont {Bautista}, \citenamefont {Blomqvist}, \citenamefont {Busca}, \citenamefont {Dawson}, \citenamefont {Etourneau} \emph {et~al.}}]{chabanier_one-dimensional_2019}%
  \BibitemOpen
  \bibfield  {author} {\bibinfo {author} {\bibfnamefont {S.}~\bibnamefont {Chabanier}}, \bibinfo {author} {\bibfnamefont {N.}~\bibnamefont {{Palanque-Delabrouille}}}, \bibinfo {author} {\bibfnamefont {C.}~\bibnamefont {Y{\`e}che}}, \bibinfo {author} {\bibfnamefont {J.-M.~L.}\ \bibnamefont {Goff}}, \bibinfo {author} {\bibfnamefont {E.}~\bibnamefont {Armengaud}}, \bibinfo {author} {\bibfnamefont {J.}~\bibnamefont {Bautista}}, \bibinfo {author} {\bibfnamefont {M.}~\bibnamefont {Blomqvist}}, \bibinfo {author} {\bibfnamefont {N.}~\bibnamefont {Busca}}, \bibinfo {author} {\bibfnamefont {K.}~\bibnamefont {Dawson}}, \bibinfo {author} {\bibfnamefont {T.}~\bibnamefont {Etourneau}}, \emph {et~al.},\ }\href {https://doi.org/10.1088/1475-7516/2019/07/017} {\bibfield  {journal} {\bibinfo  {journal} {J. Cosmol. Astropart. Phys.}\ }\textbf {\bibinfo {volume} {2019}}\bibfield  {number} {\bibinfo  {number} { (07)},\ \bibinfo {pages} {017}},\ }\Eprint {https://arxiv.org/abs/1812.03554} {arXiv:1812.03554} \BibitemShut {NoStop}%
\bibitem [{\citenamefont {{Bird}}\ \emph {et~al.}(2023)\citenamefont {{Bird}}, \citenamefont {{Fernandez}}, \citenamefont {{Ho}}, \citenamefont {{Qezlou}}, \citenamefont {{Monadi}}, \citenamefont {{Ni}}, \citenamefont {{Chen}}, \citenamefont {{Croft}},\ and\ \citenamefont {{Di Matteo}}}]{Bird_2023}%
  \BibitemOpen
  \bibfield  {author} {\bibinfo {author} {\bibfnamefont {S.}~\bibnamefont {{Bird}}}, \bibinfo {author} {\bibfnamefont {M.}~\bibnamefont {{Fernandez}}}, \bibinfo {author} {\bibfnamefont {M.-F.}\ \bibnamefont {{Ho}}}, \bibinfo {author} {\bibfnamefont {M.}~\bibnamefont {{Qezlou}}}, \bibinfo {author} {\bibfnamefont {R.}~\bibnamefont {{Monadi}}}, \bibinfo {author} {\bibfnamefont {Y.}~\bibnamefont {{Ni}}}, \bibinfo {author} {\bibfnamefont {N.}~\bibnamefont {{Chen}}}, \bibinfo {author} {\bibfnamefont {R.}~\bibnamefont {{Croft}}},\ and\ \bibinfo {author} {\bibfnamefont {T.}~\bibnamefont {{Di Matteo}}},\ }\href {https://doi.org/10.1088/1475-7516/2023/10/037} {\bibfield  {journal} {\bibinfo  {journal} {J. Cosmol. Astropart. Phys.}\ }\textbf {\bibinfo {volume} {2023}}\bibfield  {number} {\bibinfo  {number} { (10)},\ \bibinfo {eid} {037}},\ }\Eprint {https://arxiv.org/abs/2306.05471} {arXiv:2306.05471 [astro-ph.CO]} \BibitemShut {NoStop}%
\bibitem [{\citenamefont {{Fernandez}}\ \emph {et~al.}(2024)\citenamefont {{Fernandez}}, \citenamefont {{Bird}},\ and\ \citenamefont {{Ho}}}]{Fernandez_2024}%
  \BibitemOpen
  \bibfield  {author} {\bibinfo {author} {\bibfnamefont {M.~A.}\ \bibnamefont {{Fernandez}}}, \bibinfo {author} {\bibfnamefont {S.}~\bibnamefont {{Bird}}},\ and\ \bibinfo {author} {\bibfnamefont {M.-F.}\ \bibnamefont {{Ho}}},\ }\href {https://doi.org/10.1088/1475-7516/2024/07/029} {\bibfield  {journal} {\bibinfo  {journal} {J. Cosmol. Astropart. Phys.}\ }\textbf {\bibinfo {volume} {2024}}\bibfield  {number} {\bibinfo  {number} { (7)},\ \bibinfo {eid} {029}},\ }\Eprint {https://arxiv.org/abs/2309.03943} {arXiv:2309.03943 [astro-ph.CO]} \BibitemShut {NoStop}%
\bibitem [{\citenamefont {{Walther}}\ \emph {et~al.}(2024)\citenamefont {{Walther}}, \citenamefont {{Sch{\"o}neberg}}, \citenamefont {{Chabanier}}, \citenamefont {{Armengaud}}, \citenamefont {{Sexton}}, \citenamefont {{Y{\`e}che}}, \citenamefont {{Lesgourgues}}, \citenamefont {{Mosbech}}, \citenamefont {{Ravoux}}, \citenamefont {{Palanque-Delabrouille}},\ and\ \citenamefont {{Luki{\'c}}}}]{walther_emulating_2024}%
  \BibitemOpen
  \bibfield  {author} {\bibinfo {author} {\bibfnamefont {M.}~\bibnamefont {{Walther}}}, \bibinfo {author} {\bibfnamefont {N.}~\bibnamefont {{Sch{\"o}neberg}}}, \bibinfo {author} {\bibfnamefont {S.}~\bibnamefont {{Chabanier}}}, \bibinfo {author} {\bibfnamefont {E.}~\bibnamefont {{Armengaud}}}, \bibinfo {author} {\bibfnamefont {J.}~\bibnamefont {{Sexton}}}, \bibinfo {author} {\bibfnamefont {C.}~\bibnamefont {{Y{\`e}che}}}, \bibinfo {author} {\bibfnamefont {J.}~\bibnamefont {{Lesgourgues}}}, \bibinfo {author} {\bibfnamefont {M.~R.}\ \bibnamefont {{Mosbech}}}, \bibinfo {author} {\bibfnamefont {C.}~\bibnamefont {{Ravoux}}}, \bibinfo {author} {\bibfnamefont {N.}~\bibnamefont {{Palanque-Delabrouille}}},\ and\ \bibinfo {author} {\bibfnamefont {Z.}~\bibnamefont {{Luki{\'c}}}},\ }\href@noop {} {\bibfield  {journal} {\bibinfo  {journal} {arXiv e-prints}\ } (\bibinfo {year} {2024})},\ \Eprint {https://arxiv.org/abs/2412.05372} {arXiv:2412.05372 [astro-ph.CO]} \BibitemShut {NoStop}%
\bibitem [{\citenamefont {{Ivanov}}\ \emph {et~al.}(2025)\citenamefont {{Ivanov}}, \citenamefont {{Toomey}},\ and\ \citenamefont {{Kara{\c{c}}ayl{\i}}}}]{ivanov_fundamental_2025}%
  \BibitemOpen
  \bibfield  {author} {\bibinfo {author} {\bibfnamefont {M.~M.}\ \bibnamefont {{Ivanov}}}, \bibinfo {author} {\bibfnamefont {M.~W.}\ \bibnamefont {{Toomey}}},\ and\ \bibinfo {author} {\bibfnamefont {N.~G.}\ \bibnamefont {{Kara{\c{c}}ayl{\i}}}},\ }\href {https://doi.org/10.1103/PhysRevLett.134.091001} {\bibfield  {journal} {\bibinfo  {journal} {\prl}\ }\textbf {\bibinfo {volume} {134}},\ \bibinfo {eid} {091001} (\bibinfo {year} {2025})},\ \Eprint {https://arxiv.org/abs/2405.13208} {arXiv:2405.13208 [astro-ph.CO]} \BibitemShut {NoStop}%
\bibitem [{\citenamefont {{Rogers}}\ and\ \citenamefont {{Poulin}}(2025)}]{rogers_5_2024}%
  \BibitemOpen
  \bibfield  {author} {\bibinfo {author} {\bibfnamefont {K.~K.}\ \bibnamefont {{Rogers}}}\ and\ \bibinfo {author} {\bibfnamefont {V.}~\bibnamefont {{Poulin}}},\ }\href {https://doi.org/10.1103/PhysRevResearch.7.L012018} {\bibfield  {journal} {\bibinfo  {journal} {Physical Review Research}\ }\textbf {\bibinfo {volume} {7}},\ \bibinfo {eid} {L012018} (\bibinfo {year} {2025})},\ \Eprint {https://arxiv.org/abs/2311.16377} {arXiv:2311.16377 [astro-ph.CO]} \BibitemShut {NoStop}%
\bibitem [{\citenamefont {He}\ \emph {et~al.}(2025)\citenamefont {He}, \citenamefont {Ivanov}, \citenamefont {Bird}, \citenamefont {An},\ and\ \citenamefont {Gluscevic}}]{He_Fresh_2025}%
  \BibitemOpen
  \bibfield  {author} {\bibinfo {author} {\bibfnamefont {A.}~\bibnamefont {He}}, \bibinfo {author} {\bibfnamefont {M.~M.}\ \bibnamefont {Ivanov}}, \bibinfo {author} {\bibfnamefont {S.}~\bibnamefont {Bird}}, \bibinfo {author} {\bibfnamefont {R.}~\bibnamefont {An}},\ and\ \bibinfo {author} {\bibfnamefont {V.}~\bibnamefont {Gluscevic}},\ }\href@noop {} {\bibfield  {journal} {\bibinfo  {journal} {arXiv e-prints}\ } (\bibinfo {year} {2025})},\ \Eprint {https://arxiv.org/abs/2503.15592} {arXiv:2503.15592 [astro-ph.CO]} \BibitemShut {NoStop}%
\bibitem [{\citenamefont {{Abe}}\ \emph {et~al.}(2015)\citenamefont {{Abe}}, \citenamefont {{Kobayashi}},\ and\ \citenamefont {{Otsuka}}}]{Abe_natural_inflation_2015}%
  \BibitemOpen
  \bibfield  {author} {\bibinfo {author} {\bibfnamefont {H.}~\bibnamefont {{Abe}}}, \bibinfo {author} {\bibfnamefont {T.}~\bibnamefont {{Kobayashi}}},\ and\ \bibinfo {author} {\bibfnamefont {H.}~\bibnamefont {{Otsuka}}},\ }\href {https://doi.org/10.1007/JHEP04(2015)160} {\bibfield  {journal} {\bibinfo  {journal} {Journal of High Energy Physics}\ }\textbf {\bibinfo {volume} {2015}},\ \bibinfo {eid} {160} (\bibinfo {year} {2015})},\ \Eprint {https://arxiv.org/abs/1411.4768} {arXiv:1411.4768 [hep-th]} \BibitemShut {NoStop}%
\bibitem [{\citenamefont {{Kappl}}\ \emph {et~al.}(2016)\citenamefont {{Kappl}}, \citenamefont {{Nilles}},\ and\ \citenamefont {{Winkler}}}]{kappl_2016}%
  \BibitemOpen
  \bibfield  {author} {\bibinfo {author} {\bibfnamefont {R.}~\bibnamefont {{Kappl}}}, \bibinfo {author} {\bibfnamefont {H.~P.}\ \bibnamefont {{Nilles}}},\ and\ \bibinfo {author} {\bibfnamefont {M.~W.}\ \bibnamefont {{Winkler}}},\ }\href {https://doi.org/10.1016/j.physletb.2015.12.073} {\bibfield  {journal} {\bibinfo  {journal} {Physics Letters B}\ }\textbf {\bibinfo {volume} {753}},\ \bibinfo {pages} {653} (\bibinfo {year} {2016})},\ \Eprint {https://arxiv.org/abs/1511.05560} {arXiv:1511.05560} \BibitemShut {NoStop}%
\bibitem [{\citenamefont {{McDonough}}\ and\ \citenamefont {{Scalisi}}(2023)}]{mcdonough_towards_2023}%
  \BibitemOpen
  \bibfield  {author} {\bibinfo {author} {\bibfnamefont {E.}~\bibnamefont {{McDonough}}}\ and\ \bibinfo {author} {\bibfnamefont {M.}~\bibnamefont {{Scalisi}}},\ }\href {https://doi.org/10.1007/JHEP10(2023)118} {\bibfield  {journal} {\bibinfo  {journal} {Journal of High Energy Physics}\ }\textbf {\bibinfo {volume} {2023}},\ \bibinfo {eid} {118} (\bibinfo {year} {2023})},\ \Eprint {https://arxiv.org/abs/2209.00011} {arXiv:2209.00011} \BibitemShut {NoStop}%
\bibitem [{\citenamefont {{Co}}\ and\ \citenamefont {{Harigaya}}(2020)}]{co_axiogenesis_2020}%
  \BibitemOpen
  \bibfield  {author} {\bibinfo {author} {\bibfnamefont {R.~T.}\ \bibnamefont {{Co}}}\ and\ \bibinfo {author} {\bibfnamefont {K.}~\bibnamefont {{Harigaya}}},\ }\href {https://doi.org/10.1103/PhysRevLett.124.111602} {\bibfield  {journal} {\bibinfo  {journal} {\prl}\ }\textbf {\bibinfo {volume} {124}},\ \bibinfo {eid} {111602} (\bibinfo {year} {2020})},\ \Eprint {https://arxiv.org/abs/1910.02080} {arXiv:1910.02080 [hep-ph]} \BibitemShut {NoStop}%
\bibitem [{\citenamefont {Blas}\ \emph {et~al.}(2011)\citenamefont {Blas}, \citenamefont {Lesgourgues},\ and\ \citenamefont {Tram}}]{blas_cosmic_2011}%
  \BibitemOpen
  \bibfield  {author} {\bibinfo {author} {\bibfnamefont {D.}~\bibnamefont {Blas}}, \bibinfo {author} {\bibfnamefont {J.}~\bibnamefont {Lesgourgues}},\ and\ \bibinfo {author} {\bibfnamefont {T.}~\bibnamefont {Tram}},\ }\href {https://doi.org/10.1088/1475-7516/2011/07/034} {\bibfield  {journal} {\bibinfo  {journal} {J. Cosmol. Astropart. Phys.}\ }\textbf {\bibinfo {volume} {2011}}\bibfield  {number} {\bibinfo  {number} { (07)},\ \bibinfo {pages} {034}},\ }\Eprint {https://arxiv.org/abs/1104.2933} {arXiv:1104.2933} \BibitemShut {NoStop}%
\bibitem [{\citenamefont {{Ballesteros}}\ and\ \citenamefont {{Lesgourgues}}(2010)}]{ballesteros_dark_2010}%
  \BibitemOpen
  \bibfield  {author} {\bibinfo {author} {\bibfnamefont {G.}~\bibnamefont {{Ballesteros}}}\ and\ \bibinfo {author} {\bibfnamefont {J.}~\bibnamefont {{Lesgourgues}}},\ }\href {https://doi.org/10.1088/1475-7516/2010/10/014} {\bibfield  {journal} {\bibinfo  {journal} {J. Cosmol. Astropart. Phys.}\ }\textbf {\bibinfo {volume} {2010}}\bibfield  {number} {\bibinfo  {number} { (10)},\ \bibinfo {eid} {014}},\ }\Eprint {https://arxiv.org/abs/1004.5509} {arXiv:1004.5509} \BibitemShut {NoStop}%
\bibitem [{\citenamefont {Hu}\ and\ \citenamefont {Sugiyama}(1996)}]{hu_small_1996}%
  \BibitemOpen
  \bibfield  {author} {\bibinfo {author} {\bibfnamefont {W.}~\bibnamefont {Hu}}\ and\ \bibinfo {author} {\bibfnamefont {N.}~\bibnamefont {Sugiyama}},\ }\href {https://doi.org/10.1086/177989} {\bibfield  {journal} {\bibinfo  {journal} {ApJ}\ }\textbf {\bibinfo {volume} {471}},\ \bibinfo {pages} {542} (\bibinfo {year} {1996})},\ \Eprint {https://arxiv.org/abs/astro-ph/9510117} {arXiv:astro-ph/9510117} \BibitemShut {NoStop}%
\bibitem [{\citenamefont {Atek}\ \emph {et~al.}(2022)\citenamefont {Atek}, \citenamefont {Shuntov}, \citenamefont {Furtak}, \citenamefont {Richard}, \citenamefont {Kneib}, \citenamefont {Mahler}, \citenamefont {Zitrin}, \citenamefont {McCracken}, \citenamefont {Charlot}, \citenamefont {Chevallard},\ and\ \citenamefont {Chemerynska}}]{atek_revealing_2022}%
  \BibitemOpen
  \bibfield  {author} {\bibinfo {author} {\bibfnamefont {H.}~\bibnamefont {Atek}}, \bibinfo {author} {\bibfnamefont {M.}~\bibnamefont {Shuntov}}, \bibinfo {author} {\bibfnamefont {L.~J.}\ \bibnamefont {Furtak}}, \bibinfo {author} {\bibfnamefont {J.}~\bibnamefont {Richard}}, \bibinfo {author} {\bibfnamefont {J.-P.}\ \bibnamefont {Kneib}}, \bibinfo {author} {\bibfnamefont {G.}~\bibnamefont {Mahler}}, \bibinfo {author} {\bibfnamefont {A.}~\bibnamefont {Zitrin}}, \bibinfo {author} {\bibfnamefont {H.~J.}\ \bibnamefont {McCracken}}, \bibinfo {author} {\bibfnamefont {S.}~\bibnamefont {Charlot}}, \bibinfo {author} {\bibfnamefont {J.}~\bibnamefont {Chevallard}},\ and\ \bibinfo {author} {\bibfnamefont {I.}~\bibnamefont {Chemerynska}},\ }\href {https://doi.org/10.1093/mnras/stac3144} {\bibfield  {journal} {\bibinfo  {journal} {Monthly Notices of the Royal Astronomical Society}\ }\textbf {\bibinfo {volume} {519}},\ \bibinfo {pages} {1201} (\bibinfo {year} {2022})},\ \Eprint {https://arxiv.org/abs/2207.12338}
  {arXiv:2207.12338} \BibitemShut {NoStop}%
\bibitem [{\citenamefont {Finkelstein}\ \emph {et~al.}(2022)\citenamefont {Finkelstein}, \citenamefont {Bagley}, \citenamefont {Haro}, \citenamefont {Dickinson}, \citenamefont {Ferguson}, \citenamefont {Kartaltepe}, \citenamefont {Papovich}, \citenamefont {Burgarella}, \citenamefont {Kocevski}, \citenamefont {{Huertas-Company}} \emph {et~al.}}]{finkelstein_long_2022}%
  \BibitemOpen
  \bibfield  {author} {\bibinfo {author} {\bibfnamefont {S.~L.}\ \bibnamefont {Finkelstein}}, \bibinfo {author} {\bibfnamefont {M.~B.}\ \bibnamefont {Bagley}}, \bibinfo {author} {\bibfnamefont {P.~A.}\ \bibnamefont {Haro}}, \bibinfo {author} {\bibfnamefont {M.}~\bibnamefont {Dickinson}}, \bibinfo {author} {\bibfnamefont {H.~C.}\ \bibnamefont {Ferguson}}, \bibinfo {author} {\bibfnamefont {J.~S.}\ \bibnamefont {Kartaltepe}}, \bibinfo {author} {\bibfnamefont {C.}~\bibnamefont {Papovich}}, \bibinfo {author} {\bibfnamefont {D.}~\bibnamefont {Burgarella}}, \bibinfo {author} {\bibfnamefont {D.~D.}\ \bibnamefont {Kocevski}}, \bibinfo {author} {\bibfnamefont {M.}~\bibnamefont {{Huertas-Company}}}, \emph {et~al.},\ }\href {https://doi.org/10.3847/2041-8213/ac966e} {\bibfield  {journal} {\bibinfo  {journal} {The Astrophysical Journal Letters}\ }\textbf {\bibinfo {volume} {940}},\ \bibinfo {eid} {L55} (\bibinfo {year} {2022})},\ \Eprint {https://arxiv.org/abs/2207.12474} {arXiv:2207.12474} \BibitemShut {NoStop}%
\bibitem [{\citenamefont {Harikane}\ \emph {et~al.}(2023)\citenamefont {Harikane}, \citenamefont {Ouchi}, \citenamefont {Oguri}, \citenamefont {Ono}, \citenamefont {Nakajima}, \citenamefont {Isobe}, \citenamefont {Umeda}, \citenamefont {Mawatari},\ and\ \citenamefont {Zhang}}]{harikane_comprehensive_2023}%
  \BibitemOpen
  \bibfield  {author} {\bibinfo {author} {\bibfnamefont {Y.}~\bibnamefont {Harikane}}, \bibinfo {author} {\bibfnamefont {M.}~\bibnamefont {Ouchi}}, \bibinfo {author} {\bibfnamefont {M.}~\bibnamefont {Oguri}}, \bibinfo {author} {\bibfnamefont {Y.}~\bibnamefont {Ono}}, \bibinfo {author} {\bibfnamefont {K.}~\bibnamefont {Nakajima}}, \bibinfo {author} {\bibfnamefont {Y.}~\bibnamefont {Isobe}}, \bibinfo {author} {\bibfnamefont {H.}~\bibnamefont {Umeda}}, \bibinfo {author} {\bibfnamefont {K.}~\bibnamefont {Mawatari}},\ and\ \bibinfo {author} {\bibfnamefont {Y.}~\bibnamefont {Zhang}},\ }\href {https://doi.org/10.3847/1538-4365/acaaa9} {\bibfield  {journal} {\bibinfo  {journal} {ApJS}\ }\textbf {\bibinfo {volume} {265}},\ \bibinfo {pages} {5} (\bibinfo {year} {2023})},\ \Eprint {https://arxiv.org/abs/2208.01612} {arXiv:2208.01612} \BibitemShut {NoStop}%
\bibitem [{\citenamefont {Naidu}\ \emph {et~al.}(2022)\citenamefont {Naidu}, \citenamefont {Oesch}, \citenamefont {Setton}, \citenamefont {Matthee}, \citenamefont {Conroy}, \citenamefont {Johnson}, \citenamefont {Weaver}, \citenamefont {Bouwens}, \citenamefont {Brammer}, \citenamefont {Dayal} \emph {et~al.}}]{naidu_schrodingers_2022-1}%
  \BibitemOpen
  \bibfield  {author} {\bibinfo {author} {\bibfnamefont {R.~P.}\ \bibnamefont {Naidu}}, \bibinfo {author} {\bibfnamefont {P.~A.}\ \bibnamefont {Oesch}}, \bibinfo {author} {\bibfnamefont {D.~J.}\ \bibnamefont {Setton}}, \bibinfo {author} {\bibfnamefont {J.}~\bibnamefont {Matthee}}, \bibinfo {author} {\bibfnamefont {C.}~\bibnamefont {Conroy}}, \bibinfo {author} {\bibfnamefont {B.~D.}\ \bibnamefont {Johnson}}, \bibinfo {author} {\bibfnamefont {J.~R.}\ \bibnamefont {Weaver}}, \bibinfo {author} {\bibfnamefont {R.~J.}\ \bibnamefont {Bouwens}}, \bibinfo {author} {\bibfnamefont {G.~B.}\ \bibnamefont {Brammer}}, \bibinfo {author} {\bibfnamefont {P.}~\bibnamefont {Dayal}}, \emph {et~al.},\ }\bibfield  {journal} {\bibinfo  {journal} {arXiv e-prints}\ }\href {https://doi.org/10.48550/arXiv.2208.02794} {10.48550/arXiv.2208.02794} (\bibinfo {year} {2022}),\ \Eprint {https://arxiv.org/abs/2208.02794} {arXiv:2208.02794} \BibitemShut {NoStop}%
\bibitem [{\citenamefont {Labbe}\ \emph {et~al.}(2023)\citenamefont {Labbe}, \citenamefont {{van Dokkum}}, \citenamefont {Nelson}, \citenamefont {Bezanson}, \citenamefont {Suess}, \citenamefont {Leja}, \citenamefont {Brammer}, \citenamefont {Whitaker}, \citenamefont {Mathews}, \citenamefont {Stefanon},\ and\ \citenamefont {Wang}}]{labbe_population_2023}%
  \BibitemOpen
  \bibfield  {author} {\bibinfo {author} {\bibfnamefont {I.}~\bibnamefont {Labbe}}, \bibinfo {author} {\bibfnamefont {P.}~\bibnamefont {{van Dokkum}}}, \bibinfo {author} {\bibfnamefont {E.}~\bibnamefont {Nelson}}, \bibinfo {author} {\bibfnamefont {R.}~\bibnamefont {Bezanson}}, \bibinfo {author} {\bibfnamefont {K.}~\bibnamefont {Suess}}, \bibinfo {author} {\bibfnamefont {J.}~\bibnamefont {Leja}}, \bibinfo {author} {\bibfnamefont {G.}~\bibnamefont {Brammer}}, \bibinfo {author} {\bibfnamefont {K.}~\bibnamefont {Whitaker}}, \bibinfo {author} {\bibfnamefont {E.}~\bibnamefont {Mathews}}, \bibinfo {author} {\bibfnamefont {M.}~\bibnamefont {Stefanon}},\ and\ \bibinfo {author} {\bibfnamefont {B.}~\bibnamefont {Wang}},\ }\href {https://doi.org/10.1038/s41586-023-05786-2} {\bibfield  {journal} {\bibinfo  {journal} {Nature}\ }\textbf {\bibinfo {volume} {616}},\ \bibinfo {pages} {266} (\bibinfo {year} {2023})},\ \Eprint {https://arxiv.org/abs/2207.12446} {arXiv:2207.12446} \BibitemShut {NoStop}%
\bibitem [{\citenamefont {Yan}\ \emph {et~al.}(2023)\citenamefont {Yan}, \citenamefont {Ma}, \citenamefont {Ling}, \citenamefont {Cheng},\ and\ \citenamefont {Huang}}]{yan_first_2023}%
  \BibitemOpen
  \bibfield  {author} {\bibinfo {author} {\bibfnamefont {H.}~\bibnamefont {Yan}}, \bibinfo {author} {\bibfnamefont {Z.}~\bibnamefont {Ma}}, \bibinfo {author} {\bibfnamefont {C.}~\bibnamefont {Ling}}, \bibinfo {author} {\bibfnamefont {C.}~\bibnamefont {Cheng}},\ and\ \bibinfo {author} {\bibfnamefont {J.-s.}\ \bibnamefont {Huang}},\ }\href {https://doi.org/10.3847/2041-8213/aca80c} {\bibfield  {journal} {\bibinfo  {journal} {ApJL}\ }\textbf {\bibinfo {volume} {942}},\ \bibinfo {pages} {L9} (\bibinfo {year} {2023})},\ \Eprint {https://arxiv.org/abs/2207.11558} {arXiv:2207.11558} \BibitemShut {NoStop}%
\bibitem [{\citenamefont {Hill}\ \emph {et~al.}(2020)\citenamefont {Hill}, \citenamefont {McDonough}, \citenamefont {Toomey},\ and\ \citenamefont {Alexander}}]{hill_early_2020}%
  \BibitemOpen
  \bibfield  {author} {\bibinfo {author} {\bibfnamefont {J.~C.}\ \bibnamefont {Hill}}, \bibinfo {author} {\bibfnamefont {E.}~\bibnamefont {McDonough}}, \bibinfo {author} {\bibfnamefont {M.~W.}\ \bibnamefont {Toomey}},\ and\ \bibinfo {author} {\bibfnamefont {S.}~\bibnamefont {Alexander}},\ }\href {https://doi.org/10.1103/PhysRevD.102.043507} {\bibfield  {journal} {\bibinfo  {journal} {Phys. Rev. D}\ }\textbf {\bibinfo {volume} {102}},\ \bibinfo {pages} {043507} (\bibinfo {year} {2020})},\ \Eprint {https://arxiv.org/abs/2003.07355} {arXiv:2003.07355} \BibitemShut {NoStop}%
\bibitem [{\citenamefont {{Riess}}\ \emph {et~al.}(2019)\citenamefont {{Riess}}, \citenamefont {{Casertano}}, \citenamefont {{Yuan}}, \citenamefont {{Macri}},\ and\ \citenamefont {{Scolnic}}}]{Riess_large_2019}%
  \BibitemOpen
  \bibfield  {author} {\bibinfo {author} {\bibfnamefont {A.~G.}\ \bibnamefont {{Riess}}}, \bibinfo {author} {\bibfnamefont {S.}~\bibnamefont {{Casertano}}}, \bibinfo {author} {\bibfnamefont {W.}~\bibnamefont {{Yuan}}}, \bibinfo {author} {\bibfnamefont {L.~M.}\ \bibnamefont {{Macri}}},\ and\ \bibinfo {author} {\bibfnamefont {D.}~\bibnamefont {{Scolnic}}},\ }\href {https://doi.org/10.3847/1538-4357/ab1422} {\bibfield  {journal} {\bibinfo  {journal} {\apj}\ }\textbf {\bibinfo {volume} {876}},\ \bibinfo {eid} {85} (\bibinfo {year} {2019})},\ \Eprint {https://arxiv.org/abs/1903.07603} {arXiv:1903.07603 [astro-ph.CO]} \BibitemShut {NoStop}%
\bibitem [{\citenamefont {{Niedermann}}\ and\ \citenamefont {{Sloth}}(2020)}]{niedermann_resolving_2020}%
  \BibitemOpen
  \bibfield  {author} {\bibinfo {author} {\bibfnamefont {F.}~\bibnamefont {{Niedermann}}}\ and\ \bibinfo {author} {\bibfnamefont {M.~S.}\ \bibnamefont {{Sloth}}},\ }\href {https://doi.org/10.1103/PhysRevD.102.063527} {\bibfield  {journal} {\bibinfo  {journal} {\prd}\ }\textbf {\bibinfo {volume} {102}},\ \bibinfo {eid} {063527} (\bibinfo {year} {2020})},\ \Eprint {https://arxiv.org/abs/2006.06686} {arXiv:2006.06686} \BibitemShut {NoStop}%
\bibitem [{\citenamefont {Liddle}\ and\ \citenamefont {Scherrer}(1998)}]{liddle_classification_1998}%
  \BibitemOpen
  \bibfield  {author} {\bibinfo {author} {\bibfnamefont {A.~R.}\ \bibnamefont {Liddle}}\ and\ \bibinfo {author} {\bibfnamefont {R.~J.}\ \bibnamefont {Scherrer}},\ }\href {https://doi.org/10.1103/PhysRevD.59.023509} {\bibfield  {journal} {\bibinfo  {journal} {Phys. Rev. D}\ }\textbf {\bibinfo {volume} {59}},\ \bibinfo {pages} {023509} (\bibinfo {year} {1998})},\ \Eprint {https://arxiv.org/abs/astro-ph/9809272} {arXiv:astro-ph/9809272} \BibitemShut {NoStop}%
\bibitem [{\citenamefont {Lozanov}\ and\ \citenamefont {Amin}(2017)}]{lozanov_equation_2017}%
  \BibitemOpen
  \bibfield  {author} {\bibinfo {author} {\bibfnamefont {K.~D.}\ \bibnamefont {Lozanov}}\ and\ \bibinfo {author} {\bibfnamefont {M.~A.}\ \bibnamefont {Amin}},\ }\href {https://doi.org/10.1103/PhysRevLett.119.061301} {\bibfield  {journal} {\bibinfo  {journal} {Phys. Rev. Lett.}\ }\textbf {\bibinfo {volume} {119}},\ \bibinfo {pages} {061301} (\bibinfo {year} {2017})},\ \Eprint {https://arxiv.org/abs/1608.01213} {arXiv:1608.01213} \BibitemShut {NoStop}%
\bibitem [{\citenamefont {Lozanov}\ and\ \citenamefont {Amin}(2018)}]{lozanov_self-resonance_2018}%
  \BibitemOpen
  \bibfield  {author} {\bibinfo {author} {\bibfnamefont {K.~D.}\ \bibnamefont {Lozanov}}\ and\ \bibinfo {author} {\bibfnamefont {M.~A.}\ \bibnamefont {Amin}},\ }\href {https://doi.org/10.1103/PhysRevD.97.023533} {\bibfield  {journal} {\bibinfo  {journal} {Phys. Rev. D}\ }\textbf {\bibinfo {volume} {97}},\ \bibinfo {pages} {023533} (\bibinfo {year} {2018})},\ \Eprint {https://arxiv.org/abs/1710.06851} {arXiv:1710.06851} \BibitemShut {NoStop}%
\bibitem [{\citenamefont {Dodelson}\ and\ \citenamefont {Schmidt}(2020)}]{dodelson_scott_modern_2020}%
  \BibitemOpen
  \bibfield  {author} {\bibinfo {author} {\bibfnamefont {S.}~\bibnamefont {Dodelson}}\ and\ \bibinfo {author} {\bibfnamefont {F.}~\bibnamefont {Schmidt}},\ }\href@noop {} {\emph {\bibinfo {title} {Modern {{Cosmology}}}}},\ \bibinfo {edition} {2nd}\ ed.\ (\bibinfo  {publisher} {Academic Press},\ \bibinfo {year} {2020})\BibitemShut {NoStop}%
\bibitem [{\citenamefont {Aloni}\ \emph {et~al.}(2022)\citenamefont {Aloni}, \citenamefont {Berlin}, \citenamefont {Joseph}, \citenamefont {Schmaltz},\ and\ \citenamefont {Weiner}}]{aloni_step_2022-2}%
  \BibitemOpen
  \bibfield  {author} {\bibinfo {author} {\bibfnamefont {D.}~\bibnamefont {Aloni}}, \bibinfo {author} {\bibfnamefont {A.}~\bibnamefont {Berlin}}, \bibinfo {author} {\bibfnamefont {M.}~\bibnamefont {Joseph}}, \bibinfo {author} {\bibfnamefont {M.}~\bibnamefont {Schmaltz}},\ and\ \bibinfo {author} {\bibfnamefont {N.}~\bibnamefont {Weiner}},\ }\href {https://doi.org/10.1103/PhysRevD.105.123516} {\bibfield  {journal} {\bibinfo  {journal} {Phys. Rev. D}\ }\textbf {\bibinfo {volume} {105}},\ \bibinfo {pages} {123516} (\bibinfo {year} {2022})},\ \Eprint {https://arxiv.org/abs/2111.00014} {arXiv:2111.00014} \BibitemShut {NoStop}%
\bibitem [{\citenamefont {{McKeen}}\ and\ \citenamefont {{Omar}}(2024)}]{Mckeen_early_2024}%
  \BibitemOpen
  \bibfield  {author} {\bibinfo {author} {\bibfnamefont {D.}~\bibnamefont {{McKeen}}}\ and\ \bibinfo {author} {\bibfnamefont {A.}~\bibnamefont {{Omar}}},\ }\href {https://doi.org/10.48550/arXiv.2407.03508} {\bibfield  {journal} {\bibinfo  {journal} {arXiv e-prints}\ ,\ \bibinfo {eid} {arXiv:2407.03508}} (\bibinfo {year} {2024})},\ \Eprint {https://arxiv.org/abs/2407.03508} {arXiv:2407.03508 [hep-ph]} \BibitemShut {NoStop}%
\bibitem [{\citenamefont {Gariazzo}\ \emph {et~al.}(2022)\citenamefont {Gariazzo}, \citenamefont {{de Salas}}, \citenamefont {Pisanti},\ and\ \citenamefont {Consiglio}}]{gariazzo_parthenope_2022}%
  \BibitemOpen
  \bibfield  {author} {\bibinfo {author} {\bibfnamefont {S.}~\bibnamefont {Gariazzo}}, \bibinfo {author} {\bibfnamefont {P.~F.}\ \bibnamefont {{de Salas}}}, \bibinfo {author} {\bibfnamefont {O.}~\bibnamefont {Pisanti}},\ and\ \bibinfo {author} {\bibfnamefont {R.}~\bibnamefont {Consiglio}},\ }\href {https://doi.org/10.1016/j.cpc.2021.108205} {\bibfield  {journal} {\bibinfo  {journal} {Computer Physics Communications}\ }\textbf {\bibinfo {volume} {271}},\ \bibinfo {pages} {108205} (\bibinfo {year} {2022})},\ \Eprint {https://arxiv.org/abs/2103.05027} {arXiv:2103.05027} \BibitemShut {NoStop}%
\bibitem [{\citenamefont {Audren}\ \emph {et~al.}(2013)\citenamefont {Audren}, \citenamefont {Lesgourgues}, \citenamefont {Benabed},\ and\ \citenamefont {Prunet}}]{audren_conservative_2013}%
  \BibitemOpen
  \bibfield  {author} {\bibinfo {author} {\bibfnamefont {B.}~\bibnamefont {Audren}}, \bibinfo {author} {\bibfnamefont {J.}~\bibnamefont {Lesgourgues}}, \bibinfo {author} {\bibfnamefont {K.}~\bibnamefont {Benabed}},\ and\ \bibinfo {author} {\bibfnamefont {S.}~\bibnamefont {Prunet}},\ }\href {https://doi.org/10.1088/1475-7516/2013/02/001} {\bibfield  {journal} {\bibinfo  {journal} {J. Cosmol. Astropart. Phys.}\ }\textbf {\bibinfo {volume} {2013}}\bibfield  {number} {\bibinfo  {number} { (02)},\ \bibinfo {pages} {001}},\ }\Eprint {https://arxiv.org/abs/1210.7183} {arXiv:1210.7183} \BibitemShut {NoStop}%
\bibitem [{\citenamefont {Brinckmann}\ and\ \citenamefont {Lesgourgues}(2019)}]{brinckmann_montepython_2018-1}%
  \BibitemOpen
  \bibfield  {author} {\bibinfo {author} {\bibfnamefont {T.}~\bibnamefont {Brinckmann}}\ and\ \bibinfo {author} {\bibfnamefont {J.}~\bibnamefont {Lesgourgues}},\ }\bibfield  {journal} {\bibinfo  {journal} {Physics of the Dark Universe}\ }\href {https://doi.org/10.1016/j.dark.2018.100260} {10.1016/j.dark.2018.100260} (\bibinfo {year} {2019}),\ \Eprint {https://arxiv.org/abs/1804.07261} {arXiv:1804.07261} \BibitemShut {NoStop}%
\bibitem [{\citenamefont {Gelman}\ and\ \citenamefont {Rubin}(1992)}]{gelman_inference_1992}%
  \BibitemOpen
  \bibfield  {author} {\bibinfo {author} {\bibfnamefont {A.}~\bibnamefont {Gelman}}\ and\ \bibinfo {author} {\bibfnamefont {D.~B.}\ \bibnamefont {Rubin}},\ }\href {https://doi.org/10.1214/ss/1177011136} {\bibfield  {journal} {\bibinfo  {journal} {Statistical Science}\ }\textbf {\bibinfo {volume} {7}},\ \bibinfo {pages} {457} (\bibinfo {year} {1992})}\BibitemShut {NoStop}%
\bibitem [{\citenamefont {Lewis}(2019)}]{lewis_getdist_2019}%
  \BibitemOpen
  \bibfield  {author} {\bibinfo {author} {\bibfnamefont {A.}~\bibnamefont {Lewis}},\ }\href@noop {} {\bibfield  {journal} {\bibinfo  {journal} {arXiv e-prints}\ } (\bibinfo {year} {2019})},\ \Eprint {https://arxiv.org/abs/1910.13970} {arXiv:1910.13970} \BibitemShut {NoStop}%
\bibitem [{\citenamefont {{Planck Collaboration}}\ \emph {et~al.}(2020{\natexlab{b}})\citenamefont {{Planck Collaboration}}, \citenamefont {Aghanim} \emph {et~al.}}]{planck_collaboration_likelihoods}%
  \BibitemOpen
  \bibfield  {author} {\bibinfo {author} {\bibnamefont {{Planck Collaboration}}}, \bibinfo {author} {\bibfnamefont {N.}~\bibnamefont {Aghanim}}, \emph {et~al.},\ }\href {https://doi.org/10.1051/0004-6361/201936386} {\bibfield  {journal} {\bibinfo  {journal} {A\&A}\ }\textbf {\bibinfo {volume} {641}},\ \bibinfo {pages} {A5} (\bibinfo {year} {2020}{\natexlab{b}})},\ \Eprint {https://arxiv.org/abs/1907.12875} {arXiv:1907.12875} \BibitemShut {NoStop}%
\bibitem [{\citenamefont {Beutler}\ \emph {et~al.}(2011)\citenamefont {Beutler}, \citenamefont {Blake}, \citenamefont {Colless}, \citenamefont {Jones}, \citenamefont {{Staveley-Smith}}, \citenamefont {Campbell}, \citenamefont {Parker}, \citenamefont {Saunders},\ and\ \citenamefont {Watson}}]{beutler_6df_2011-1}%
  \BibitemOpen
  \bibfield  {author} {\bibinfo {author} {\bibfnamefont {F.}~\bibnamefont {Beutler}}, \bibinfo {author} {\bibfnamefont {C.}~\bibnamefont {Blake}}, \bibinfo {author} {\bibfnamefont {M.}~\bibnamefont {Colless}}, \bibinfo {author} {\bibfnamefont {D.~H.}\ \bibnamefont {Jones}}, \bibinfo {author} {\bibfnamefont {L.}~\bibnamefont {{Staveley-Smith}}}, \bibinfo {author} {\bibfnamefont {L.}~\bibnamefont {Campbell}}, \bibinfo {author} {\bibfnamefont {Q.}~\bibnamefont {Parker}}, \bibinfo {author} {\bibfnamefont {W.}~\bibnamefont {Saunders}},\ and\ \bibinfo {author} {\bibfnamefont {F.}~\bibnamefont {Watson}},\ }\href {https://doi.org/10.1111/j.1365-2966.2011.19250.x} {\bibfield  {journal} {\bibinfo  {journal} {Monthly Notices of the Royal Astronomical Society}\ }\textbf {\bibinfo {volume} {416}},\ \bibinfo {pages} {3017} (\bibinfo {year} {2011})},\ \Eprint {https://arxiv.org/abs/1106.3366} {arXiv:1106.3366} \BibitemShut {NoStop}%
\bibitem [{\citenamefont {Howlett}\ \emph {et~al.}(2015)\citenamefont {Howlett}, \citenamefont {Ross}, \citenamefont {Samushia}, \citenamefont {Percival},\ and\ \citenamefont {Manera}}]{howlett_clustering_2015}%
  \BibitemOpen
  \bibfield  {author} {\bibinfo {author} {\bibfnamefont {C.}~\bibnamefont {Howlett}}, \bibinfo {author} {\bibfnamefont {A.~J.}\ \bibnamefont {Ross}}, \bibinfo {author} {\bibfnamefont {L.}~\bibnamefont {Samushia}}, \bibinfo {author} {\bibfnamefont {W.~J.}\ \bibnamefont {Percival}},\ and\ \bibinfo {author} {\bibfnamefont {M.}~\bibnamefont {Manera}},\ }\href {https://doi.org/10.1093/mnras/stu2693} {\bibfield  {journal} {\bibinfo  {journal} {Monthly Notices of the Royal Astronomical Society}\ }\textbf {\bibinfo {volume} {449}},\ \bibinfo {pages} {848} (\bibinfo {year} {2015})},\ \Eprint {https://arxiv.org/abs/1409.3238} {arXiv:1409.3238} \BibitemShut {NoStop}%
\bibitem [{\citenamefont {Alam}\ \emph {et~al.}(2017)\citenamefont {Alam}, \citenamefont {Ata}, \citenamefont {Bailey}, \citenamefont {Beutler}, \citenamefont {Bizyaev}, \citenamefont {Blazek}, \citenamefont {Bolton}, \citenamefont {Brownstein}, \citenamefont {Burden}, \citenamefont {Chuang} \emph {et~al.}}]{alam_clustering_2017-1}%
  \BibitemOpen
  \bibfield  {author} {\bibinfo {author} {\bibfnamefont {S.}~\bibnamefont {Alam}}, \bibinfo {author} {\bibfnamefont {M.}~\bibnamefont {Ata}}, \bibinfo {author} {\bibfnamefont {S.}~\bibnamefont {Bailey}}, \bibinfo {author} {\bibfnamefont {F.}~\bibnamefont {Beutler}}, \bibinfo {author} {\bibfnamefont {D.}~\bibnamefont {Bizyaev}}, \bibinfo {author} {\bibfnamefont {J.~A.}\ \bibnamefont {Blazek}}, \bibinfo {author} {\bibfnamefont {A.~S.}\ \bibnamefont {Bolton}}, \bibinfo {author} {\bibfnamefont {J.~R.}\ \bibnamefont {Brownstein}}, \bibinfo {author} {\bibfnamefont {A.}~\bibnamefont {Burden}}, \bibinfo {author} {\bibfnamefont {C.-H.}\ \bibnamefont {Chuang}}, \emph {et~al.},\ }\href {https://doi.org/10.1093/mnras/stx721} {\bibfield  {journal} {\bibinfo  {journal} {Monthly Notices of the Royal Astronomical Society}\ }\textbf {\bibinfo {volume} {470}},\ \bibinfo {pages} {2617} (\bibinfo {year} {2017})},\ \Eprint {https://arxiv.org/abs/1607.03155} {arXiv:1607.03155} \BibitemShut {NoStop}%
\bibitem [{\citenamefont {Bautista}\ \emph {et~al.}(2020)\citenamefont {Bautista}, \citenamefont {Paviot}, \citenamefont {Maga{\~n}a}, \citenamefont {{de la Torre}}, \citenamefont {Fromenteau}, \citenamefont {{Gil-Mar{\'i}n}}, \citenamefont {Ross}, \citenamefont {Burtin}, \citenamefont {Dawson}, \citenamefont {Hou} \emph {et~al.}}]{bautista_completed_2020}%
  \BibitemOpen
  \bibfield  {author} {\bibinfo {author} {\bibfnamefont {J.~E.}\ \bibnamefont {Bautista}}, \bibinfo {author} {\bibfnamefont {R.}~\bibnamefont {Paviot}}, \bibinfo {author} {\bibfnamefont {M.~V.}\ \bibnamefont {Maga{\~n}a}}, \bibinfo {author} {\bibfnamefont {S.}~\bibnamefont {{de la Torre}}}, \bibinfo {author} {\bibfnamefont {S.}~\bibnamefont {Fromenteau}}, \bibinfo {author} {\bibfnamefont {H.}~\bibnamefont {{Gil-Mar{\'i}n}}}, \bibinfo {author} {\bibfnamefont {A.~J.}\ \bibnamefont {Ross}}, \bibinfo {author} {\bibfnamefont {E.}~\bibnamefont {Burtin}}, \bibinfo {author} {\bibfnamefont {K.~S.}\ \bibnamefont {Dawson}}, \bibinfo {author} {\bibfnamefont {J.}~\bibnamefont {Hou}}, \emph {et~al.},\ }\href {https://doi.org/10.1093/mnras/staa2800} {\bibfield  {journal} {\bibinfo  {journal} {Monthly Notices of the Royal Astronomical Society}\ }\textbf {\bibinfo {volume} {500}},\ \bibinfo {pages} {736} (\bibinfo {year} {2020})},\ \Eprint {https://arxiv.org/abs/2007.08993} {arXiv:2007.08993} \BibitemShut {NoStop}%
\bibitem [{\citenamefont {{Gil-Mar{\'i}n}}\ \emph {et~al.}(2020)\citenamefont {{Gil-Mar{\'i}n}}, \citenamefont {Bautista}, \citenamefont {Paviot}, \citenamefont {{Vargas-Maga{\~n}a}}, \citenamefont {{de la Torre}}, \citenamefont {Fromenteau}, \citenamefont {Alam}, \citenamefont {{\'A}vila}, \citenamefont {Burtin}, \citenamefont {Chuang} \emph {et~al.}}]{gil-marin_completed_2020}%
  \BibitemOpen
  \bibfield  {author} {\bibinfo {author} {\bibfnamefont {H.}~\bibnamefont {{Gil-Mar{\'i}n}}}, \bibinfo {author} {\bibfnamefont {J.~E.}\ \bibnamefont {Bautista}}, \bibinfo {author} {\bibfnamefont {R.}~\bibnamefont {Paviot}}, \bibinfo {author} {\bibfnamefont {M.}~\bibnamefont {{Vargas-Maga{\~n}a}}}, \bibinfo {author} {\bibfnamefont {S.}~\bibnamefont {{de la Torre}}}, \bibinfo {author} {\bibfnamefont {S.}~\bibnamefont {Fromenteau}}, \bibinfo {author} {\bibfnamefont {S.}~\bibnamefont {Alam}}, \bibinfo {author} {\bibfnamefont {S.}~\bibnamefont {{\'A}vila}}, \bibinfo {author} {\bibfnamefont {E.}~\bibnamefont {Burtin}}, \bibinfo {author} {\bibfnamefont {C.-H.}\ \bibnamefont {Chuang}}, \emph {et~al.},\ }\href {https://doi.org/10.1093/mnras/staa2455} {\bibfield  {journal} {\bibinfo  {journal} {Monthly Notices of the Royal Astronomical Society}\ }\textbf {\bibinfo {volume} {498}},\ \bibinfo {pages} {2492} (\bibinfo {year} {2020})},\ \Eprint {https://arxiv.org/abs/2007.08994} {arXiv:2007.08994} \BibitemShut {NoStop}%
\bibitem [{\citenamefont {Hou}\ \emph {et~al.}(2020)\citenamefont {Hou}, \citenamefont {S{\'a}nchez}, \citenamefont {Ross}, \citenamefont {Smith}, \citenamefont {Neveux}, \citenamefont {Bautista}, \citenamefont {Burtin}, \citenamefont {Zhao}, \citenamefont {Scoccimarro}, \citenamefont {Dawson} \emph {et~al.}}]{hou_completed_2020}%
  \BibitemOpen
  \bibfield  {author} {\bibinfo {author} {\bibfnamefont {J.}~\bibnamefont {Hou}}, \bibinfo {author} {\bibfnamefont {A.~G.}\ \bibnamefont {S{\'a}nchez}}, \bibinfo {author} {\bibfnamefont {A.~J.}\ \bibnamefont {Ross}}, \bibinfo {author} {\bibfnamefont {A.}~\bibnamefont {Smith}}, \bibinfo {author} {\bibfnamefont {R.}~\bibnamefont {Neveux}}, \bibinfo {author} {\bibfnamefont {J.}~\bibnamefont {Bautista}}, \bibinfo {author} {\bibfnamefont {E.}~\bibnamefont {Burtin}}, \bibinfo {author} {\bibfnamefont {C.}~\bibnamefont {Zhao}}, \bibinfo {author} {\bibfnamefont {R.}~\bibnamefont {Scoccimarro}}, \bibinfo {author} {\bibfnamefont {K.~S.}\ \bibnamefont {Dawson}}, \emph {et~al.},\ }\href {https://doi.org/10.1093/mnras/staa3234} {\bibfield  {journal} {\bibinfo  {journal} {Monthly Notices of the Royal Astronomical Society}\ }\textbf {\bibinfo {volume} {500}},\ \bibinfo {pages} {1201} (\bibinfo {year} {2020})},\ \Eprint {https://arxiv.org/abs/2007.08998} {arXiv:2007.08998} \BibitemShut {NoStop}%
\bibitem [{\citenamefont {Neveux}\ \emph {et~al.}(2020)\citenamefont {Neveux}, \citenamefont {Burtin}, \citenamefont {{de Mattia}}, \citenamefont {Smith}, \citenamefont {Ross}, \citenamefont {Hou}, \citenamefont {Bautista}, \citenamefont {Brinkmann}, \citenamefont {Chuang}, \citenamefont {Dawson} \emph {et~al.}}]{neveux_completed_2020}%
  \BibitemOpen
  \bibfield  {author} {\bibinfo {author} {\bibfnamefont {R.}~\bibnamefont {Neveux}}, \bibinfo {author} {\bibfnamefont {E.}~\bibnamefont {Burtin}}, \bibinfo {author} {\bibfnamefont {A.}~\bibnamefont {{de Mattia}}}, \bibinfo {author} {\bibfnamefont {A.}~\bibnamefont {Smith}}, \bibinfo {author} {\bibfnamefont {A.~J.}\ \bibnamefont {Ross}}, \bibinfo {author} {\bibfnamefont {J.}~\bibnamefont {Hou}}, \bibinfo {author} {\bibfnamefont {J.}~\bibnamefont {Bautista}}, \bibinfo {author} {\bibfnamefont {J.}~\bibnamefont {Brinkmann}}, \bibinfo {author} {\bibfnamefont {C.-H.}\ \bibnamefont {Chuang}}, \bibinfo {author} {\bibfnamefont {K.~S.}\ \bibnamefont {Dawson}}, \emph {et~al.},\ }\href {https://doi.org/10.1093/mnras/staa2780} {\bibfield  {journal} {\bibinfo  {journal} {Monthly Notices of the Royal Astronomical Society}\ }\textbf {\bibinfo {volume} {499}},\ \bibinfo {pages} {210} (\bibinfo {year} {2020})},\ \Eprint {https://arxiv.org/abs/2007.08999} {arXiv:2007.08999} \BibitemShut {NoStop}%
\bibitem [{\citenamefont {des Bourboux}\ \emph {et~al.}(2020)\citenamefont {des Bourboux}, \citenamefont {Rich}, \citenamefont {{Font-Ribera}}, \citenamefont {Agathe}, \citenamefont {Farr}, \citenamefont {Etourneau}, \citenamefont {Goff}, \citenamefont {Cuceu}, \citenamefont {Balland}, \citenamefont {Bautista} \emph {et~al.}}]{bourboux_completed_2020}%
  \BibitemOpen
  \bibfield  {author} {\bibinfo {author} {\bibfnamefont {H.~d.~M.}\ \bibnamefont {des Bourboux}}, \bibinfo {author} {\bibfnamefont {J.}~\bibnamefont {Rich}}, \bibinfo {author} {\bibfnamefont {A.}~\bibnamefont {{Font-Ribera}}}, \bibinfo {author} {\bibfnamefont {V.~d.~S.}\ \bibnamefont {Agathe}}, \bibinfo {author} {\bibfnamefont {J.}~\bibnamefont {Farr}}, \bibinfo {author} {\bibfnamefont {T.}~\bibnamefont {Etourneau}}, \bibinfo {author} {\bibfnamefont {J.-M.~L.}\ \bibnamefont {Goff}}, \bibinfo {author} {\bibfnamefont {A.}~\bibnamefont {Cuceu}}, \bibinfo {author} {\bibfnamefont {C.}~\bibnamefont {Balland}}, \bibinfo {author} {\bibfnamefont {J.~E.}\ \bibnamefont {Bautista}}, \emph {et~al.},\ }\href {https://doi.org/10.3847/1538-4357/abb085} {\bibfield  {journal} {\bibinfo  {journal} {ApJ}\ }\textbf {\bibinfo {volume} {901}},\ \bibinfo {pages} {153} (\bibinfo {year} {2020})},\ \Eprint {https://arxiv.org/abs/2007.08995} {arXiv:2007.08995} \BibitemShut {NoStop}%
\bibitem [{\citenamefont {{eBOSS Collaboration}}\ \emph {et~al.}(2021)\citenamefont {{eBOSS Collaboration}}, \citenamefont {Alam} \emph {et~al.}}]{eboss_collaboration_completed_2021-2}%
  \BibitemOpen
  \bibfield  {author} {\bibinfo {author} {\bibnamefont {{eBOSS Collaboration}}}, \bibinfo {author} {\bibfnamefont {S.}~\bibnamefont {Alam}}, \emph {et~al.},\ }\href {https://doi.org/10.1103/PhysRevD.103.083533} {\bibfield  {journal} {\bibinfo  {journal} {Phys. Rev. D}\ }\textbf {\bibinfo {volume} {103}},\ \bibinfo {pages} {083533} (\bibinfo {year} {2021})},\ \Eprint {https://arxiv.org/abs/2007.08991} {arXiv:2007.08991} \BibitemShut {NoStop}%
\bibitem [{\citenamefont {Brout}\ \emph {et~al.}(2022)\citenamefont {Brout}, \citenamefont {Scolnic}, \citenamefont {Popovic}, \citenamefont {Riess}, \citenamefont {Zuntz}, \citenamefont {Kessler}, \citenamefont {Carr}, \citenamefont {Davis}, \citenamefont {Hinton}, \citenamefont {Jones}, \citenamefont {Kenworthy} \emph {et~al.}}]{brout_pantheon_2022}%
  \BibitemOpen
  \bibfield  {author} {\bibinfo {author} {\bibfnamefont {D.}~\bibnamefont {Brout}}, \bibinfo {author} {\bibfnamefont {D.}~\bibnamefont {Scolnic}}, \bibinfo {author} {\bibfnamefont {B.}~\bibnamefont {Popovic}}, \bibinfo {author} {\bibfnamefont {A.~G.}\ \bibnamefont {Riess}}, \bibinfo {author} {\bibfnamefont {J.}~\bibnamefont {Zuntz}}, \bibinfo {author} {\bibfnamefont {R.}~\bibnamefont {Kessler}}, \bibinfo {author} {\bibfnamefont {A.}~\bibnamefont {Carr}}, \bibinfo {author} {\bibfnamefont {T.~M.}\ \bibnamefont {Davis}}, \bibinfo {author} {\bibfnamefont {S.}~\bibnamefont {Hinton}}, \bibinfo {author} {\bibfnamefont {D.}~\bibnamefont {Jones}}, \bibinfo {author} {\bibfnamefont {W.~D.}\ \bibnamefont {Kenworthy}}, \emph {et~al.},\ }\href {https://doi.org/10.3847/1538-4357/ac8e04} {\bibfield  {journal} {\bibinfo  {journal} {ApJ}\ }\textbf {\bibinfo {volume} {938}},\ \bibinfo {pages} {110} (\bibinfo {year} {2022})},\ \Eprint {https://arxiv.org/abs/2202.04077} {arXiv:2202.04077} \BibitemShut {NoStop}%
\bibitem [{\citenamefont {Scolnic}\ \emph {et~al.}(2022)\citenamefont {Scolnic}, \citenamefont {Brout}, \citenamefont {Carr}, \citenamefont {Riess}, \citenamefont {Davis}, \citenamefont {Dwomoh}, \citenamefont {Jones}, \citenamefont {Ali}, \citenamefont {Charvu}, \citenamefont {Chen} \emph {et~al.}}]{scolnic_pantheon_2022}%
  \BibitemOpen
  \bibfield  {author} {\bibinfo {author} {\bibfnamefont {D.}~\bibnamefont {Scolnic}}, \bibinfo {author} {\bibfnamefont {D.}~\bibnamefont {Brout}}, \bibinfo {author} {\bibfnamefont {A.}~\bibnamefont {Carr}}, \bibinfo {author} {\bibfnamefont {A.~G.}\ \bibnamefont {Riess}}, \bibinfo {author} {\bibfnamefont {T.~M.}\ \bibnamefont {Davis}}, \bibinfo {author} {\bibfnamefont {A.}~\bibnamefont {Dwomoh}}, \bibinfo {author} {\bibfnamefont {D.~O.}\ \bibnamefont {Jones}}, \bibinfo {author} {\bibfnamefont {N.}~\bibnamefont {Ali}}, \bibinfo {author} {\bibfnamefont {P.}~\bibnamefont {Charvu}}, \bibinfo {author} {\bibfnamefont {R.}~\bibnamefont {Chen}}, \emph {et~al.},\ }\href {https://doi.org/10.3847/1538-4357/ac8b7a} {\bibfield  {journal} {\bibinfo  {journal} {ApJ}\ }\textbf {\bibinfo {volume} {938}},\ \bibinfo {pages} {113} (\bibinfo {year} {2022})},\ \Eprint {https://arxiv.org/abs/2112.03863} {arXiv:2112.03863} \BibitemShut {NoStop}%
\bibitem [{\citenamefont {Cooke}\ \emph {et~al.}(2018)\citenamefont {Cooke}, \citenamefont {Pettini},\ and\ \citenamefont {Steidel}}]{cooke_one_2018}%
  \BibitemOpen
  \bibfield  {author} {\bibinfo {author} {\bibfnamefont {R.}~\bibnamefont {Cooke}}, \bibinfo {author} {\bibfnamefont {M.}~\bibnamefont {Pettini}},\ and\ \bibinfo {author} {\bibfnamefont {C.~C.}\ \bibnamefont {Steidel}},\ }\href {https://doi.org/10.3847/1538-4357/aaab53} {\bibfield  {journal} {\bibinfo  {journal} {ApJ}\ }\textbf {\bibinfo {volume} {855}},\ \bibinfo {pages} {102} (\bibinfo {year} {2018})},\ \Eprint {https://arxiv.org/abs/1710.11129} {arXiv:1710.11129} \BibitemShut {NoStop}%
\bibitem [{\citenamefont {Chown}\ \emph {et~al.}(2018)\citenamefont {Chown}, \citenamefont {Omori}, \citenamefont {Aylor}, \citenamefont {Benson}, \citenamefont {Bleem}, \citenamefont {Carlstrom}, \citenamefont {Chang}, \citenamefont {Cho}, \citenamefont {Crawford}, \citenamefont {Crites} \emph {et~al.}}]{chown_maps_2018}%
  \BibitemOpen
  \bibfield  {author} {\bibinfo {author} {\bibfnamefont {R.}~\bibnamefont {Chown}}, \bibinfo {author} {\bibfnamefont {Y.}~\bibnamefont {Omori}}, \bibinfo {author} {\bibfnamefont {K.}~\bibnamefont {Aylor}}, \bibinfo {author} {\bibfnamefont {B.~A.}\ \bibnamefont {Benson}}, \bibinfo {author} {\bibfnamefont {L.~E.}\ \bibnamefont {Bleem}}, \bibinfo {author} {\bibfnamefont {J.~E.}\ \bibnamefont {Carlstrom}}, \bibinfo {author} {\bibfnamefont {C.~L.}\ \bibnamefont {Chang}}, \bibinfo {author} {\bibfnamefont {H.-M.}\ \bibnamefont {Cho}}, \bibinfo {author} {\bibfnamefont {T.}~\bibnamefont {Crawford}}, \bibinfo {author} {\bibfnamefont {A.~T.}\ \bibnamefont {Crites}}, \emph {et~al.},\ }\href {https://doi.org/10.3847/1538-4365/aae694} {\bibfield  {journal} {\bibinfo  {journal} {ApJS}\ }\textbf {\bibinfo {volume} {239}},\ \bibinfo {pages} {10} (\bibinfo {year} {2018})},\ \Eprint {https://arxiv.org/abs/1803.10682} {arXiv:1803.10682} \BibitemShut {NoStop}%
\bibitem [{\citenamefont {Dutcher}\ \emph {et~al.}(2021)\citenamefont {Dutcher}, \citenamefont {Balkenhol}, \citenamefont {Ade}, \citenamefont {Ahmed}, \citenamefont {Anderes}, \citenamefont {Anderson}, \citenamefont {Archipley}, \citenamefont {Avva}, \citenamefont {Aylor}, \citenamefont {Barry} \emph {et~al.}}]{dutcher_measurements_2021}%
  \BibitemOpen
  \bibfield  {author} {\bibinfo {author} {\bibfnamefont {D.}~\bibnamefont {Dutcher}}, \bibinfo {author} {\bibfnamefont {L.}~\bibnamefont {Balkenhol}}, \bibinfo {author} {\bibfnamefont {P.~A.~R.}\ \bibnamefont {Ade}}, \bibinfo {author} {\bibfnamefont {Z.}~\bibnamefont {Ahmed}}, \bibinfo {author} {\bibfnamefont {E.}~\bibnamefont {Anderes}}, \bibinfo {author} {\bibfnamefont {A.~J.}\ \bibnamefont {Anderson}}, \bibinfo {author} {\bibfnamefont {M.}~\bibnamefont {Archipley}}, \bibinfo {author} {\bibfnamefont {J.~S.}\ \bibnamefont {Avva}}, \bibinfo {author} {\bibfnamefont {K.}~\bibnamefont {Aylor}}, \bibinfo {author} {\bibfnamefont {P.~S.}\ \bibnamefont {Barry}}, \emph {et~al.},\ }\href {https://doi.org/10.1103/PhysRevD.104.022003} {\bibfield  {journal} {\bibinfo  {journal} {Phys. Rev. D}\ }\textbf {\bibinfo {volume} {104}},\ \bibinfo {pages} {022003} (\bibinfo {year} {2021})},\ \Eprint {https://arxiv.org/abs/2101.01684} {arXiv:2101.01684} \BibitemShut {NoStop}%
\bibitem [{\citenamefont {Balkenhol}\ \emph {et~al.}(2023)\citenamefont {Balkenhol}, \citenamefont {Dutcher}, \citenamefont {Mancini}, \citenamefont {Doussot}, \citenamefont {Benabed}, \citenamefont {Galli}, \citenamefont {Ade}, \citenamefont {Anderson}, \citenamefont {Ansarinejad} \emph {et~al.}}]{balkenhol_measurement_2022}%
  \BibitemOpen
  \bibfield  {author} {\bibinfo {author} {\bibfnamefont {L.}~\bibnamefont {Balkenhol}}, \bibinfo {author} {\bibfnamefont {D.}~\bibnamefont {Dutcher}}, \bibinfo {author} {\bibfnamefont {A.~S.}\ \bibnamefont {Mancini}}, \bibinfo {author} {\bibfnamefont {A.}~\bibnamefont {Doussot}}, \bibinfo {author} {\bibfnamefont {K.}~\bibnamefont {Benabed}}, \bibinfo {author} {\bibfnamefont {S.}~\bibnamefont {Galli}}, \bibinfo {author} {\bibfnamefont {P.~A.~R.}\ \bibnamefont {Ade}}, \bibinfo {author} {\bibfnamefont {A.~J.}\ \bibnamefont {Anderson}}, \bibinfo {author} {\bibfnamefont {B.}~\bibnamefont {Ansarinejad}}, \emph {et~al.},\ }\href {https://doi.org/10.1103/PhysRevD.108.023510} {\bibfield  {journal} {\bibinfo  {journal} {\prd}\ }\textbf {\bibinfo {volume} {108}},\ \bibinfo {eid} {023510} (\bibinfo {year} {2023})},\ \Eprint {https://arxiv.org/abs/2212.05642} {arXiv:2212.05642} \BibitemShut {NoStop}%
\bibitem [{\citenamefont {{Smith}}\ \emph {et~al.}(2021)\citenamefont {{Smith}}, \citenamefont {{Poulin}}, \citenamefont {{Bernal}}, \citenamefont {{Boddy}}, \citenamefont {{Kamionkowski}},\ and\ \citenamefont {{Murgia}}}]{Smith_2021}%
  \BibitemOpen
  \bibfield  {author} {\bibinfo {author} {\bibfnamefont {T.~L.}\ \bibnamefont {{Smith}}}, \bibinfo {author} {\bibfnamefont {V.}~\bibnamefont {{Poulin}}}, \bibinfo {author} {\bibfnamefont {J.~L.}\ \bibnamefont {{Bernal}}}, \bibinfo {author} {\bibfnamefont {K.~K.}\ \bibnamefont {{Boddy}}}, \bibinfo {author} {\bibfnamefont {M.}~\bibnamefont {{Kamionkowski}}},\ and\ \bibinfo {author} {\bibfnamefont {R.}~\bibnamefont {{Murgia}}},\ }\href {https://doi.org/10.1103/PhysRevD.103.123542} {\bibfield  {journal} {\bibinfo  {journal} {\prd}\ }\textbf {\bibinfo {volume} {103}},\ \bibinfo {eid} {123542} (\bibinfo {year} {2021})},\ \Eprint {https://arxiv.org/abs/2009.10740} {arXiv:2009.10740 [astro-ph.CO]} \BibitemShut {NoStop}%
\bibitem [{\citenamefont {{G{\'o}mez-Valent}}(2022)}]{Gomez_Valent_2022}%
  \BibitemOpen
  \bibfield  {author} {\bibinfo {author} {\bibfnamefont {A.}~\bibnamefont {{G{\'o}mez-Valent}}},\ }\href {https://doi.org/10.1103/PhysRevD.106.063506} {\bibfield  {journal} {\bibinfo  {journal} {\prd}\ }\textbf {\bibinfo {volume} {106}},\ \bibinfo {eid} {063506} (\bibinfo {year} {2022})},\ \Eprint {https://arxiv.org/abs/2203.16285} {arXiv:2203.16285 [astro-ph.CO]} \BibitemShut {NoStop}%
\bibitem [{\citenamefont {{Herold}}\ \emph {et~al.}(2022)\citenamefont {{Herold}}, \citenamefont {{Ferreira}},\ and\ \citenamefont {{Komatsu}}}]{Herold_2022}%
  \BibitemOpen
  \bibfield  {author} {\bibinfo {author} {\bibfnamefont {L.}~\bibnamefont {{Herold}}}, \bibinfo {author} {\bibfnamefont {E.~G.~M.}\ \bibnamefont {{Ferreira}}},\ and\ \bibinfo {author} {\bibfnamefont {E.}~\bibnamefont {{Komatsu}}},\ }\href {https://doi.org/10.3847/2041-8213/ac63a3} {\bibfield  {journal} {\bibinfo  {journal} {The Astrophysical Journal Letters}\ }\textbf {\bibinfo {volume} {929}},\ \bibinfo {eid} {L16} (\bibinfo {year} {2022})},\ \Eprint {https://arxiv.org/abs/2112.12140} {arXiv:2112.12140 [astro-ph.CO]} \BibitemShut {NoStop}%
\bibitem [{\citenamefont {Karwal}\ \emph {et~al.}(2024)\citenamefont {Karwal}, \citenamefont {Patel}, \citenamefont {Bartlett}, \citenamefont {Poulin}, \citenamefont {Smith},\ and\ \citenamefont {Pfeffer}}]{Karwal:2024qpt}%
  \BibitemOpen
  \bibfield  {author} {\bibinfo {author} {\bibfnamefont {T.}~\bibnamefont {Karwal}}, \bibinfo {author} {\bibfnamefont {Y.}~\bibnamefont {Patel}}, \bibinfo {author} {\bibfnamefont {A.}~\bibnamefont {Bartlett}}, \bibinfo {author} {\bibfnamefont {V.}~\bibnamefont {Poulin}}, \bibinfo {author} {\bibfnamefont {T.~L.}\ \bibnamefont {Smith}},\ and\ \bibinfo {author} {\bibfnamefont {D.~N.}\ \bibnamefont {Pfeffer}},\ }\href@noop {} {\bibfield  {journal} {\bibinfo  {journal} {arXiv e-prints}\ } (\bibinfo {year} {2024})},\ \Eprint {https://arxiv.org/abs/2401.14225} {arXiv:2401.14225 [astro-ph.CO]} \BibitemShut {NoStop}%
\bibitem [{\citenamefont {Gilman}\ \emph {et~al.}(2022)\citenamefont {Gilman}, \citenamefont {Benson}, \citenamefont {Bovy}, \citenamefont {Birrer}, \citenamefont {Treu},\ and\ \citenamefont {Nierenberg}}]{gilman_primordial_2022}%
  \BibitemOpen
  \bibfield  {author} {\bibinfo {author} {\bibfnamefont {D.}~\bibnamefont {Gilman}}, \bibinfo {author} {\bibfnamefont {A.}~\bibnamefont {Benson}}, \bibinfo {author} {\bibfnamefont {J.}~\bibnamefont {Bovy}}, \bibinfo {author} {\bibfnamefont {S.}~\bibnamefont {Birrer}}, \bibinfo {author} {\bibfnamefont {T.}~\bibnamefont {Treu}},\ and\ \bibinfo {author} {\bibfnamefont {A.}~\bibnamefont {Nierenberg}},\ }\href {https://doi.org/10.1093/mnras/stac670} {\bibfield  {journal} {\bibinfo  {journal} {Monthly Notices of the Royal Astronomical Society}\ }\textbf {\bibinfo {volume} {512}},\ \bibinfo {pages} {3163} (\bibinfo {year} {2022})},\ \Eprint {https://arxiv.org/abs/2112.03293} {arXiv:2112.03293} \BibitemShut {NoStop}%
\bibitem [{\citenamefont {Dekker}\ and\ \citenamefont {Kravtsov}(2024)}]{dekker_constraints_2024}%
  \BibitemOpen
  \bibfield  {author} {\bibinfo {author} {\bibfnamefont {A.}~\bibnamefont {Dekker}}\ and\ \bibinfo {author} {\bibfnamefont {A.}~\bibnamefont {Kravtsov}},\ }\bibfield  {journal} {\bibinfo  {journal} {arXiv e-prints}\ }\href {https://doi.org/10.48550/arXiv.2407.04198} {10.48550/arXiv.2407.04198} (\bibinfo {year} {2024}),\ \Eprint {https://arxiv.org/abs/2407.04198} {arXiv:2407.04198} \BibitemShut {NoStop}%
\bibitem [{\citenamefont {Nadler}\ \emph {et~al.}(2019)\citenamefont {Nadler}, \citenamefont {Gluscevic}, \citenamefont {Boddy},\ and\ \citenamefont {Wechsler}}]{nadler_constraints_2019}%
  \BibitemOpen
  \bibfield  {author} {\bibinfo {author} {\bibfnamefont {E.~O.}\ \bibnamefont {Nadler}}, \bibinfo {author} {\bibfnamefont {V.}~\bibnamefont {Gluscevic}}, \bibinfo {author} {\bibfnamefont {K.~K.}\ \bibnamefont {Boddy}},\ and\ \bibinfo {author} {\bibfnamefont {R.~H.}\ \bibnamefont {Wechsler}},\ }\href {https://doi.org/10.3847/2041-8213/ab1eb2} {\bibfield  {journal} {\bibinfo  {journal} {ApJL}\ }\textbf {\bibinfo {volume} {878}},\ \bibinfo {pages} {L32} (\bibinfo {year} {2019})},\ \Eprint {https://arxiv.org/abs/1904.10000} {arXiv:1904.10000} \BibitemShut {NoStop}%
\bibitem [{\citenamefont {Nadler}\ \emph {et~al.}(2021)\citenamefont {Nadler}, \citenamefont {{Drlica-Wagner}}, \citenamefont {Bechtol}, \citenamefont {Mau}, \citenamefont {Wechsler}, \citenamefont {Gluscevic}, \citenamefont {Boddy}, \citenamefont {Pace}, \citenamefont {Li}, \citenamefont {McNanna} \emph {et~al.}}]{nadler_milky_2021}%
  \BibitemOpen
  \bibfield  {author} {\bibinfo {author} {\bibfnamefont {E.~O.}\ \bibnamefont {Nadler}}, \bibinfo {author} {\bibfnamefont {A.}~\bibnamefont {{Drlica-Wagner}}}, \bibinfo {author} {\bibfnamefont {K.}~\bibnamefont {Bechtol}}, \bibinfo {author} {\bibfnamefont {S.}~\bibnamefont {Mau}}, \bibinfo {author} {\bibfnamefont {R.~H.}\ \bibnamefont {Wechsler}}, \bibinfo {author} {\bibfnamefont {V.}~\bibnamefont {Gluscevic}}, \bibinfo {author} {\bibfnamefont {K.}~\bibnamefont {Boddy}}, \bibinfo {author} {\bibfnamefont {A.~B.}\ \bibnamefont {Pace}}, \bibinfo {author} {\bibfnamefont {T.~S.}\ \bibnamefont {Li}}, \bibinfo {author} {\bibfnamefont {M.}~\bibnamefont {McNanna}}, \emph {et~al.},\ }\href {https://doi.org/10.1103/PhysRevLett.126.091101} {\bibfield  {journal} {\bibinfo  {journal} {Phys. Rev. Lett.}\ }\textbf {\bibinfo {volume} {126}},\ \bibinfo {pages} {091101} (\bibinfo {year} {2021})},\ \Eprint {https://arxiv.org/abs/2008.00022} {arXiv:2008.00022} \BibitemShut {NoStop}%
\bibitem [{\citenamefont {{Press}}\ and\ \citenamefont {{Schechter}}(1974)}]{press-schechter}%
  \BibitemOpen
  \bibfield  {author} {\bibinfo {author} {\bibfnamefont {W.~H.}\ \bibnamefont {{Press}}}\ and\ \bibinfo {author} {\bibfnamefont {P.}~\bibnamefont {{Schechter}}},\ }\href {https://doi.org/10.1086/152650} {\bibfield  {journal} {\bibinfo  {journal} {\apj}\ }\textbf {\bibinfo {volume} {187}},\ \bibinfo {pages} {425} (\bibinfo {year} {1974})}\BibitemShut {NoStop}%
\bibitem [{\citenamefont {Gomez-Navarro}\ \emph {et~al.}(2023)\citenamefont {Gomez-Navarro}, \citenamefont {Aviles},\ and\ \citenamefont {de~la Macorra}}]{gomez-navarro_impact_2023}%
  \BibitemOpen
  \bibfield  {author} {\bibinfo {author} {\bibfnamefont {D.~V.}\ \bibnamefont {Gomez-Navarro}}, \bibinfo {author} {\bibfnamefont {A.}~\bibnamefont {Aviles}},\ and\ \bibinfo {author} {\bibfnamefont {A.}~\bibnamefont {de~la Macorra}},\ }\bibfield  {journal} {\bibinfo  {journal} {Physical Review D}\ }\textbf {\bibinfo {volume} {108}},\ \href {https://doi.org/10.1103/physrevd.108.043541} {10.1103/physrevd.108.043541} (\bibinfo {year} {2023})\BibitemShut {NoStop}%
\bibitem [{\citenamefont {Ivezi{\'c}}\ \emph {et~al.}(2019)\citenamefont {Ivezi{\'c}}, \citenamefont {Kahn}, \citenamefont {Tyson}, \citenamefont {Abel}, \citenamefont {Acosta}, \citenamefont {Allsman}, \citenamefont {Alonso}, \citenamefont {AlSayyad}, \citenamefont {Anderson}, \citenamefont {Andrew} \emph {et~al.}}]{ivezic_lsst_2019}%
  \BibitemOpen
  \bibfield  {author} {\bibinfo {author} {\bibfnamefont {{\v Z}.}~\bibnamefont {Ivezi{\'c}}}, \bibinfo {author} {\bibfnamefont {S.~M.}\ \bibnamefont {Kahn}}, \bibinfo {author} {\bibfnamefont {J.~A.}\ \bibnamefont {Tyson}}, \bibinfo {author} {\bibfnamefont {B.}~\bibnamefont {Abel}}, \bibinfo {author} {\bibfnamefont {E.}~\bibnamefont {Acosta}}, \bibinfo {author} {\bibfnamefont {R.}~\bibnamefont {Allsman}}, \bibinfo {author} {\bibfnamefont {D.}~\bibnamefont {Alonso}}, \bibinfo {author} {\bibfnamefont {Y.}~\bibnamefont {AlSayyad}}, \bibinfo {author} {\bibfnamefont {S.~F.}\ \bibnamefont {Anderson}}, \bibinfo {author} {\bibfnamefont {J.}~\bibnamefont {Andrew}}, \emph {et~al.},\ }\href {https://doi.org/10.3847/1538-4357/ab042c} {\bibfield  {journal} {\bibinfo  {journal} {ApJ}\ }\textbf {\bibinfo {volume} {873}},\ \bibinfo {pages} {111} (\bibinfo {year} {2019})},\ \Eprint {https://arxiv.org/abs/0805.2366} {arXiv:0805.2366} \BibitemShut {NoStop}%
\bibitem [{\citenamefont {Nadler}\ \emph {et~al.}(2024)\citenamefont {Nadler}, \citenamefont {Gluscevic}, \citenamefont {Driskell}, \citenamefont {Wechsler}, \citenamefont {Moustakas}, \citenamefont {Benson},\ and\ \citenamefont {Mao}}]{nadler_forecasts_2024}%
  \BibitemOpen
  \bibfield  {author} {\bibinfo {author} {\bibfnamefont {E.~O.}\ \bibnamefont {Nadler}}, \bibinfo {author} {\bibfnamefont {V.}~\bibnamefont {Gluscevic}}, \bibinfo {author} {\bibfnamefont {T.}~\bibnamefont {Driskell}}, \bibinfo {author} {\bibfnamefont {R.~H.}\ \bibnamefont {Wechsler}}, \bibinfo {author} {\bibfnamefont {L.~A.}\ \bibnamefont {Moustakas}}, \bibinfo {author} {\bibfnamefont {A.}~\bibnamefont {Benson}},\ and\ \bibinfo {author} {\bibfnamefont {Y.-Y.}\ \bibnamefont {Mao}},\ }\href {https://doi.org/10.3847/1538-4357/ad3bb1} {\bibfield  {journal} {\bibinfo  {journal} {ApJ}\ }\textbf {\bibinfo {volume} {967}},\ \bibinfo {pages} {61} (\bibinfo {year} {2024})},\ \Eprint {https://arxiv.org/abs/2401.10318} {arXiv:2401.10318} \BibitemShut {NoStop}%
\bibitem [{\citenamefont {Banik}\ \emph {et~al.}(2021{\natexlab{a}})\citenamefont {Banik}, \citenamefont {Bovy}, \citenamefont {Bertone}, \citenamefont {Erkal},\ and\ \citenamefont {{de Boer}}}]{banik_evidence_2021}%
  \BibitemOpen
  \bibfield  {author} {\bibinfo {author} {\bibfnamefont {N.}~\bibnamefont {Banik}}, \bibinfo {author} {\bibfnamefont {J.}~\bibnamefont {Bovy}}, \bibinfo {author} {\bibfnamefont {G.}~\bibnamefont {Bertone}}, \bibinfo {author} {\bibfnamefont {D.}~\bibnamefont {Erkal}},\ and\ \bibinfo {author} {\bibfnamefont {T.~J.~L.}\ \bibnamefont {{de Boer}}},\ }\href {https://doi.org/10.1093/mnras/stab210} {\bibfield  {journal} {\bibinfo  {journal} {Monthly Notices of the Royal Astronomical Society}\ }\textbf {\bibinfo {volume} {502}},\ \bibinfo {pages} {2364} (\bibinfo {year} {2021}{\natexlab{a}})},\ \Eprint {https://arxiv.org/abs/1911.02662} {arXiv:1911.02662} \BibitemShut {NoStop}%
\bibitem [{\citenamefont {Banik}\ \emph {et~al.}(2021{\natexlab{b}})\citenamefont {Banik}, \citenamefont {Bovy}, \citenamefont {Bertone}, \citenamefont {Erkal},\ and\ \citenamefont {{de Boer}}}]{banik_novel_2021}%
  \BibitemOpen
  \bibfield  {author} {\bibinfo {author} {\bibfnamefont {N.}~\bibnamefont {Banik}}, \bibinfo {author} {\bibfnamefont {J.}~\bibnamefont {Bovy}}, \bibinfo {author} {\bibfnamefont {G.}~\bibnamefont {Bertone}}, \bibinfo {author} {\bibfnamefont {D.}~\bibnamefont {Erkal}},\ and\ \bibinfo {author} {\bibfnamefont {T.~J.~L.}\ \bibnamefont {{de Boer}}},\ }\href {https://doi.org/10.1088/1475-7516/2021/10/043} {\bibfield  {journal} {\bibinfo  {journal} {J. Cosmol. Astropart. Phys.}\ }\textbf {\bibinfo {volume} {2021}}\bibfield  {number} {\bibinfo  {number} { (10)},\ \bibinfo {pages} {043}},\ }\Eprint {https://arxiv.org/abs/1911.02663} {arXiv:1911.02663} \BibitemShut {NoStop}%
\bibitem [{\citenamefont {Bonaca}\ and\ \citenamefont {{Price-Whelan}}(2024)}]{bonaca_stellar_2024}%
  \BibitemOpen
  \bibfield  {author} {\bibinfo {author} {\bibfnamefont {A.}~\bibnamefont {Bonaca}}\ and\ \bibinfo {author} {\bibfnamefont {A.~M.}\ \bibnamefont {{Price-Whelan}}},\ }\bibfield  {journal} {\bibinfo  {journal} {arXiv e-prints}\ }\href {https://doi.org/10.48550/arXiv.2405.19410} {10.48550/arXiv.2405.19410} (\bibinfo {year} {2024}),\ \Eprint {https://arxiv.org/abs/2405.19410} {arXiv:2405.19410} \BibitemShut {NoStop}%
\bibitem [{\citenamefont {Aganze}\ \emph {et~al.}(2024)\citenamefont {Aganze}, \citenamefont {Pearson}, \citenamefont {Starkenburg}, \citenamefont {Contardo}, \citenamefont {Johnston}, \citenamefont {Tavangar}, \citenamefont {{Price-Whelan}},\ and\ \citenamefont {Burgasser}}]{aganze_prospects_2024}%
  \BibitemOpen
  \bibfield  {author} {\bibinfo {author} {\bibfnamefont {C.}~\bibnamefont {Aganze}}, \bibinfo {author} {\bibfnamefont {S.}~\bibnamefont {Pearson}}, \bibinfo {author} {\bibfnamefont {T.}~\bibnamefont {Starkenburg}}, \bibinfo {author} {\bibfnamefont {G.}~\bibnamefont {Contardo}}, \bibinfo {author} {\bibfnamefont {K.~V.}\ \bibnamefont {Johnston}}, \bibinfo {author} {\bibfnamefont {K.}~\bibnamefont {Tavangar}}, \bibinfo {author} {\bibfnamefont {A.~M.}\ \bibnamefont {{Price-Whelan}}},\ and\ \bibinfo {author} {\bibfnamefont {A.~J.}\ \bibnamefont {Burgasser}},\ }\href {https://doi.org/10.3847/1538-4357/ad159c} {\bibfield  {journal} {\bibinfo  {journal} {ApJ}\ }\textbf {\bibinfo {volume} {962}},\ \bibinfo {pages} {151} (\bibinfo {year} {2024})},\ \Eprint {https://arxiv.org/abs/2305.12045} {arXiv:2305.12045} \BibitemShut {NoStop}%
\bibitem [{\citenamefont {McDonald}\ \emph {et~al.}(2005)\citenamefont {McDonald}, \citenamefont {Seljak}, \citenamefont {Cen}, \citenamefont {Shih}, \citenamefont {Weinberg}, \citenamefont {Burles}, \citenamefont {Schneider}, \citenamefont {Schlegel}, \citenamefont {Bahcall}, \citenamefont {Briggs} \emph {et~al.}}]{mcdonald_linear_2005}%
  \BibitemOpen
  \bibfield  {author} {\bibinfo {author} {\bibfnamefont {P.}~\bibnamefont {McDonald}}, \bibinfo {author} {\bibfnamefont {U.}~\bibnamefont {Seljak}}, \bibinfo {author} {\bibfnamefont {R.}~\bibnamefont {Cen}}, \bibinfo {author} {\bibfnamefont {D.}~\bibnamefont {Shih}}, \bibinfo {author} {\bibfnamefont {D.~H.}\ \bibnamefont {Weinberg}}, \bibinfo {author} {\bibfnamefont {S.}~\bibnamefont {Burles}}, \bibinfo {author} {\bibfnamefont {D.~P.}\ \bibnamefont {Schneider}}, \bibinfo {author} {\bibfnamefont {D.~J.}\ \bibnamefont {Schlegel}}, \bibinfo {author} {\bibfnamefont {N.~A.}\ \bibnamefont {Bahcall}}, \bibinfo {author} {\bibfnamefont {J.~W.}\ \bibnamefont {Briggs}}, \emph {et~al.},\ }\href {https://doi.org/10.1086/497563} {\bibfield  {journal} {\bibinfo  {journal} {ApJ}\ }\textbf {\bibinfo {volume} {635}},\ \bibinfo {pages} {761} (\bibinfo {year} {2005})},\ \Eprint {https://arxiv.org/abs/astro-ph/0407377} {arXiv:astro-ph/0407377} \BibitemShut {NoStop}%
\bibitem [{\citenamefont {Pedersen}\ \emph {et~al.}(2023)\citenamefont {Pedersen}, \citenamefont {{Font-Ribera}},\ and\ \citenamefont {Gnedin}}]{pedersen_compressing_2023}%
  \BibitemOpen
  \bibfield  {author} {\bibinfo {author} {\bibfnamefont {C.}~\bibnamefont {Pedersen}}, \bibinfo {author} {\bibfnamefont {A.}~\bibnamefont {{Font-Ribera}}},\ and\ \bibinfo {author} {\bibfnamefont {N.~Y.}\ \bibnamefont {Gnedin}},\ }\href {https://doi.org/10.3847/1538-4357/acb433} {\bibfield  {journal} {\bibinfo  {journal} {ApJ}\ }\textbf {\bibinfo {volume} {944}},\ \bibinfo {pages} {223} (\bibinfo {year} {2023})},\ \Eprint {https://arxiv.org/abs/2209.09895} {arXiv:2209.09895} \BibitemShut {NoStop}%
\bibitem [{\citenamefont {{He}}\ \emph {et~al.}(2024)\citenamefont {{He}}, \citenamefont {{An}}, \citenamefont {{Ivanov}},\ and\ \citenamefont {{Gluscevic}}}]{he_self-interacting_2024}%
  \BibitemOpen
  \bibfield  {author} {\bibinfo {author} {\bibfnamefont {A.}~\bibnamefont {{He}}}, \bibinfo {author} {\bibfnamefont {R.}~\bibnamefont {{An}}}, \bibinfo {author} {\bibfnamefont {M.~M.}\ \bibnamefont {{Ivanov}}},\ and\ \bibinfo {author} {\bibfnamefont {V.}~\bibnamefont {{Gluscevic}}},\ }\href {https://doi.org/10.1103/PhysRevD.109.103527} {\bibfield  {journal} {\bibinfo  {journal} {\prd}\ }\textbf {\bibinfo {volume} {109}},\ \bibinfo {eid} {103527} (\bibinfo {year} {2024})},\ \Eprint {https://arxiv.org/abs/2309.03956} {arXiv:2309.03956} \BibitemShut {NoStop}%
\bibitem [{\citenamefont {{Mutlu-Pakdil}}\ \emph {et~al.}(2021)\citenamefont {{Mutlu-Pakdil}}, \citenamefont {{Sand}}, \citenamefont {{Crnojevi{\'c}}}, \citenamefont {{Drlica-Wagner}}, \citenamefont {{Caldwell}}, \citenamefont {{Guhathakurta}}, \citenamefont {{Seth}}, \citenamefont {{Simon}}, \citenamefont {{Strader}},\ and\ \citenamefont {{Toloba}}}]{Mutlu-Pakdil_resolved_2021}%
  \BibitemOpen
  \bibfield  {author} {\bibinfo {author} {\bibfnamefont {B.}~\bibnamefont {{Mutlu-Pakdil}}}, \bibinfo {author} {\bibfnamefont {D.~J.}\ \bibnamefont {{Sand}}}, \bibinfo {author} {\bibfnamefont {D.}~\bibnamefont {{Crnojevi{\'c}}}}, \bibinfo {author} {\bibfnamefont {A.}~\bibnamefont {{Drlica-Wagner}}}, \bibinfo {author} {\bibfnamefont {N.}~\bibnamefont {{Caldwell}}}, \bibinfo {author} {\bibfnamefont {P.}~\bibnamefont {{Guhathakurta}}}, \bibinfo {author} {\bibfnamefont {A.~C.}\ \bibnamefont {{Seth}}}, \bibinfo {author} {\bibfnamefont {J.~D.}\ \bibnamefont {{Simon}}}, \bibinfo {author} {\bibfnamefont {J.}~\bibnamefont {{Strader}}},\ and\ \bibinfo {author} {\bibfnamefont {E.}~\bibnamefont {{Toloba}}},\ }\href {https://doi.org/10.3847/1538-4357/ac0db8} {\bibfield  {journal} {\bibinfo  {journal} {\apj}\ }\textbf {\bibinfo {volume} {918}},\ \bibinfo {eid} {88} (\bibinfo {year} {2021})},\ \Eprint {https://arxiv.org/abs/2105.01658} {arXiv:2105.01658 [astro-ph.GA]} \BibitemShut {NoStop}%
\end{thebibliography}%

\end{document}